\documentclass[12pt]{article}
\usepackage{amssymb}
\usepackage{amsmath}
\usepackage{latexsym}
\usepackage[usenames]{color}
\usepackage{fancybox}
\usepackage{simplewick}
\usepackage{cite}
\usepackage{framed}
\definecolor{shadecolor}{rgb}{0.9,0.9,0.95}
\usepackage{setspace}
\usepackage{graphicx} 
\usepackage{epsfig}

\usepackage[setpagesize=false,pagebackref=false, linktocpage, bookmarksopen=true, colorlinks=true, linkcolor=blue,citecolor=blue,urlcolor=blue]{hyperref}
\def\ket#1{|#1\rangle}
\def\comma{\,,}
\def\period{\,.}
\begin{document}
\thispagestyle{empty}
\baselineskip 7ex
\topmargin 7cm
\begin{center}
{\Large {\bf Lectures on Three-point Functions in $\mathcal{N}=4$ Supersymmetric Yang-Mills Theory}}\vspace{2cm}\\
\baselineskip 5ex
{\large {\sc Shota Komatsu}}\\
{\it School of Natural Sciences, Institute for Advanced Study, Princeton, New Jersey USA} \vspace{1.5cm}\\
{\bf Based on lectures given in Les Houches Summer School 2016}\\ {\bf Integrability: From statistical systems to gauge theory}
\end{center}
\vspace{0.5cm}
\baselineskip 3.3ex
\parskip 1ex
\newpage
\topmargin -1.3cm
\begin{center}
{\bf Abstract}
\end{center}
This is a pedagogical review on the integrability-based approach to the three-point function in $\mathcal{N}=4$ supersymmetric Yang-Mills theory. We first discuss the computation of the structure constant at weak coupling and show that the result can be recast as a sum over partitions of the rapidities of the magnons. We then introduce a non-perturbative framework, called the ``hexagon approach'', and explain how one can use the symmetries (i.e.~superconformal and gauge symmetries) and integrability to determine the structure constants. This article is based on the lectures given in Les Houches Summer School ``Integrability: From statistical systems to gauge theory'' in June 2016. 
\tableofcontents
\newpage
\section{Preface\label{sec:preface}}
In the past ten years, there has been significant development in our understanding of $\mathcal{N}=4$ supersymmetric Yang-Mills theory ($\mathcal{N}=4$ SYM). This was achieved largely by the use of integrability, a powerful technique to study a certain class of two-dimensional quantum field theories. Such study was initiated in the seminal paper by Minahan and Zarembo \cite{MZ}, in which they discovered a link between the computation of the anomalous dimension and the diagonalization of the Hamiltonians of certain integrable spin chains. Rapid progress in the subsequent years has culminated in the powerful and elegant method called the quantum spectral curve\cite{QSC}.

  In a parallel line of development, the integrability method was applied also to other observables. One notable success in this regard is the computation of the expectation value of the null polygonal Wilson loop \cite{BSV}, which, through the T-duality, is related to the scattering amplitude \cite{AM}. The basic idea of the computation is to break it down into building blocks called pentagons and characterize them as form factors in a two-dimensional integrable theory.
  
 Quite recently, a similar approach was proposed also for the three-point function \cite{BKV}. It provides a powerful framework which allows us to compute the structure constant at the non-perturbative level. The main goal of this lecture is to give a pedagogical introduction and explain necessary backgrounds for that approach. In the first lecture, we describe the computation of the three-point function at tree level and rewrite them as a sum over partitions. In the second lecture, we explain how the sum-over-partition expression can be generalized to the finite coupling and then introduce a nonperturbative approach, called the {\it hexagon approach}. We also explain how the symmetry and the integrability can be used to determine the hexagon form factor, which is the fundamental object in that approach.

\section{Lecture I: Three-point functions at weak coupling}
\subsection{Motivations}
Before plunging into the computation, let us give several motivations as to why we wish to study three-point functions of $\mathcal{N}=4$ SYM using integrability.

Firstly they are simply interesting to study. By studying them, one can appreciate the beautiful interplay of physics in two and four dimensions. For instance, it tells us how two-dimensional integrable models emerge from four-dimensional gauge theories and how four-dimensional physics is reflected in those two-dimensional models. As we will see in the second lecture, the two-dimensional integrable model arising from $\mathcal{N}=4$ SYM is rather special and is highly constrained by the fact that it comes out of a {\it gauge} theory. 

Secondly the integrabillity might be helpful for understanding the AdS/CFT correspondence. $\mathcal{N}=4$ SYM is the most typical example of the AdS/CFT correspondence and its three-point function describes the interaction of strings in the AdS spacetime. The study of such objects will undoubtedly give us deeper insight into holography. 

Thirdly solving $\mathcal{N}=4$ SYM can be an important stepping stone towards understanding four-dimensional interacting conformal field theories. As is well-known, the two-dimensional conformal field theory is highly constrained by the infinite-dimensional Virasoro algebra. On the other hand, not much is known about theories in higher dimensions partly because the conformal group is finite-dimensional. The hope is that $\mathcal{N}=4$ SYM would give us important insights, with which we can start exploring more general CFT's in higher dimensions. 

Let us make an additional remark on this point: In recent years, the conformal bootstrap is applied to higher-dimensional conformal field theories and has been yielding impressive results\cite{3dIsing}. The conformal bootstrap and the integrability-based approach share a common feature that both of them are non-perturbative approaches, but they are, of course, different in numerous respects. One obvious difference is that, while the integrability-based method works only for special theories in the large $N_c$ limit, the conformal bootstrap can work, in principle, for any theories. However, a more important difference which we wish to emphasize is the role of the gauge symmetry: In the conformal bootstrap, one always deals with the gauge-invariant quantities and the gauge symmetry is completely invisible (and unnecessary). In the integrability-based approach on the other hand, the gauge symmetry plays an essential role as we will see in what follows.
\subsection{Brief review of $\mathcal{N}=4$ SYM\label{sec:recap}}
Let us start with a lightening review of the basic properties of $\mathcal{N}=4$ SYM.
\subsubsection*{Symmetries and field contents}
$\mathcal{N}=4$ SYM has the following global symmetries:
\begin{itemize}
\item As the name suggests, it has four sets of supercharges,
\begin{align}
\{Q_{\alpha}^{A}\,, \quad \bar{Q}_{\dot{\alpha}A}\} \qquad (\alpha,\dot{\alpha}=1,2,\quad A=1,\ldots,4)\,. 
 \end{align}
 Here $\alpha$ and $\dot{\alpha}$ are the spinor indices which takes values $1$ or $2$, and $A$ is the index distinguishing different sets of supercharges and runs from $1$ to $4$. In total it has $4\times 4=16$ supercharges.
 \item The theory also has the global SU(4) $\simeq$ SO(6) symmetry which rotates these four sets of supercharges.
\item In addition, the theory is known to have conformal symmetry SO(4,2) $\simeq$ SU(2,2), whose generators are given by the dilatation $D$, special conformal transformations $K_{\mu}$, translations $P^{\mu}$, and Lorentz transformations $L_{\mu\nu}$.
\item The commutators of the supersymmetry generators and the special conformal generators yield extra 16 fermionic generators $[K, Q] \sim S$, which are superconformal generators:
\begin{align}
\{S^{\alpha}_{A}\,, \quad \bar{S}^{\dot{\alpha}A}\} \qquad (\alpha,\dot{\alpha}=1,2,\quad A=1,\ldots,4)\,.
\end{align}
\end{itemize}
These symmetries combine into the $\mathcal{N}=4$ superconformal group\footnote{For those who want to know more about the superconformal symmetry, we recommend the paper \cite{Minwalla}.}, which is isomorphic to the supergroup PSU(2,2$|$4).

\begin{figure}[t]
\begin{center}
\includegraphics[clip,height=4.5cm]{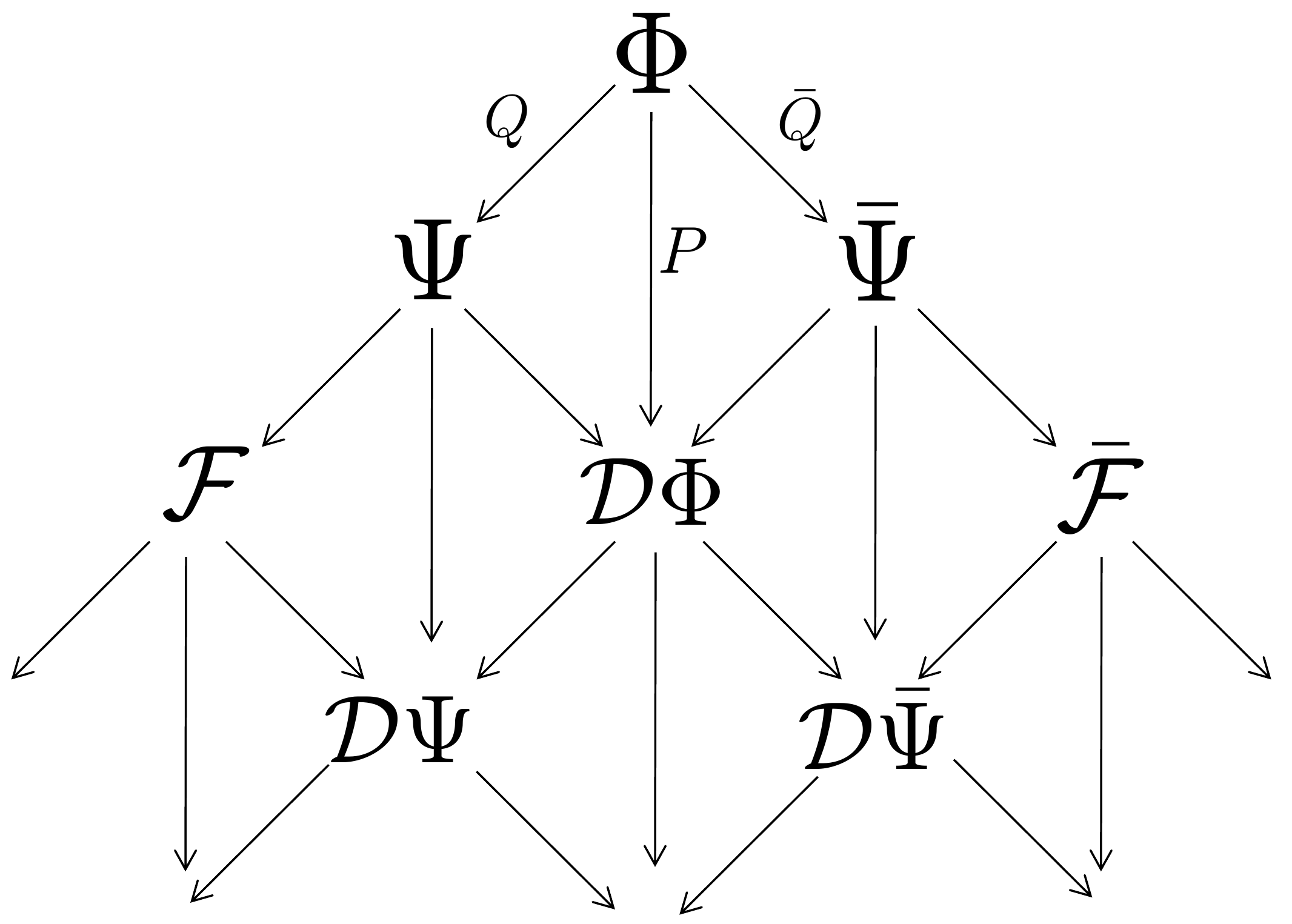}
\end{center}
\caption{The structure of the field mutiplet in $\mathcal{N}=4$ SYM. $\mathcal{D}$ denotes the covariant derivative and $\mathcal{F}$ and $\bar{\mathcal{F}}$ are the field strength.\label{fig:multiplet}}
\end{figure}

Since the $\mathcal{N}=4$ superconformal group is sufficiently large, the fields appearing in the Lagrangian all fit in a single superconformal multiplet. It consists of $6$ scalars, $4$ Weyl fermions and $1$ gauge field (see figure \ref{fig:multiplet} for the structure of the multiplet):
\begin{align}
\begin{aligned}
&\Phi_I &&(I=1,\ldots, 6)\\
&\Psi_{\alpha}^{A},\,\,\,\bar{\Psi}_{\dot{\alpha}A}&& (\alpha=1,2,\quad  A=1,\ldots ,4)\\
&A_{\mu}
\end{aligned}
\end{align}
As a consequence, all the fields belong to the adjoint representation of the gauge group SU($N_c$).
\subsubsection*{Two- and three-point functions}
In conformal field theories, the basic objects to study are the correlation functions of local operators. The most fundamental among them are two- and three-point functions since any higher-point correlation functions can be decomposed into two- and three-point functions using the operator product expansion. 

The space-time dependence of the two- and three-point functions are constrained by the conformal symmetry. For instance, the correlators of the scalar operators\footnote{Here we also assume that the operators are conformal primaries.} take the following form:
\begin{align}\label{twothree}
\begin{aligned}
\langle \mathcal{O}_i (x_1) \mathcal{O}_j (x_2)\rangle &= \frac{\delta_{ij}\mathcal{N}_i}{|x_1 -x_2|^{2\Delta_i}}\,,\\
\langle \mathcal{O}_i (x_1)\mathcal{O}_j (x_2) \mathcal{O}_k (x_3)\rangle &= \frac{C_{ijk}\sqrt{\mathcal{N}_i\mathcal{N}_j\mathcal{N}_k}}{|x_1-x_2|^{\Delta_i+\Delta_j-\Delta_k}|x_2-x_3|^{\Delta_j+\Delta_k-\Delta_i}|x_3-x_1|^{\Delta_k+\Delta_i-\Delta_j}}
\end{aligned}
\end{align}
Here $\Delta_i$ is called the conformal dimension and is the eigenvalue of the dilatation operator,
\begin{align}
D \cdot \mathcal{O}_i =\Delta_i \mathcal{O}_i\,.
\end{align} 
whereas $C_{ijk}$ is the structure constant of the operator product expansion, 
\begin{align}
\mathcal{O}_i(x_1) \mathcal{O}_j(x_2)\sim \frac{C_{ijk}}{|x_1-x_2|^{\Delta_i+\Delta_j-\Delta_k}}\mathcal{O}_k (x_2) +\cdots\,.\label{OPE}
\end{align}
The constants $\mathcal{N}_i$'s denote the normalization factors of the operators. They can always be set to $1$ by re-normalizing the operators as $\mathcal{O}_i\to \mathcal{O}_i/\sqrt{\mathcal{N}_i}$. For actual computation however, it is sometimes useful to keep such factors. 
\subsubsection*{Single-trace operators and the spin chain}
As mentioned above, all the fields in $\mathcal{N}=4$ SYM belong to the adjoint representation of the gauge group SU($N_c$). Therefore, the gauge invariant quantities can be built easily by multiplying the fields and taking traces.

In the large $N_c$ limit, the most important set of local operators are {\it single-trace operators}, the operators which consist of a single trace:
\begin{align}
{\rm tr} \left(\Phi \Psi \Phi F_{\mu\nu} \cdots \right)\quad \text{etc}\,.
\end{align}
In addition to the single-trace operators, there exist other operators which consist of multiple traces. However, in the large $N_c$ limit, the conformal dimension of a multi-trace operator is simply given by a sum of conformal dimensions of the constituent single-trace operators:
\begin{align}
\mathcal{O}=\underbrace{{\rm tr}\left(\cdots \right)}_{\mathcal{O}_1}\underbrace{{\rm tr}\left( \cdots\right)}_{\mathcal{O}_2}\quad \longrightarrow\quad  \Delta_{\mathcal{O}}=\Delta_{\mathcal{O}_1}+\Delta_{\mathcal{O}_2}+O(1/N_c)\,.
\end{align}
Because of this property, the single-trace operators are the most basic quantity in the large $N_c$ theories.

As found by Minahan and Zarembo \cite{MZ}, there is a nontrivial link between the single-trace operators in the large $N_c$ limit and the two-dimensional integrable system. We will not review it in detail here but the upshot is that one can identify the dilatation operator acting on the single-trace operators with the Hamiltonian of a certain integrable spin chain. Under this identification, the conformal dimension is mapped to the energy of the integrable system. We can then use various techniques of integrability, such as the Bethe ansatz and the Thermodynamic Bethe ansatz, to determine the conformal dimension:
\begin{align}
\begin{aligned}
D &\leftrightarrow H_{\rm 2d}\,,\qquad \Delta&\leftrightarrow E_{\rm 2d}\,.
\end{aligned}
\end{align}

\begin{figure}[t]
\begin{center}
\includegraphics[clip,height=2cm]{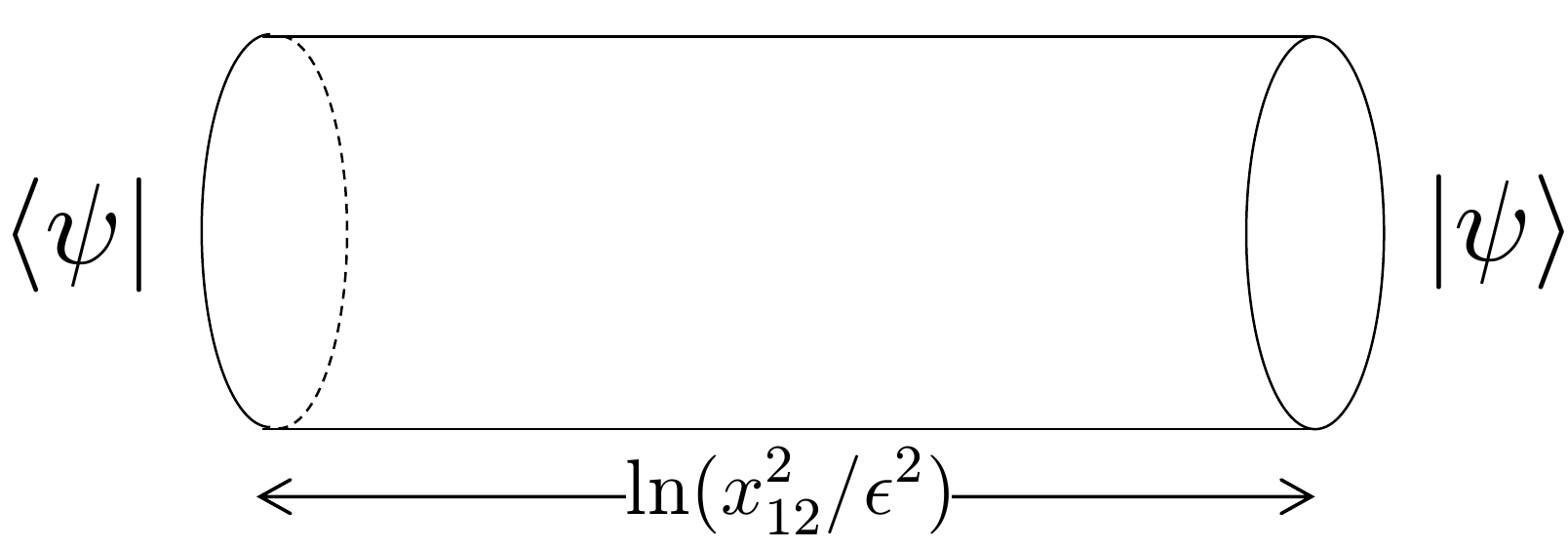}
\end{center}
\caption{The cylinder partition function with length $\ln \left(x_{12}^2/\epsilon^2 \right)$ corresponds to the two-point function in $\mathcal{N}=4$ SYM.\label{fig:cylinder}}
\end{figure}

It is also possible to relate the two-point function itself, not just the conformal dimension, to a 2d integrable system. The idea is to consider a (Euclidean) two-dimensional cylinder with length $\ln (x_{12}^2 /\epsilon^2)$ as depicted in figure \ref{fig:cylinder}. The partition function of this cylinder can be computed as\footnote{Here $\epsilon (\ll 1)$ is the cut-off we introduced.}
\begin{align}
\langle \psi_{\rm 2d}| e^{-H_{\rm 2d} \ln (x_{12}^2/\epsilon^2)}|\psi_{\rm 2d}\rangle = \left(\frac{\epsilon}{x_{12}}\right)^{2\Delta}\,.
\end{align}
Obviously this reproduces the correct two-point function including the space-time dependence. This map, however, may seem too ad-hoc since it gives an impression that we intentionally chose the length of the cylinder so that the final result comes out correctly. However, it actually has natural physical interpretation. To further clarify this point, it is useful to work out the following small exercise.\vspace{5pt}

\begin{itemize}
\item[] {\bf Exercise}: Consider two operators inserted at $x_1^{\mu}=(0,0,0,0)$ and $x_2^{\mu}=(a,0,0,0)$ respectively and cut out small spheres with radius $\epsilon$ around these points (see figure \ref{fig:special}). Our claim is that $\ln (x_{12}^2/\epsilon^2)$ is related to how much ``dilatation transformation'' we need in order to bring one sphere to the other. In order to make this statement precise, we perform a special conformal transformation and bring $x_2$ to infinity. This is because the dilatation transformation does not map points near $x_1$ to points near $x_2$. Instead it maps points near the origin to the points near infinity. 
\begin{figure}[h]
\begin{center}
\includegraphics[clip,height=5.2cm]{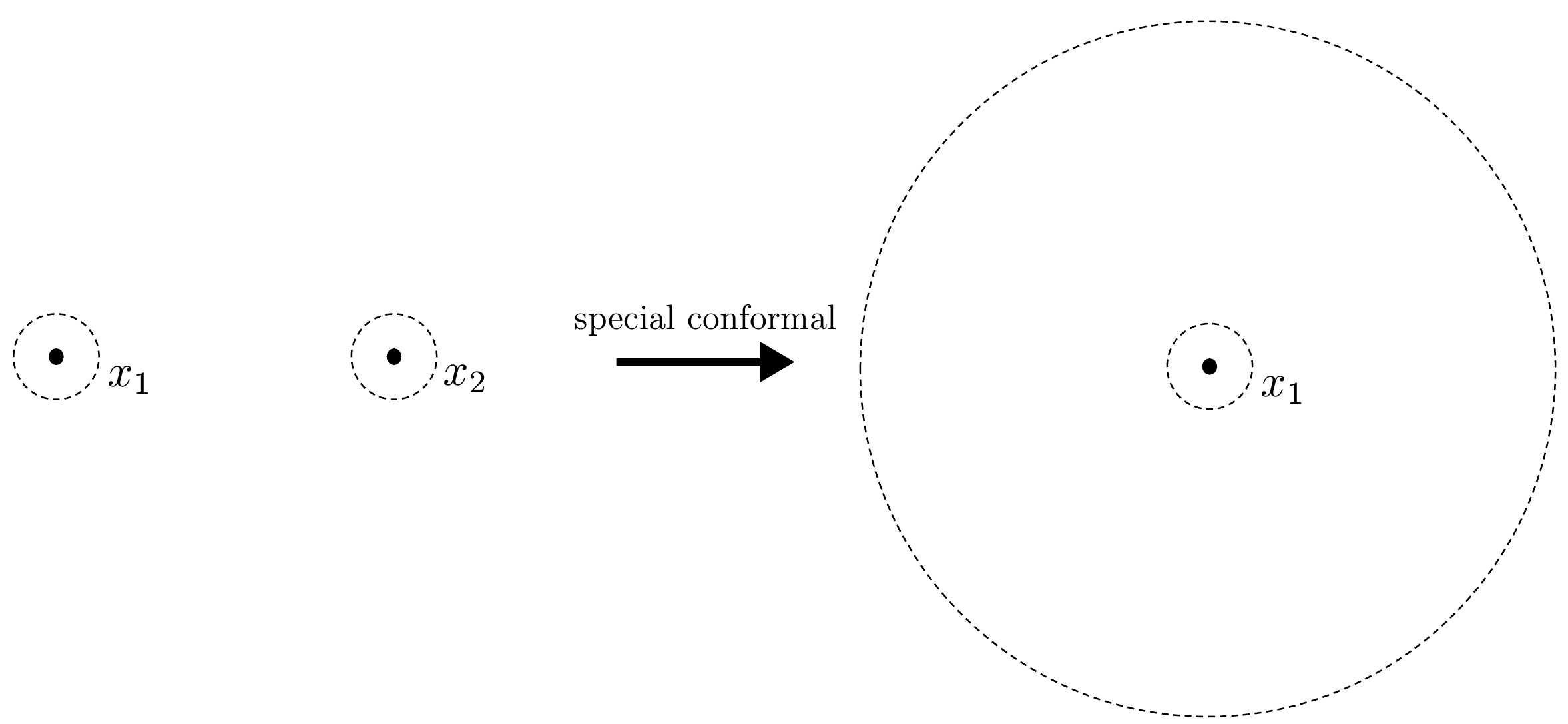}
\end{center}\vspace{-0.5cm}
\caption{The action of the special conformal transformation.\label{fig:special}}
\end{figure}

After the special conformal transformation, we get a small sphere around the origin and a large sphere near infinity. What this exercise requires you is to figure out the radius of these spheres after the transformation and read off how much dilatation you need in order to map them to each other. You will then find that the length $\ln (x_{12}^2/\epsilon^2)$ appears naturally.

\end{itemize}

\subsection{BPS correlators at tree level}
Having reviewed the minimal backgrounds, we now start studying the correlation functions at weak coupling. 

Among the single-trace operators, the simplest ones are $1/2$ BPS operators. Those are the operators of the following form,
\begin{align}
{\rm tr} \left( (P^{I} \Phi_I)^{L}\right)=:{\rm tr}\left( (P\cdot \Phi)^{L}\right)\,,
\end{align}
where $P^{I}$ is a complex six-dimensional null vector, called the {\it polarization vector}, satisfying\footnote{This condition is necessary to make the operator BPS.} $P^2 =0$. For instance, if we choose $P^{I}$ to be $(1,i,0,0,0,0)$, we get
\begin{align}
{\rm tr}\left(Z^{L}\right)\,, \qquad Z:= \Phi_1 +i \Phi_2\,,
\end{align}
which is the operator commonly used in the literature. The $1/2$ BPS operators have several important properties:
\begin{itemize}
\item They are annihilated by all 16 superconformal charges $S$ and 8 supercharges $Q$. The combination of the supercharges which annihilates the operator depends on the choice of $P^{I}$.
\item It belongs to the short multiplet of the superconformal group and its conformal dimension is fixed to be $\Delta =L$. Namely, they do not receive quantum corrections. 
\end{itemize}

\subsubsection*{Two-point function}
\begin{figure}[t]
\centering
\begin{minipage}{0.35\hsize}
\centering
\includegraphics[clip,height=0.8cm]{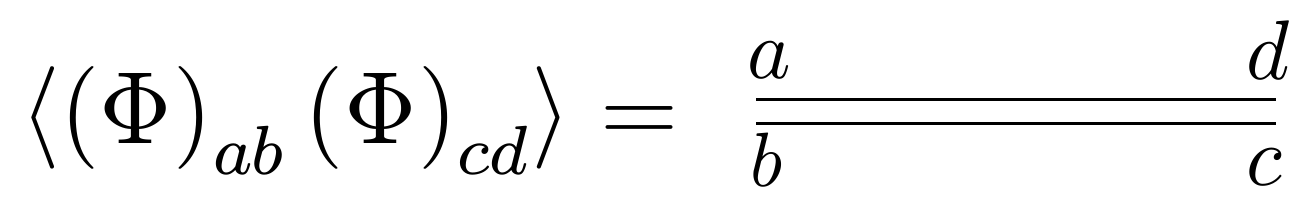}
\end{minipage}
\begin{minipage}{0.4\hsize}
\centering
\includegraphics[clip,height=1.8cm]{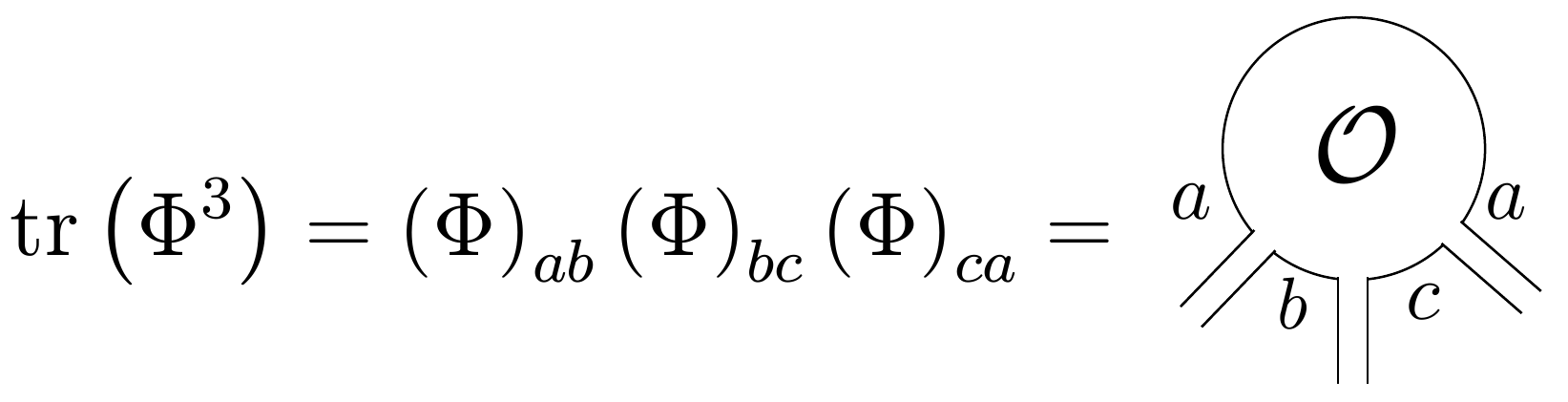}
\end{minipage}\begin{minipage}{0.2\hsize}
\centering
\includegraphics[clip,height=4cm]{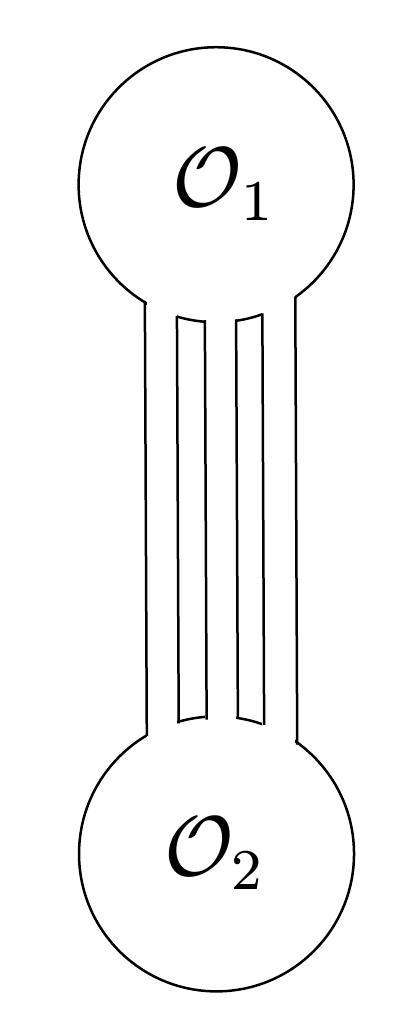}
\end{minipage}\\
\begin{minipage}{0.33\hsize}
\centering
$(a)$
\end{minipage}
\begin{minipage}{0.4\hsize}
\centering
$(b)$
\end{minipage}
\begin{minipage}{0.2\hsize}
\centering
$(c)$
\end{minipage}
\caption{The double-line notations. $(a)$: The double-line notation for the propagator of the scalar fields. Each line signifies the Kronecker delta, $\delta_{ad}$ etc. $(b)$: The double-line representation of the single-trance operator. $(c)$: The planar Wick contraction for the two-point function.\label{fig:doubleline}}
\end{figure}

Let us now compute the two-point function of the BPS operators:
\begin{align}
\langle {\rm tr}\left( (P_1\cdot \Phi)^{L}\right)(x_1)  \quad {\rm tr}\left( (P_2\cdot \Phi)^{L}\right)(x_2)\rangle
\end{align}
At tree level, we just need to perform the Wick contraction. For two scalar fields, the Wick contraction reads
\begin{align}\label{contract}
\begin{aligned}
&\langle \Phi_I (x_1) \Phi_J(x_2) \rangle = \frac{\delta_{IJ}}{|x_1-x_2|^{2}}\\
&\Rightarrow \langle (P_1\cdot\Phi) (x_1) (P_2\cdot\Phi) (x_2) \rangle = \frac{(P_1\cdot P_2)}{|x_1-x_2|^{2}}
\end{aligned}
\end{align}
In \eqref{contract}, we suppressed the indices for the gauge group SU($N_c$), which are given by\footnote{To derive \eqref{matrix}, we first rewrite the kinetic term ${\rm tr} \left(\partial_{\mu}\Phi \partial^{\mu}\Phi\right)$ in terms of the orthogonal basis as $\Phi=\sum_A\Phi_A T^{A}$ with ${\rm tr}\left(T^AT^{B} \right)=\delta^{AB}$. Then, the kinetic term and the Wick contraction read
\begin{align}
\sum_{A}\partial_{\mu} \Phi_A \partial^{\mu} \Phi_A \,, \qquad  \langle\Phi_A \Phi_B \rangle\propto \delta_{AB}\,.
\end{align}
To go back to the matrix representation, we simply multiply $T^{A}$ and $T^{B}$ and use the relation
\begin{align}
\sum_A \left( T^{A}\right)_{ab}\left( T^{A}\right)_{cd}\propto \delta_{ad}\delta_{bc}-\frac{1}{N_c}\delta_{ab}\delta_{cd}\,. 
\end{align}
In the large $N_c$ limit, the second term can be neglected and we get
\begin{align}
\langle (\Phi)_{ab}(\Phi)_{cd}\rangle=\sum_{A,B}\left(T^A\right)_{ab} \left(T^B\right)_{cd} \langle\Phi_A \Phi_B \rangle \propto \delta_{ad}\delta_{bc}\,.
\end{align}
},
\begin{align}\label{matrix}
\langle (\Phi)_{ab}(\Phi)_{cd}\rangle \propto \delta_{ad}\delta_{bc}\,.
\end{align}
As is well-known, the best way to take into account the index structure of SU($N_c$) is to use the double-line notation as depicted in figure \ref{fig:doubleline}-$(a)$.

Now, using the double-line notation, each single trace operator can be denoted as in figure \ref{fig:doubleline}-$(b)$. Then the two-point function can be computed by drawing all the possible planar diagrams connecting two operators. For the length $L$ operator, there are $L$ different ways of planar contractions which are related to each other by the cyclic permutation of one of the operators. As a result, we get
\begin{align}
\langle {\rm tr}\left( (P_1\cdot \Phi)^{L}\right)(x_1)\quad {\rm tr}\left( (P_2\cdot \Phi)^{L}\right)(x_2)\rangle \sim L N_c^{L}\frac{(P_1\cdot P_2)^{L}}{|x_1-x_2|^{2L}}
\end{align}
The dependence on $N_c$ can be easily deduced from figure \ref{fig:doubleline}-$(c)$.
In order to normalize the two-point function, we should redefine the operator as
\begin{align}\label{normalizeBPS}
\mathcal{O}_i \to \frac{1}{N_c^{L_i/2}\sqrt{L_i}}\mathcal{O}_i\,.
\end{align}
\subsubsection*{Three-point function}
\begin{figure}[t]
\centering
\includegraphics[clip,height=5.5cm]{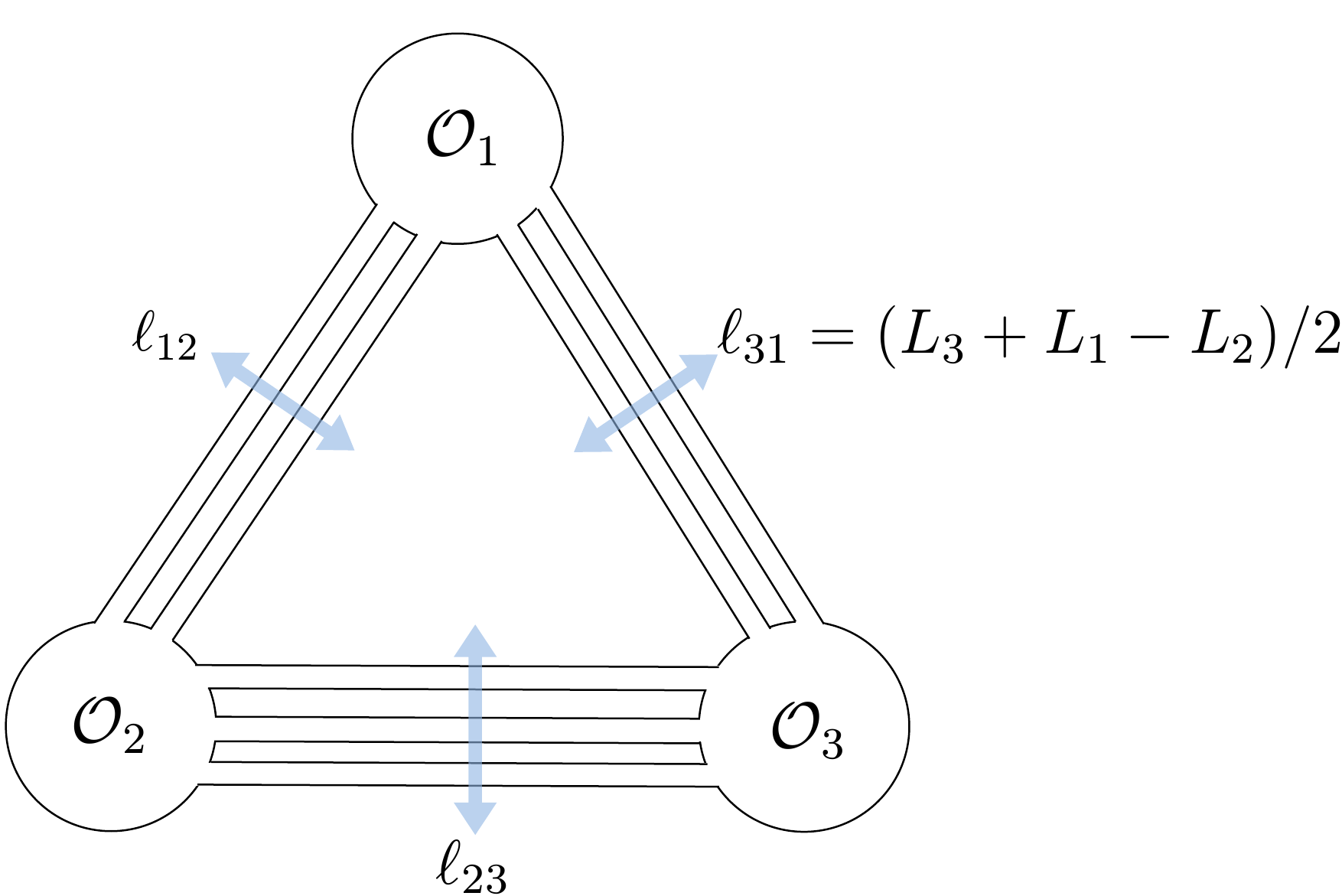}
\caption{Three-point functions at tree level. $L_i$ denotes the length of the operator $\mathcal{O}_i$ and $\ell_{ij}$ denotes the number of Wick contractions between operators $\mathcal{O}_i$ and $\mathcal{O}_j$.\label{fig:3pt}}
\end{figure}

We now move onto the computation of the three-point function,
\begin{align}
\langle{\rm tr}\left((P_1\cdot \Phi)^{L_1} \right)(x_1)\quad {\rm tr}\left((P_2\cdot \Phi)^{L_2} \right)(x_2)\quad {\rm tr}\left((P_3\cdot \Phi)^{L_3} \right)(x_3) \rangle\,.
\end{align}
An example of the planar diagram connecting all three operators is given in figure \ref{fig:3pt}. In the case of three-point functions, there are $L_1 L_2 L_3$ inequivalent ways of planar contractions which are related to each other by the cyclic permutations of the individual operators. Therefore the result reads
\begin{align}\label{bare3pt}
\begin{aligned}
&\langle{\rm tr}\left((P_1\cdot \Phi)^{L_1} \right)(x_1)\quad {\rm tr}\left((P_2\cdot \Phi)^{L_2} \right)(x_2)\quad {\rm tr}\left((P_3\cdot \Phi)^{L_3} \right)(x_3) \rangle \\
&= N_c^{\frac{L_1+L_2+L_3}{2}-1} L_1 L_2 L_3\frac{(P_1\cdot P_2)^{\ell_{12}}(P_2\cdot P_3)^{\ell_{23}}(P_3\cdot P_1)^{\ell_{31}}}{|x_1-x_2|^{2 \ell_{12}}|x_2-x_3|^{2 \ell_{23}}|x_3-x_1|^{2 \ell_{31}}}
\end{aligned}
\end{align}
Here, $\ell_{ij}=(L_i+L_j-L_k)/2$ denotes the number of the Wick contractions between the operators $\mathcal{O}_i$ and $\mathcal{O}_j$. In what follows, we call it the {\it bridge length}.

After the normalization \eqref{normalizeBPS}, the three-point function is given by
\begin{align}\label{3is1/N}
\frac{ \eqref{bare3pt}}{N_c^{L_1/2}\sqrt{L_1}N_c^{L_2/2}\sqrt{L_2}N_c^{L_3/2}\sqrt{L_3}}\quad \propto \quad \frac{\sqrt{L_1 L_2 L_3}}{N_c}
\end{align}
As is manifest from \eqref{3is1/N}, the three-point function is $O(1/N_c)$. In this sense, studying the three-point function already requires us to go slightly beyond the strict large $N_c$ limit. This is easier to understand on the dual string-theory side: As mentioned in section \ref{sec:preface}, the three-point function corresponds to the splitting and joining processes of the strings in AdS and therefore it must come with the string coupling constant $g_s \sim 1/N_c$. 

Since the factor \eqref{3is1/N} is common to any three-point functions, it is convenient to strip it off from the structure constant as
\begin{align}
C_{123}= \frac{\sqrt{L_1 L_2 L_3}}{N_c} c_{123}\,.
\end{align}
In the next section, we compute $c_{123}$ using the perturbation theory.

 \begin{itemize}
 \item[] {\bf Exercise}: Confirm the $N_c$ dependence of the three-point function \eqref{bare3pt}. 
 \end{itemize}
 
Before finishing this section, let us make a short remark on the BPS three-point function. As with the case of the two-point functions, the BPS three-point functions turns out to be a protected quantity. This can be shown\footnote{For details, see \cite{BBP}. It provides a simple and concise argument on this point.} by taking a derivative of the three-point function with respect to the coupling constant,
\begin{align}
\frac{\partial}{\partial g_{\rm YM}^2} \langle \mathcal{O}_1 \mathcal{O}_2 \mathcal{O}_3\rangle\,.
\end{align}
In the path-integral formulation, this brings down one Lagrangian from the action,
\begin{align}
\begin{aligned}
\frac{\partial}{\partial g_{\rm YM}^2} \langle \mathcal{O}_1 \mathcal{O}_2 \mathcal{O}_3\rangle&=\frac{\partial}{\partial g_{\rm YM}^2}\int [DA_{\mu}]  \mathcal{O}_1 \mathcal{O}_2 \mathcal{O}_3 \,e^{-\frac{1}{g_{\rm YM}^2}\int d^4 x \mathcal{L}}\\
&\propto \int [DA_{\mu}]  \mathcal{O}_1 \mathcal{O}_2 \mathcal{O}_3 \left(\int d^4 x \mathcal{L}\right)\,e^{-\frac{1}{g_{\rm YM}^2}\int d^4 x \mathcal{L}}\,.
\end{aligned}
\end{align}
We thus obtain a relation,
 \begin{align}
 \frac{\partial}{\partial g_{\rm YM}^2} \langle \mathcal{O}_1 \mathcal{O}_2 \mathcal{O}_3\rangle\propto\int d^4 x \langle\mathcal{O}_1 \mathcal{O}_2 \mathcal{O}_3 \mathcal{L} \rangle\,.
 \end{align}
In $\mathcal{N}=4$ SYM, the Lagrangian is a superconformal descendant of the length 2 BPS operator $\hat{\mathcal{O}}$,
 \begin{align}
 \mathcal{L}=(Q)^4\hat{\mathcal{O}} \,,
 \end{align}
 where $(Q)^4$ denotes an appropriate linear combination of products of four supercharges. Using this property, we can express the Lagrangian as
 \begin{align}
 \mathcal{L}=\tilde{Q} \tilde{\mathcal{O}}
 \end{align}
with $\tilde{Q}$ being a certain linear combination of the supercharges\footnote{An example of such supercharges will be given explicitly in the second lecture.} which annihilates all three operators $\tilde{Q}\mathcal{O}_i=0$.
Now, using the Ward identity, we can prove
\begin{align}
\langle \mathcal{O}_1\, \mathcal{O}_2 \,\mathcal{O}_3\,\mathcal{L}\rangle=\langle \mathcal{O}_1\, \mathcal{O}_2 \,\mathcal{O}_3\,\tilde{Q}\tilde{\mathcal{O}}\rangle=-\langle \tilde{Q}\left(\mathcal{O}_1 \mathcal{O}_2 \mathcal{O}_3\right)\,\,\tilde{\mathcal{O}}\rangle=0
\end{align}
This shows that the three-point function of BPS operator does not depend on the coupling constant.
  \subsection{Non-BPS three-point functions at tree level}
We now set out to the computation of the non-BPS three-point functions\footnote{See also \cite{RoibanVolovich} and \cite{Tailoring} for the formulation of the tree-level three-point functions in terms of spin chains.}. 
\subsubsection*{Set-up}
\begin{figure}[t]
  \centering
  \includegraphics[clip,height=5cm]{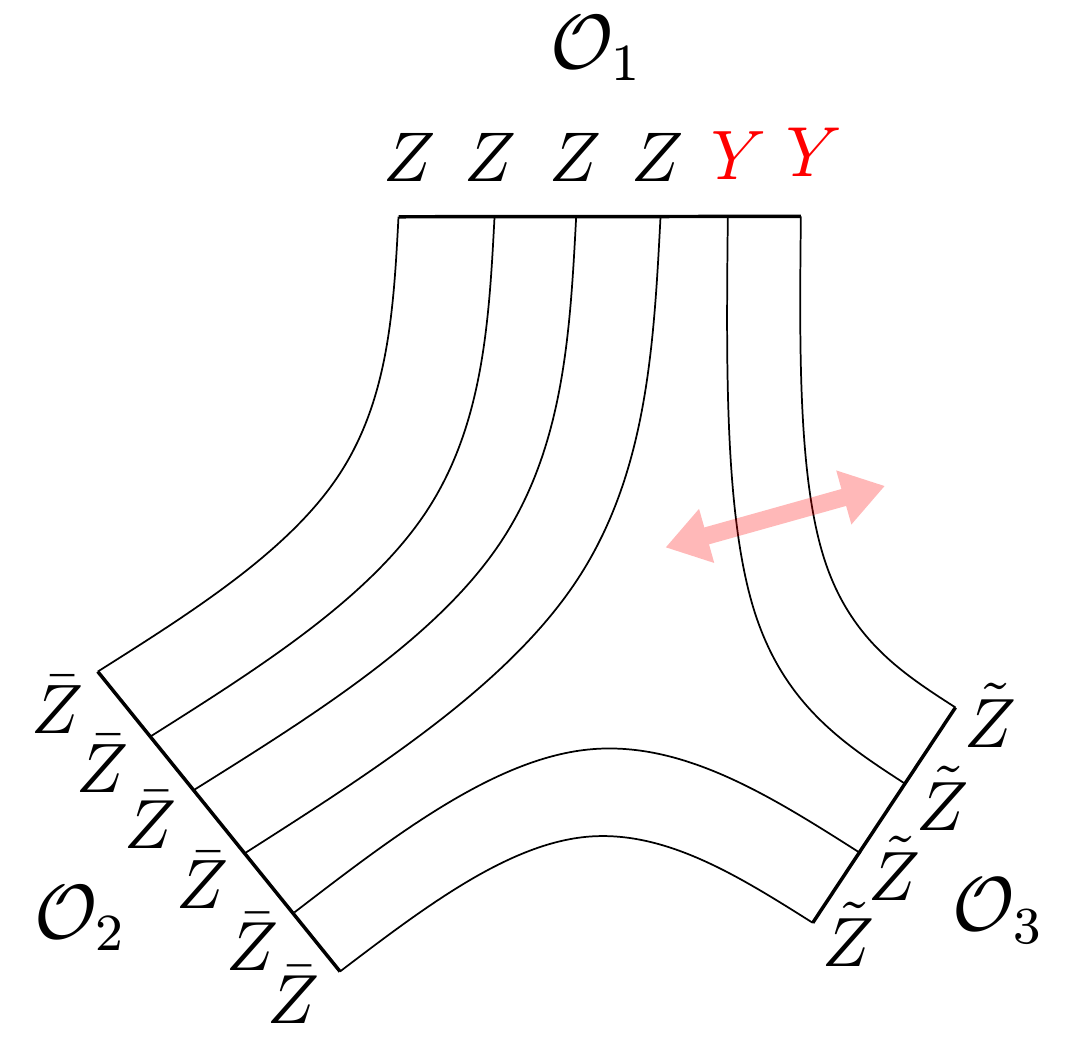}
  \caption{The non-BPS three-point function we discuss in the main text. Here $\tilde{Z}$ denotes a linear combination of fields $\tilde{Z}=Z+\bar{Z}+Y-\bar{Y}$. A simplifying feature of this configuration is that all the $Y$'s in $\mathcal{O}_1$ are contracted with the operator $\mathcal{O}_3$ (not with $\mathcal{O}_2$); namely $Y$'s can only live in the region indicated by the red arrow.\label{fig:reservoir}}
  \end{figure}

To illustrate the basic physical picture avoiding unnecessary complication, we focus on the following configurations, where one operator is a non-BPS operator in the so-called SU(2) sector and the rest are BPS (see also figure \ref{fig:reservoir}): 
  \begin{align}
  \begin{aligned}
  \mathcal{O}_1 &= \left(ZYZZY\cdots \right)+\cdots\,,\quad
  \mathcal{O}_2 = {\rm tr} \left( \bar{Z}^{L_2}\right)\,, \quad
  \mathcal{O}_3 = {\rm tr} \left((Z+\bar{Z}+Y-\bar{Y})^{L_3} \right)\,.
  \end{aligned}
  \end{align}
  Here $Z$ and $Y$ are given by $Z=\Phi_1+i\Phi_2$ and $Y=\Phi_3+i\Phi_4$ respectively. In our notation, it amounts to choosing $P_2= (1,-i,0,0,0,0)$ and $P_3=(1,0,0,i,0,0)$. A simplifying feature of this configuration is that all $Y$ fields contained in $\mathcal{O}_1$ contract only with $\mathcal{O}_3$ since $Y$ cannot contract with $\bar{Z}$. This substantially simplifies the combinatorics of the Wick contraction.
  
To perform the contraction, we need to know the precise form of $\mathcal{O}_1$. It is determined by the condition that $\mathcal{O}_1$ is the eigenvector of the dilatation operator:
\begin{align}
D\cdot \mathcal{O}_1 =\Delta_1 \mathcal{O}_1\,.
\end{align}
By expanding both sides with respect to the coupling constant, we obtain
\begin{align}
\begin{aligned}
( D^{(0)} +\lambda D^{(1)}+\cdots) (\mathcal{O}_1^{(0)}+\lambda\mathcal{O}_1^{(1)}+\cdots)=(\Delta^{(0)}_1+\lambda \Delta_1^{(1)}+\cdots )(\mathcal{O}_1^{(0)}+\lambda\mathcal{O}_1^{(1)}+\cdots)\,.\nonumber
\end{aligned}
\end{align}
The leading term in the expansion yields the relation $D^{(0)}\mathcal{O}_1^{(0)}=\Delta_1^{(0)}\mathcal{O}_1^{(0)}$. This relation, however, is not constraining enough to fix the form of $\mathcal{O}_1^{(0)}$: The tree-level dilatation operator $D^{(0)}$ only counts the number of fields and therefore it is proportional to the identity operator if we restrict ourselves to the space of operators with a fixed number of fields. Obviously, there is no way to determine the eigenvector of the identity operator uniquely. The situation is quite analogous to the degenerate perturbation theory in quantum mechanics. As in that case, what we need to do is to consider the next term in the expansion to lift the degeneracy\footnote{To understand this point, it might be helpful to consider the eigenvectors of the following matrix:
 \begin{align}
 M= \left(\begin{array}{cc}1&0\\0&1\end{array} \right)+\lambda \left(\begin{array}{cc}a&b\\c&d\end{array} \right)\,.
 \end{align}
 Clearly, the eigenvectors of the matrix $M$ is determined by the second term on the right hand side, not the first term. In addition, although the matrix $M$ depends linearly on $\lambda$, the eigenvectors are independent of $\lambda$. In other words, the ``one-loop part'' of $M$ determines the ``tree-level part''
 of the eigenvector.}. In the present case, this amounts to considering the eigenvector of the {\it one-loop} dilatation operator:
 \begin{align}
 D^{(1)} \cdot\mathcal{O}^{(0)}_1 =\Delta_1^{(1)}\mathcal{O}_1^{(0)}\,.
 \end{align}

As mentioned in section \ref{sec:recap}, the one-loop dilatation operator can be identified with the integrable spin-chain Hamiltonian. In the present case, the map between the operator and the spin is
\begin{align}
Z \,\mapsto \,\uparrow \,,\qquad Y\,\mapsto \,\downarrow\,.
\end{align}
and the Hamiltonian is given by
\begin{align}
H_{\rm Heisenberg}\propto \sum_j (I_{j,j+1}-P_{j,j+1})\,, 
\end{align}
where $I_{j,j+1}$ and $P_{j,j+1}$ are the identity and the permutation operators acting on the $j$-th and $(j+1)$-th spins.
Once we map the problem to the spin chain,  we can systematically construct the eigenvector using the coordinate Bethe ansatz as we see below\footnote{A concise summary of the coordinate Bethe ansatz of the Heisenberg spin chain is given in section 2 of \cite{Tailoring}. For a more pedagogical introduction, see section 3 of \cite{Takahashi}.}.
\subsubsection*{Single-magnon state}
Let us first consider the operator with a single $Y$ (called ``magnon'' in the spin-chain terminology) in the sea of $Z$'s. In this case, the eigenvector is simply given by the plane wave:
\begin{align}
|\psi\rangle = \sum_{n=1}^{L_1} e^{i pn}|\cdots \underset{n}{\downarrow}\cdots\rangle
\end{align}
Since $Y$ excitations can only contract with $\mathcal{O}_3$ and all different contractions yield the same space-time dependence, the structure constant $c_{123}$ is simply given by the sum of the wave function on the segment with length $\ell_{31}$:
\begin{align}
c_{123}\propto \sum_{1\leq n \leq \ell_{31}} e^{i p n}\,.
\end{align}
This is just a geometric series and one can easily perform the summation as
\begin{align}\label{sumone}
c_{123}\propto \sum_{1\leq n \leq \ell_{31}} e^{i p n}=g(p) (1-e^{ip \ell_{31}})\,,
\end{align}
with $g(p)$ defined by
\begin{align}
g(p):= \frac{1}{e^{-ip}-1}
\end{align}

To understand the physical meaning of the geometric sum, it is convenient to introduce a pictorial notation and denote \eqref{sumone} as 
\begin{align}
\sum_{1\leq n \leq \ell_{31}} \substack{\\\includegraphics[clip,width=2cm]{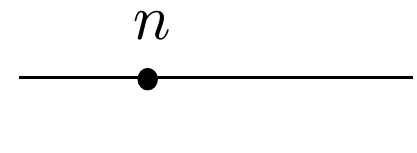}}=g(p)\left(\substack{\\\includegraphics[clip,width=2cm]{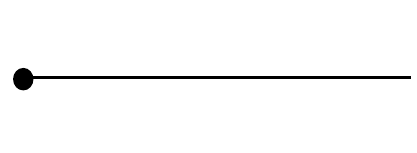}}-\substack{\\\includegraphics[clip,width=2cm]{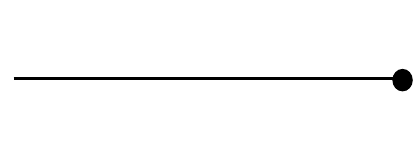}}\right)\,.
\end{align}
In this notation, the magnon is denoted by the black dot. The second term on the right hand side, in which the magnon lives at the right edge of the segment, corresponds to the factor $e^{ip \ell_{31}}$. This is the phase shift acquired by the magnon when it propagates from the left edge to the right edge. In the first term on the other hand, the magnon lives at the left edge of the segment. Therefore, it corresponds to the factor $1(=e^{ip 0})$.
\subsubsection*{Two-magnon state}
We next discuss the two-magnon state. The wave function for the two-magnon state is composed of two terms,
\begin{align}\label{2wave}
|\psi (p_1,p_2)\rangle =\sum_{1\leq n<m\leq L_1}\left(\underbrace{e^{i p_1 n +ip_2 m}}_{\psi_{12}(n,m)}+\underbrace{S(p_1,p_2)e^{i p_2 n +ip_1 m}}_{\psi_{21}(n,m)}\right)|\cdots \underset{n}{\downarrow}\cdots \underset{m}{\downarrow}\cdots\rangle
\end{align}
where $S(p_1,p_2)$ is the S-matrix of the Heisenberg spin chain. It takes a simple form in terms of the so-called rapidity variables $u_1$ and $u_2$:
\begin{align}
\begin{aligned}
&S(p_1,p_2):= \frac{u_1-u_2-i}{u_1-u_2+i}\,,\qquad\qquad  \left(\frac{u_1+i/2}{u_1-i/2}= e^{ip_1}\,,\quad \frac{u_2+i/2}{u_2-i/2}= e^{ip_2}\right)\,.
\end{aligned}
\end{align}
In what follows, we denote the first term in \eqref{2wave} by $\psi_{12}$ and the second term by $\psi_{21}$. Then the structure constant $c_{123}$ can be computed as
\begin{align}
c_{123}\propto \sum_{1\leq n <m\leq \ell_{31} }\left[\psi_{12}(n,m)+\psi_{21}(n,m)\right]\,.
\end{align}

Let us first consider the contribution from $\psi_{12}$:
\begin{align}\label{2particlefor}
\begin{aligned}
\psi_{12}:\quad &\sum_{1\leq n<m \leq \ell_{31}}e^{ip_1 n +ip_2 m}\,\, \,\,\underset{\text{sum over $n$}}{=} \,\,\,\, g(p_1)\sum_{1\leq m\leq \ell_{31}}\left(1-e^{ip_1 (m-1)} \right)e^{ip_2m}\\
&\underset{\text{sum over $m$}}{=} \,\,\,\,g(p_1)\Big[g(p_2)(1-e^{ip_2\ell_{31}})-g(p_1+p_2)\,e^{-ip_1}(1-e^{i(p_1+p_2)\ell_{31}})\Big]\,.
\end{aligned}
\end{align}
To make clear the physical picture, let us rewrite this summation using the pictorial notation introduced above:
\begin{align}\label{2particlefig}
\begin{aligned}
\sum_{1\leq n<m\leq \ell_{31}} \substack{\\\includegraphics[clip,width=2cm]{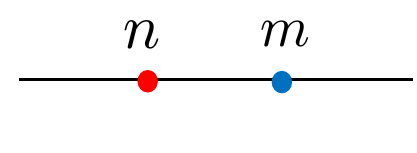}}=&g(p_1)\sum_{m}\left(\substack{\\\includegraphics[clip,width=2cm]{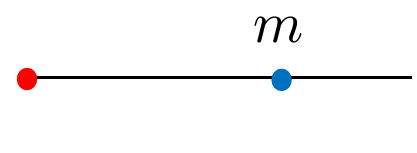}}-\substack{\\\includegraphics[clip,width=2cm]{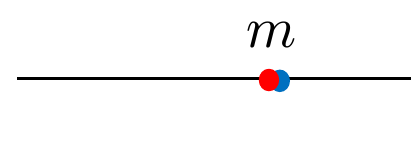}}\right)\\
=&g(p_1)\left(g(p_2)\substack{\\\includegraphics[clip,width=2cm]{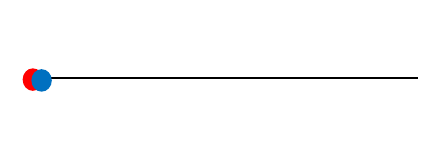}}-g(p_2)\substack{\\\includegraphics[clip,width=2cm]{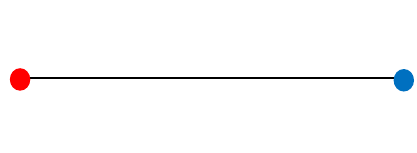}}\right.\\
&\left.-g(p_1+p_2)e^{-ip_1}\substack{\\\includegraphics[clip,width=2cm]{dgeom5.pdf}}+g(p_1+p_2)e^{-ip_1}\substack{\\\includegraphics[clip,width=2cm]{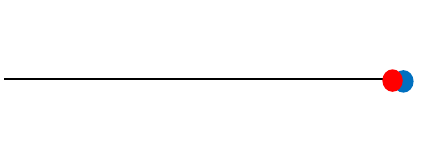}}\right)
\end{aligned}
\end{align}
As is clear from this expression, the result of the geometric sum can be written as the sum over different ways of distributing magnons between the left and the right edges of the segment. This structure persists also for the multi-magnon state and it is the origin of the sum-over-partition expression we derive at the end of this first lecture.

The contribution from $\psi_{21}$ can be computed similarly. Aside from the S-matrix factor, the result can be obtained by exchanging $p_1$ and $p_2$ in \eqref{2particlefor}. Now, by summing two contributions, we obtain the following expression:
\begin{align}
 -(u_1+i/2)(u_2+i/2) \left[h(u_1,u_2)-e^{ip_2 \ell_{31}}-S(p_1,p_2)e^{ip_1\ell_{31}}+h(u_1,u_2)e^{i (p_1+p_2)\ell_{31}}\right]\label{sumexercise}\,.
\end{align}
Here $h(u,v)$ is a function defined by
\begin{align}
h(u,v)=\frac{u-v}{u-v+i}\,.
\end{align}
\begin{itemize}
\item[] {\bf Exercise}: Check \eqref{sumexercise}.
\end{itemize}
\noindent The function $h(u,v)$ appearing in the formula satisfies an important relation
\begin{align}\label{watson0}
S(u,v)=\frac{h(u,v)}{h(v,u)}\,,
\end{align}
whose physical meaning will become clear in the second lecture.

Before moving onto the multi-magnon generalization, let us explain a simple trick to efficiently compute this geometric series. In \eqref{2particlefor}, we first summed over $n$ and then summed over $m$. In this way of summation, we obtain two different terms which do not contain a length-dependent phase shift, $g(p_1)g(p_2)$ and $-g(p_1)g(p_1+p_2)e^{-ip_1}$. However, if we instead sum over $m$ first and then over $n$, there will be only a single term without the phase shift and we can immediately get $g(p_2)g(p_1+p_2)$:
\begin{align}
\begin{aligned}
\sum_{n,m}\substack{\\\includegraphics[clip,width=2cm]{dgeom1.pdf}}&\overset{\text{sum over $m$}}{=}g(p_2) \sum_m \substack{\\\includegraphics[clip,width=2cm]{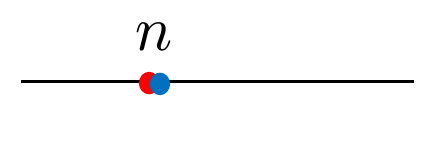}}+\cdots\\
&\overset{\text{sum over $n$}}{=}g(p_2)g(p_1+p_2)\substack{\\\includegraphics[clip,width=2cm]{dgeom5.pdf}}+\cdots
\end{aligned}
\end{align}
    In general, for each different phase shift, there exists an appropriate order of doing the summation and, if we choose that order, we can immediately get the term of interest. This trick is quite useful when we discuss the multi-magnon generalization below.  
\subsubsection*{Generalization to multi-magnon state}
\begin{figure}
\centering
\includegraphics[clip,height=3.5cm]{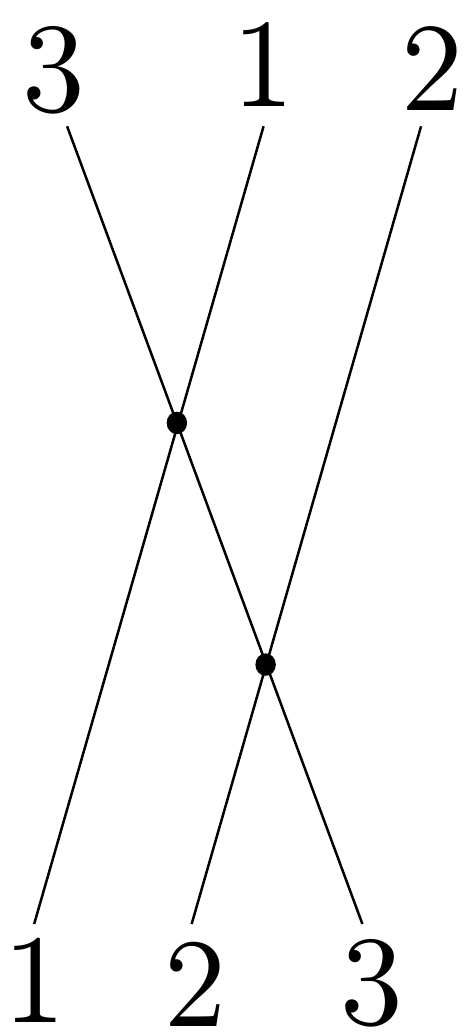}
\caption{The structure of the Bethe wave function. The S-matrix factor associated with the permutation $\{1,2,3\}\to \{3,1,2\}$ can be obtained by multiplying the S-matrices at the black dots in the figure.\label{fig:smatrix}}
\end{figure}

We are now in a position to generalize the previous results to the muti-magnon states\footnote{The derivation explained below was first worked out independetly by Naoki Kiryu and Ho Tat Lam.}. The coordinate Bethe wave function for the multi-particle state is given by a superposition of the plane waves:
\begin{align}
|\psi(p_1,\ldots, p_M)\rangle&= \sum_{n_1<\cdots <n_M} \psi(n_1,\ldots, n_M)|\cdots \underset{n_1}{\downarrow}\underset{\cdots}{\cdots}\underset{n_M}{\downarrow}\cdots \rangle\,,
\end{align}
with
\begin{align}\label{wavestr}
\psi (n_1,\ldots, n_M)=\sum_{\sigma \in S_M} \prod_{\substack{j<k\\\sigma_k<\sigma_j}}
S(p_{\sigma_k},p_{\sigma_j})\prod_{j=1}^{M}e^{ip_{\sigma_j}n_j}\,.
\end{align}
At first sight, the expression for the S-matrix factor may seem cryptic. However, there is a simple and physically intuitive way to figure out this factor. As an example, let us take a look at the three-particle wave function:
\begin{align}
\begin{split}
\psi (n_1,n_2,n_3)=&e^{i (p_1 n_1 +p_2 n_2 +p_3 n_3)} + S(p_1,p_2) e^{i(p_2n_1+p_1 n_2+p_3 n_3) } \\
&+\cdots + \textcolor[rgb]{0,0,1}{S(p_1,p_3)S(p_2,p_3)e^{i(p_3 n_1 +p_1 n_2+p_2 n_3)}}+\cdots
\end{split}
\end{align}
The term denoted in blue comes from the permutation $\{1,2,3\}\to \{3,1,2\}$. The S-matrix factor associated with this permutation can be determined by drawing a diagram as figure \ref{fig:smatrix}. For each intersection in the diagram, we associate an S-matrix and the S-matrix factor in \eqref{wavestr} can be obtained simply by multiplying such S-matrices.  

To proceed, it is useful to slightly rewrite the wave function. The idea is to rewrite the S-matrix factor using \eqref{watson0} as follows:
\begin{align}
\begin{aligned}
\prod_{\substack{j<k\\\sigma_k<\sigma_j}}
S(p_{\sigma_k},p_{\sigma_j})&=\textcolor[rgb]{0,0,1}{\left(\prod_{\substack{j<k\\\sigma_k<\sigma_j}}
h(u_{\sigma_k},u_{\sigma_j})\right)}\left(\prod_{\substack{j<k\\\sigma_k<\sigma_j}}
\frac{1}{h(u_{\sigma_j},u_{\sigma_k})}\right)\\
&=\textcolor[rgb]{0,0,1}{\left( \prod_{\sigma_k<\sigma_j}h(u_{\sigma_k}, u_{\sigma_j})\right)\left(\prod_{\substack{j>k\\\sigma_k<\sigma_j}}\frac{1}{h(u_{\sigma_k},u_{\sigma_j})}\right)}\left(\prod_{\substack{j<k\\\sigma_k<\sigma_j}}\frac{1}{h(u_{\sigma_j},u_{\sigma_k})}\right)\\
&=\left( \prod_{j<k}h(u_{j}, u_{k})\right)\left(\prod_{\substack{j<k\\\sigma_j<\sigma_k}}\frac{1}{h(u_{\sigma_j},u_{\sigma_k})}\right)\left(\prod_{\substack{j<k\\\sigma_k<\sigma_j}}\frac{1}{h(u_{\sigma_j},u_{\sigma_k})}\right)\\
&=\prod_{j<k}h(u_{j}, u_{k})\prod_{j<k}\frac{1}{h(u_{\sigma_j},u_{\sigma_k})}
\end{aligned}
\end{align}
In the second line, we decomposed the term denoted in blue into a product of two factors. In the third line, we relabelled the indices of the first and the second term. To go to the last line, we combined the second and the third terms in the line above.
After doing so, the wave function takes a concise form:
\begin{align}
\psi (n_1,\ldots, n_M)=\prod_{j<k}h(u_j,u_k)\sum_{\sigma \in S_M} \prod_{j<k}\frac{1}{h(u_{\sigma_j},u_{\sigma_k})}e^{i\sum_jp_{\sigma_j}n_j}
\end{align}

Now using this expression, we can compute the structure constant as
\begin{align}\label{naivesum}
\begin{aligned}
c_{123}&\propto \sum_{1\leq n_1 < \cdots < n_M \leq \ell_{31}} \psi (n_1 ,\ldots , n_M)\\
&= \prod_{j<k}h(u_j,u_k)\sum_{\sigma \in S_M} \left(\prod_{j<k}\frac{1}{h(u_{\sigma_j},u_{\sigma_k})}\right) M(p_{\sigma_1},\ldots ,p_{\sigma_M})
\end{aligned}
\end{align}
with
\begin{align}
M(p_1,\ldots, p_M):= \sum_{1\leq n_1 < \cdots < n_M \leq \ell_{31}} e^{i\sum_j p_{j}n_j}\,.
\end{align}
The function $M$ has the following expression in terms of a sum over (ordered) partitions:
\begin{shaded}
\noindent {\bf Lemma 1}:\\
\begin{align}\label{lemma1}
\begin{aligned}
&M(p_1,\ldots, p_M):=\\
& \sum_{\substack{\alpha=\{1,\ldots, m\}\\\bar{\alpha}=\{m+1,\ldots,M\}}}(-1)^{|\bar{\alpha}|}\left(\prod_{j\in \bar{\alpha}}e^{ip_j \ell_{31}}\right)\left(\prod_{j\in \alpha}\frac{1}{e^{-i \sum_{k=j}^{m}p_k}-1}\right)\left(\prod_{j\in \bar{\alpha}}\frac{e^{i p_j}}{1-e^{i\sum_{k=j+1}^{M}p_k}}\right)
\end{aligned}
\end{align}
Here $|\bar{\alpha}|$ is the number of elements in $\bar{\alpha}$.
\end{shaded}
\begin{itemize}
\item[]{\bf Proof}: Here we only sketch the proof and leave it for readers to fill in details. In terms of the pictorial notation we introduced, the set $\alpha$ ($\bar{\alpha}$) denotes a set of magnons living on the left (right) edge of the segment. 

The basic idea is to use the trick explained above: To compute the contribution corresponding for $\alpha=\{1,\ldots, m\}$, we should first sum over the position of the $m$-th magnon. This produces two terms, one of which corresponds to $n_m= n_{m-1}$ and the other corresponds to $n_{m}=n_{m+1}$. However, the second term cannot contribute to the partition of our interest, where the $m$-th magnon and the $(m+1)$-th are separate. Thus, we just need to keep the first term:
\begin{align}
\begin{aligned}
\sum_{n_m}\substack{\\\includegraphics[clip,width=2.5cm]{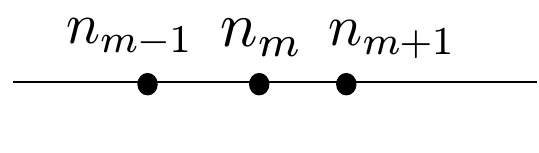}}=g(p_m)\big(\!\!\!\underbrace{\substack{\\\includegraphics[clip,width=2.5cm]{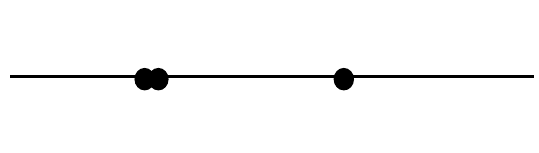}}}_{\text{Only this term contributes}}-\substack{\\\includegraphics[clip,width=2.5cm]{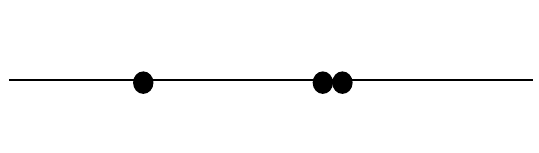}}\big)
\end{aligned}
\end{align}
  We then sum over the position of the $(m-1)$-th magnon. Also in this case, we only need to keep one of two terms that arise from the geometric sum. Performing the same analysis up until the $1$-st magnon, we obtain $\prod_{j\in \alpha}\cdots$ term in \eqref{lemma1}. 

After doing so, we then sum over the positions of the rest of magnons starting from the $(m+1)$-th magnon up to the $M$-th magnon. This reproduces the rest of the terms.\vspace{10pt}
\end{itemize}

The next task is to rewrite \eqref{naivesum} and simplify the expression. From the structure of \eqref{naivesum} and \eqref{lemma1}, it is not hard to see that the final expression is given in terms of the sum over partition $\alpha \cup \bar{\alpha}=\{1,\cdots , M\}$. To a given partition, several different permutations can contribute and we should sum over such permutations to get a simple formula. For instance, if we consider four particles, the partition $\alpha= \{14\}$, $\bar{\alpha}=\{23\}$ receives a contribution from four different permutations:
\begin{align}
\{1,2,3,4\}\mapsto \quad \{1,4,2,3\}\,,\quad\{4,1,2,3\}\,,\quad\{1,4,3,2\}\,,\quad \{4,1,3,2\}\,.
\end{align}
As is clear from this example, these permutations are related with each other by the permutation inside $\alpha$ and the permutation inside $\bar{\alpha}$. This observation leads to the following formula: 
\begin{align}
\begin{aligned}
&\sum_{\sigma \in S_M} \left(\prod_{j<k}\frac{1}{h(u_{\sigma_j},u_{\sigma_k})}\right) M(p_{\sigma_1},\ldots ,p_{\sigma_M})\\
&=\sum_{\alpha \cup \bar{\alpha}=\{1,\ldots , M\}}(-1)^{\bar{\alpha}}\left(\prod_{j\in \bar{\alpha}}e^{ip_j\ell_{31}}\right)\left(\prod_{\substack{i,j\\i\in \alpha,j\in \bar{\alpha}}}\frac{1}{h(u_i,u_j)}\right)F(\alpha)\bar{F}(\bar{\alpha})
\end{aligned}
\end{align}
with
\begin{align}
\begin{aligned}
F(1,2,\ldots,m)&:= \sum_{\sigma\in S_m}\left(\prod_{i<j}\frac{1}{h(u_{\sigma_i},u_{\sigma_j})}\right) \left( \prod_{j=1}^{m}\frac{1}{e^{-i\sum_{k=j}^{m}p_{\sigma_k}}-1} \right)\,,\\
\bar{F}(1,2,\ldots,m)&:= \sum_{\sigma\in S_m}\left(\prod_{i<j}\frac{1}{h(u_{\sigma_i},u_{\sigma_j})}\right) \left( \prod_{j=1}^{m}\frac{e^{ip_{\sigma_j}}}{1-e^{i\sum_{k=j}^{m}p_{\sigma_k}}} \right)\,.
\end{aligned}
\end{align}

Although the definitions of $F$ and $\bar{F}$ involve a sum over permutations, it turns out that they evaluate to a simple product:
\begin{shaded}
{\bf Lemma 2}:
\begin{align}\label{lemma2}
F(1,\cdots,m)=\bar{F}(1,\cdots, m)=\prod_{k=1}^{m}i (u_k+i/2)\,.
\end{align}
\end{shaded}
\begin{itemize}
\item[]{\bf Proof}: We focus on $F(1,\cdots,m)$ since the derivation for $\bar{F}(1,\cdots,m)$ is very similar. The basic strategy is to use the mathematical induction. First, it is trivial to check the relation for $m=1$. Next, suppose the relation holds for $F(1,\ldots,m-1)$. Then, we decompose $F(1,\cdots,m)$ with respect to $\sigma_m$ as follows:
\begin{align}\label{recursion}
\begin{aligned}
F(1,\cdots,m)=&\frac{1}{e^{-i \sum_{k=1}^{m}p_k}-1} \sum_{j=1}^{m} \left(\prod_{k\neq j}\frac{1}{h(u_k,u_j)}\right)\\
&\times\sum_{\sigma^{\prime} \in S_{\{1,\ldots, m\}/j}}\,\,\, \left(\prod_{1\leq i<j\leq m-1}\frac{1}{h(u_{\sigma^{\prime}_i},u_{\sigma^{\prime}_j})}\right)\left(\frac{1}{e^{-i\sum_{k=j}^{m}p_{\sigma^{\prime}_k}}-1} \right)\,.
\end{aligned}
\end{align}
It turns out that the second line in \eqref{recursion} is nothing but $F(1,\ldots,\check{j},\dots,m)$ ($F(1,\ldots,m)$ with $j$ omitted). Thus using the assumption of the induction, we obtain
\begin{align}
\begin{aligned}
F(1,\cdots,m)=&\frac{\prod_{k=1}^{m}i(u_k+i/2)}{e^{-i \sum_{k=1}^{m}p_k}-1} \sum_{j=1}^{m} \frac{1}{i (u_j+i/2)}\prod_{k\neq j}\frac{1}{h(u_k,u_j)}\\
\end{aligned}
\end{align}
To proceed, we use the following identity:
\begin{align}\label{usefulidentity}
\left( e^{-i \sum_{k=1}^{m}p_k}-1 =\right)\prod_{k=1}^{m}\frac{u_k-i/2}{u_k+i/2}-1=\sum_{j=1}^{m} \frac{1}{i (u_j+i/2)}\prod_{k\neq j}\frac{1}{h(u_k,u_j)}\,.
\end{align}
This identity can be shown by first expressing the left hand side using the contour integral as
\begin{align}
\prod_{k=1}^{m}\frac{u_k-i/2}{u_k+i/2}-1=\oint_{z=0} \frac{dz}{2\pi i}\frac{1}{z} \left(\prod_{k=1}^{m}\frac{u_k-z-i/2}{u_k-z+i/2}-1 \right)\,,
\end{align}
then deforming the contour and picking up other poles in the integrand. From this identity, the statement of the lemma follows immediately.
\end{itemize}
Now using the Lemma 2, we arrive at the following expression for the structure constant:
\begin{align}
c_{123}\propto \prod_k i (u_k+i/2)\prod_{j<k} h(u_j, u_k) \sum_{\alpha \cup \bar{\alpha}=\{1,\ldots , M\}}(-1)^{\bar{\alpha}}\left(\prod_{j\in \bar{\alpha}}e^{ip_j\ell_{31}}\right)\left(\prod_{\substack{i,j\\i\in \alpha,j\in \bar{\alpha}}}\frac{1}{h(u_i,u_j)}\right)\,.
\end{align}
To compute the structure constant precisely, we need to normalize the operators: The plane-wave state we have been using has a nontrivial norm $\langle \psi | \psi \rangle$, and we need to divide by its square root in order to get a correct structure constant. Fortunately, for integrable spin chains, there is a compact expression for the norm of the state, which is called the {\it Gaudin formula}\footnote{In this lecture, we will not discuss the Gaudin formula. For details, see for instance \cite{Gaudin}.}. Using the Gaudin formula, we find the following expression for the structure constant:
\begin{shaded}
\begin{align}\label{treefin}
\begin{aligned}
&c_{123}= \frac{\mathcal{A}}{\sqrt{\prod_{i<j}S(u_i,u_j)\det \partial_{u_j} \phi_k}}\,,\\
&\mathcal{A}=\prod_{j<k} h(u_j, u_k) \sum_{\alpha \cup \bar{\alpha}=\{1,\ldots , M\}}(-1)^{|\bar{\alpha}|}\left(\prod_{j\in \bar{\alpha}}e^{ip_j\ell_{31}}\right)\left(\prod_{\substack{i,j\\i\in \alpha,j\in \bar{\alpha}}}\frac{1}{h(u_i,u_j)}\right)\,.
\end{aligned}
\end{align}
\end{shaded}
\noindent Here $\phi_j$ is defined by the Bethe equation as
\begin{align}
e^{i \phi_j}:= e^{i p_j L}\prod_{k\neq j} S(u_j, u_k)\,.
\end{align}
\subsection{Physical interpretation}
Let us now interpret the result \eqref{treefin} physically. For this purpose, we rewrite $\mathcal{A}$ as
\begin{shaded}
\begin{align}\label{treenew}
\mathcal{A}=\sum_{\alpha \cup \bar{\alpha}=\{1,\ldots,M\}} (-1)^{|\bar{\alpha}|} \prod_{j\in \bar{\alpha}}e^{i p_j \ell_{31}} \prod_{\substack{j<k\\j\in \bar{\alpha},k\in \alpha}}S(u_i, u_j) \mathcal{H}(\alpha) \mathcal{H}(\bar{\alpha})\,,
\end{align}
with
\begin{align}\label{multihextree}
\mathcal{H} (\alpha)=\prod_{\substack{i<j\\ i, j\in \alpha}} h(u_i,u_j)\,,\qquad\mathcal{H} (\bar{\alpha})=\prod_{\substack{i<j\\ i, j\in \bar{\alpha}}} h(u_i,u_j)\,.
\end{align}
\end{shaded}
\begin{itemize}
\item[]{\bf Exercise}: Show the equivalence of \eqref{treefin} and \eqref{treenew} using the relation $S(u,v)=h(u,v)/h(v,u)$.
\end{itemize}
\begin{figure}[t]
\begin{center}
\includegraphics[clip,height=5cm]{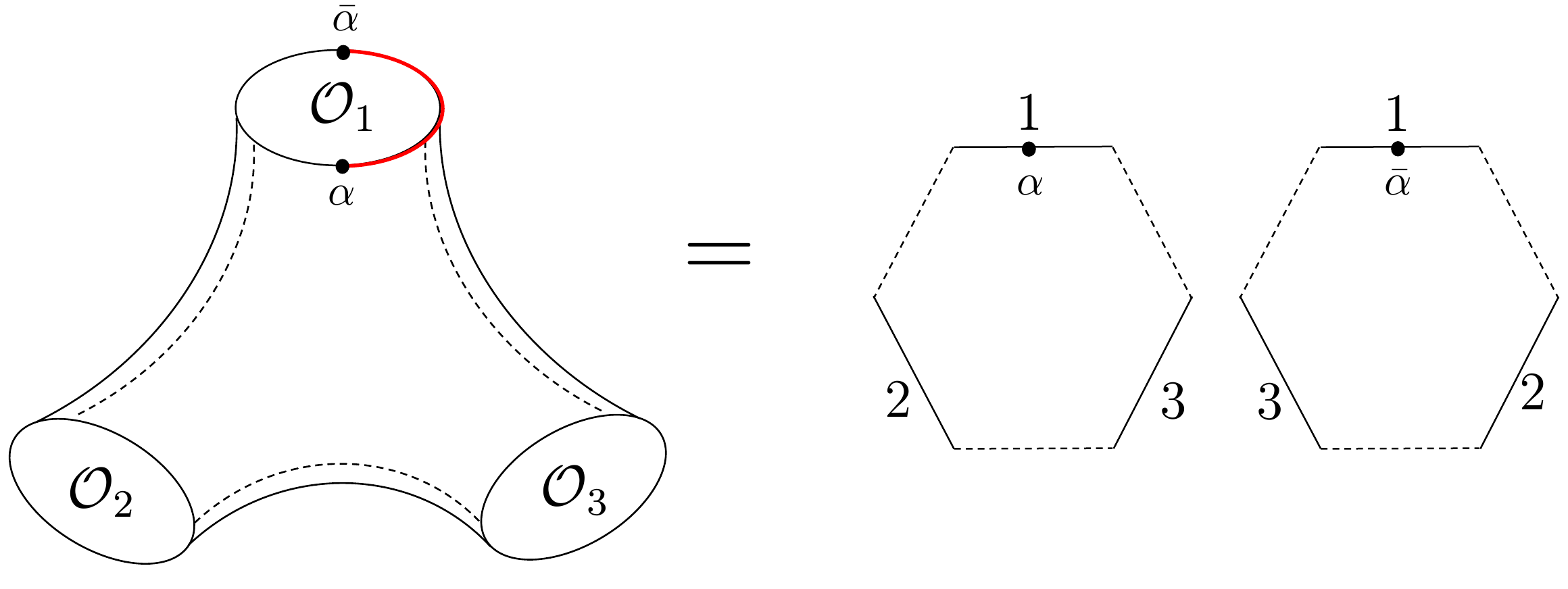}
\end{center}
\vspace{-0.5cm}
\caption{The planar diagram for the three-point function has a topology of a pair of pants. The red curve on the left hand side denotes the segment on which we summed the geometric series. To separate two ends of the segment, we cut the pair of pants into two hexagons as depicted on the right hand side.\label{fig:pants}}
\end{figure}

We now examine this expression more closely. Firstly, it contains a sum over bi-partite partitions. As explained above, the bi-partite partition comes from the geometric series and $\alpha$ ($\bar{\alpha}$) denotes a set of magnons living at the left (right) edge of the segment. If we draw the segment in the full planar diagram, which has a topology of a pair of pants, the two ends of the segment sit at the front and the back of the pair of pants respectively. This naturally leads to the interpretation that $\mathcal{H}(\alpha)$ ($\mathcal{H}(\bar{\alpha})$) corresponds to the contribution coming from the front (back) side of the pair of pants. In order to separate the front and the back sides, it is natural to cut the pair of pants along the dashed lines as denoted in figure \ref{fig:pants}. After doing so, we obtain two hexagons; the one from the front the other from the back. The expression \eqref{treenew} suggests that these hexagons (which we also call {\it hexagon form factors}) are the building blocks for three-point functions.

\begin{figure}[h]
\begin{center}
\includegraphics[clip,height=6.5cm]{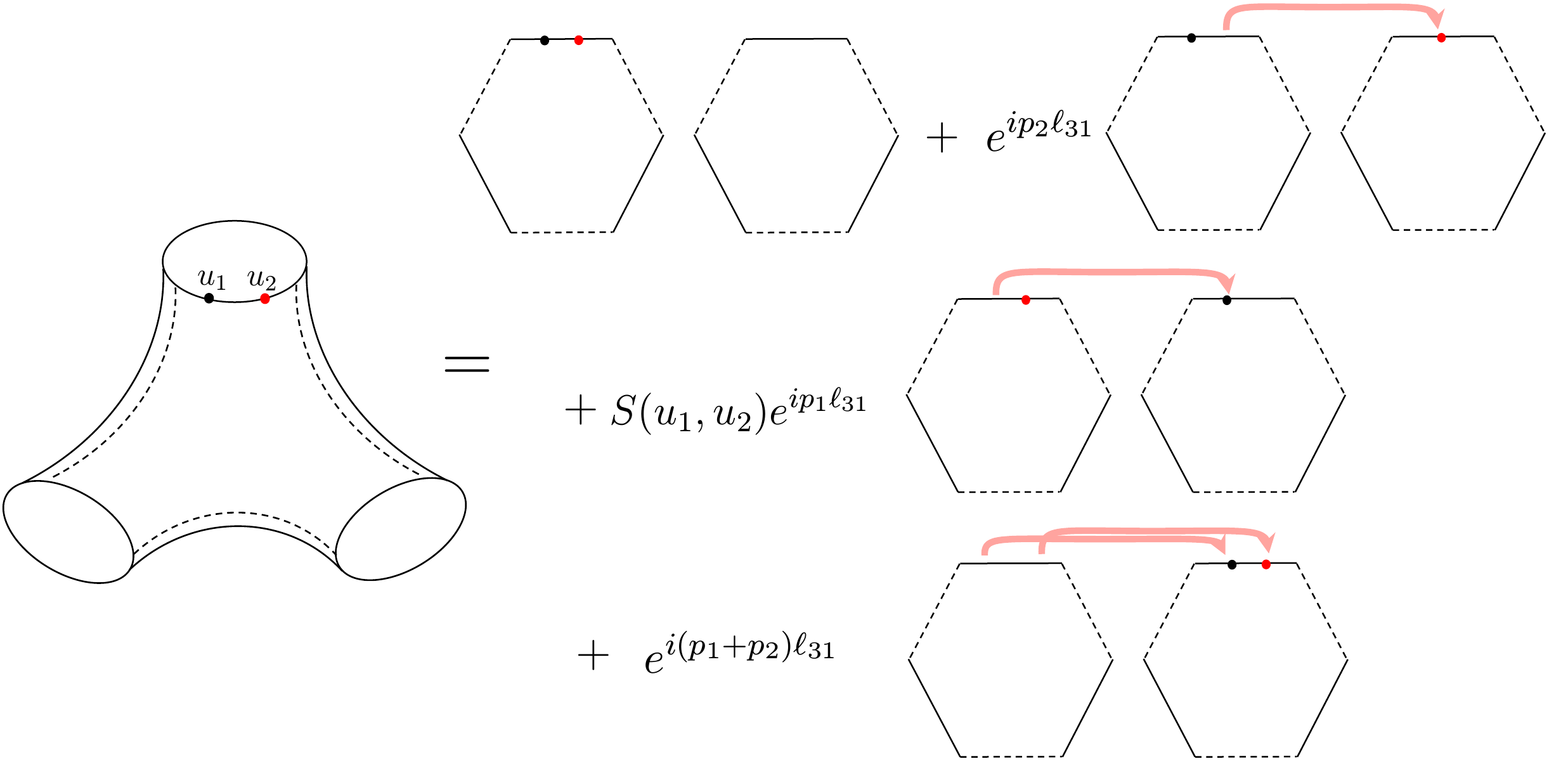}
\end{center}
\caption{A pictorial explanation of the sum over partitions. Each partition comes with the weight factor given by a product of the propagation factor ($e^{ip \ell_{31}}$) and the S-matrices.\label{fig:pants2}}
\end{figure}

The remaining factors admit natural physical interpretation as well. The factor $e^{ip_j \ell_{31}}$ is the phase shift, required for moving the magnons from the first (front) hexagon to the second (back) hexagon. When we move magnons, we sometimes need to change their orders. When this happens, the phase shift also receives the contribution from the S-matrix. The S-matrix factor present in \eqref{treenew} precisely accounts for this effect (see figure \ref{fig:pants2} for an illustrative example).

The only remaining factor is $(-1)^{|\bar{\alpha}|}$.  It is clear why such a factor arises from the sum over the wave function. Unfortunately, however, we have not yet figured out their geometric origin. It would be desirable to understand this factor more physically.

\begin{figure}[t]
\begin{center}
\includegraphics[clip,height=4cm]{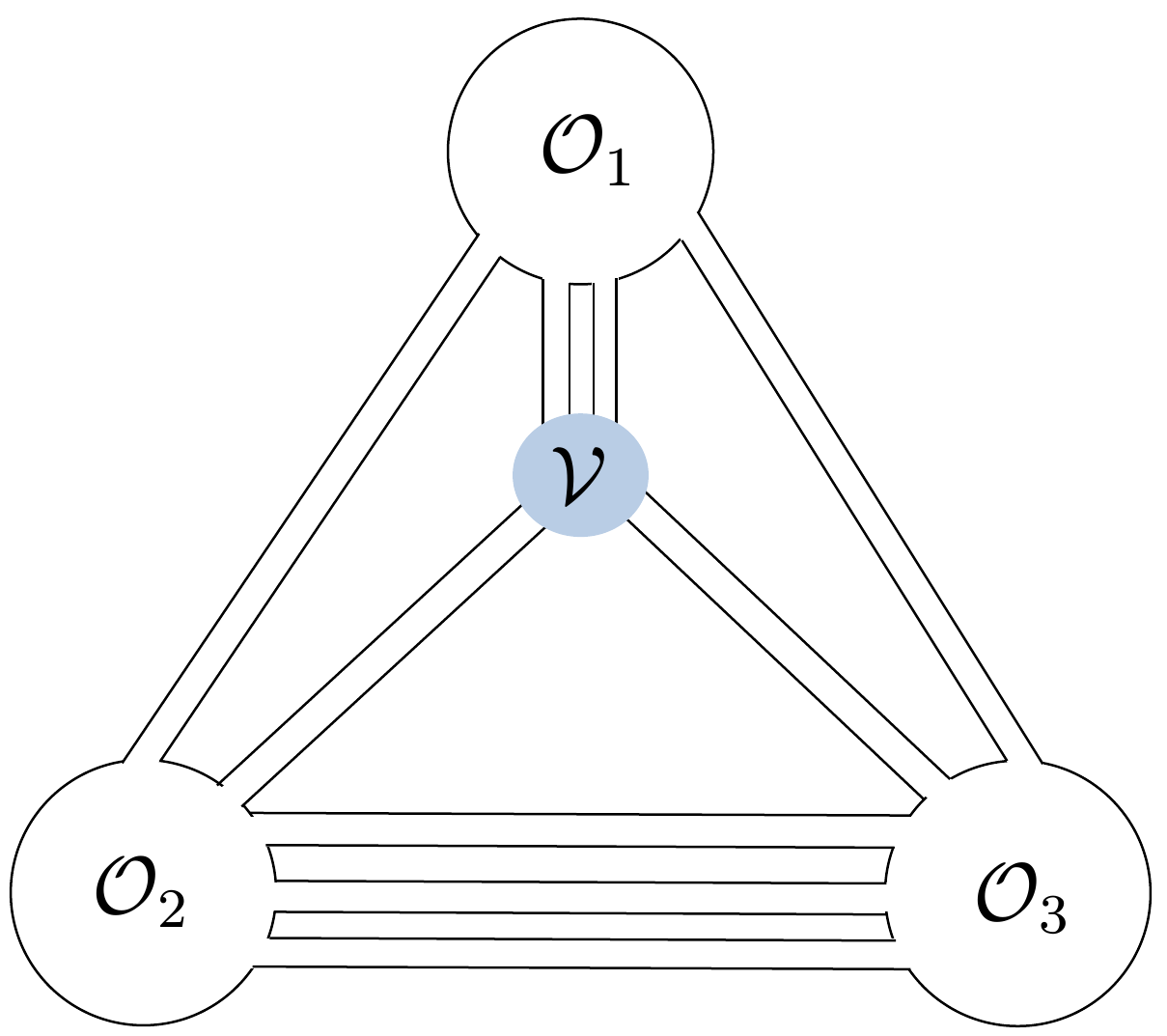}
\end{center}
\caption{One-loop correction to the three-point function. It amounts to inserting an extra spin-chain operator $\mathcal{V}$ at the positions where the operator splits into two. There are six such splitting points (two for each operator) but here we depicted only one of them.\label{fig:oneloop3pt}}
\end{figure}

This concludes our analysis at weak coupling. Before moving on, let us briefly mention what happens if we go to one loop. At one loop, there are two sources of corrections\footnote{For a detail on the one-loop computation, see \cite{OT,ADGN}}: First the wave function should now be the eigenstate of the two-loop dilatation operator. Second there are corrections coming from the one-loop planar diagrams. Although there are numerous diagrams that one can draw at one loop, it turns out that most of them do not contribute to the structure constant. (They contribute either to the anomalous dimensions or to the scheme-dependent normalization of the operators). The only diagrams that contribute to the structure constant turns out to be the ones that one can draw at the positions where the operator splits into two (see figure \ref{fig:oneloop3pt}). In terms of the spin chain, this amounts to inserting some local operator at such points. 

Each of these effects leads to a complicated correction. However, what is truly remarkable is that, if we combine these two effects, the resulting expression simplifies substantially and can be recast again into a sum over partition, now with $h(u,v)$ modified slightly \cite{QInt,TM}. This suggests that the sum over partitions and the physical picture associated with it has deeper meaning. In the second lecture, we will indeed see that this physical picture can be transformed into a non-perturbative framework which allows us to compute the structure constant at finite coupling. 
\newpage
\section{Lecture II: Hexagon approach}
Based on the observations we made in the previous lecture, we now introduce a non-perturbative framework to study the three-point function, called the hexagon approach.
\subsection{Asymptotic three-point function\label{sec:asymptotic3pt}}
Let us first state our conjecture for the finite-coupling generalization of \eqref{treenew} for correlators of one non-BPS SU(2) operator and two BPS operators.
\begin{shaded}
\noindent {\bf Claim} (Asymptotic three-point function): When the bridge lengths satisfy $\ell_{ij}\gg 1$, the sum-over-partition expression \eqref{treenew} holds also at finite coupling, under the following replacements:
\begin{align}\label{conjecturesu2}
e^{i p(u)}&=\frac{x^{+}(u)}{x^{-}(u)}\,,
&& S_{\rm su(2)}(u,v)=\frac{u-v-i}{u-v+i}\frac{1}{\sigma^2 (u,v)}\,,\nonumber\\
H(\alpha)&=\prod_{i\in \alpha} \sqrt{\mu (u_i)} \prod_{\substack{i<j\\i,j\in \alpha}}h_{\rm su(2)}(u_i,u_j)\,,\qquad &&
\mu(u)=\frac{\left(1-1/x^{+}(u)x^{-}(u)\right)^2}{\left(1-1/(x^{-}(u))^2\right)\left(1-1/(x^{+}(u))^2\right)}\nonumber\\
h_{\rm su(2)}(u_i,u_j)&=\frac{u-v}{u-v+i}\frac{1-1/x^{+}(u)x^{+}(v)}{1-1/x^{+}(u)x^{-}(v)}\frac{1-1/x^{-}(u)x^{-}(v)}{1-1/x^{-}(u)x^{+}(v)}\frac{1}{\sigma (u,v)}\,,\hspace{-10cm}&&
\end{align}
The normalization factor (the factor inside the square root in \eqref{treefin}) will also be replaced accordingly.
\end{shaded}
\noindent Here and below the notation $f^{\pm}(u)$ denotes the shift of the argument $f^{\pm}(u)=f(u\pm i/2)$ and the function $x(u)$ is called Zhukowski variable, defined by
\begin{align}\label{Zhukowdef}
u= g\left(x+\frac{1}{x}\right)\,,
\end{align}
where $g$ is related to `t Hooft coupling constant $\lambda$ as 
\begin{align}
g:= \frac{\sqrt{\lambda}}{4\pi}\,.
\end{align}
The function $\sigma(u,v)$ is called the dressing phase. We will not write down their explicit forms since they are complicated and not necessary for the purpose of these lectures. For those who want to know the definitions, see \cite{dressing phase} for instance.
Note also that $H(\alpha)$ now includes the factor $\prod \sqrt{\mu}$, which we call the {\it measure} factor.
\begin{itemize}
\item[] {\bf Exercise 1}: The energy of a magnon is defined by 
\begin{align}
E(u):=\frac{1}{2}\frac{1+1/x^{+}x^{-}}{1-1/x^{+}x^{-}}\,.
\end{align} 
Show that, at weak coupling, it gives
\begin{align}
E(u)=\frac{1}{2}+ \frac{g^2}{u^2+1/4}+O(g^4)\,.
\end{align}
\item[] {\bf Exercise 2}: Show the following useful identity:
\begin{align}
\frac{x^{+}(u)-x^{+}(v)}{1-1/x^{-}(u)x^{-}(v)}=\frac{x^{-}(u)-x^{-}(v)}{1-1/x^{+}(u)x^{+}(v)}\,.
\end{align}
\end{itemize}
Regarding the conjecture above, there are several questions that one would come up with immediately:
\begin{itemize}
\item How do we determine various coupling-dependent quantities such as $h(u,v)$?
\item What about other sectors (beyond the SU(2) sector)?
\item The function $h_{\rm su(2)}(u,v)$ still satisfies\footnote{One can verify this by using $\sigma (u,v)\sigma(v,u)=1$.} the relation $S_{\rm su(2)}(u,v)= h_{\rm su(2)}(u,v)/h_{\rm su(2)}(v,u)$. What is the meaning of this relation?
\end{itemize}
The rest of the lecture will be devoted to answering these questions. We will see that the symmetries (namely the superconformal symmetry and the gauge symmetry) and the integrability are key players in the story.
\subsection{Symmetries of two-point functions}
\begin{figure}[h]
\centering
\includegraphics[clip,height=2.5cm]{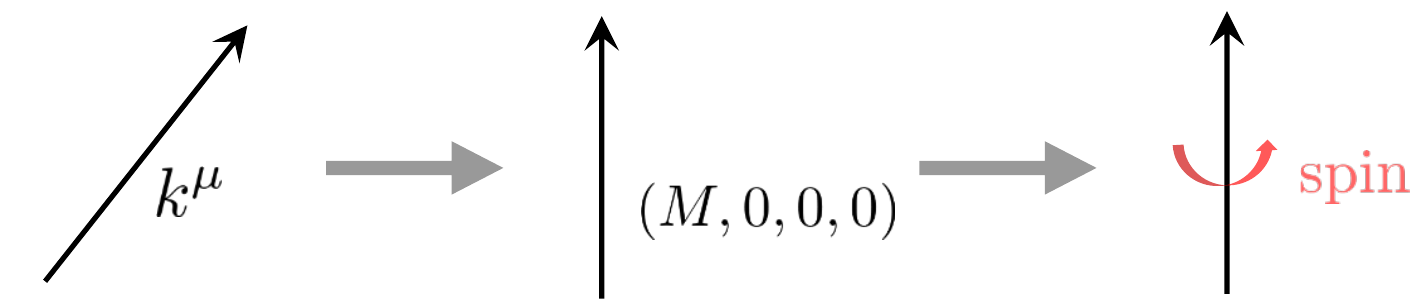}
\caption{Classification of particles using the Poincar\'{e} symmetry. To classify particles in flat space using the Poincar\'{e} symmetry, we first bring the system to a ``canonical configuration" (namely, the rest frame), and then classify the internal motion such as spin by using the little group, which leaves the canonical configuration invariant.\label{fig:littlegroup}}
\end{figure}

Understanding the symmetry is always an important starting point for solving the problem. A textbook example is the classification of particles using the Poincar\'{e} group. The key steps in that case are (see also figure \ref{fig:littlegroup})
\begin{enumerate}
 \item Fix the center of mass motion by going to the {\it rest frame}, in which the 4-momentum takes $P^{\mu}= (M,0,0,0)$.
\item Consider the {\it little group}, which leaves the momentum invariant.
\item Classify the ``internal motion" (namely spin) using the representation theory of the little group.
\end{enumerate}

It turns out that a similar idea works also for our problem: The analogue of step 1 is to consider correlation functions of BPS operators. The correlators of BPS operators often preserve a certain amount of symmetries, which play the role of the little group in step 2. The non-BPS operators can be obtained by adding magnons to BPS correlators, which correspond to exciting the ``internal motion". As in step 3, they are constrained by the little group. In what follows, we will see more in detail how it works for two-point functions. 
\subsubsection*{Global symmetry}
The ``rest frame" for two-point functions is the following correlator of two BPS operators:
\begin{align}
\langle \underbrace{{\rm tr} \left( Z^{L}\right)(0)}_{\mathcal{O}_1}\quad \underbrace{{\rm tr} \left( \bar{Z}^{L}\right)(\infty)}_{\mathcal{O}_2}\rangle \,.
\end{align}

As alluded to above, there is a nontrivial ``little group", which leaves this two-point function invariant. 
Its bosonic part is particularly easy to understand: Since the operators are inserted at the origin and at infinity, the SO$(4)$ rotation around the origin clearly leaves the correlator invariant. In addition to this, there is an SO$(4)$ R-symmetry, which  rotates the scalars $\phi_3 \ldots \phi_6$. 

To identify its fermionic part, we need to know properties of the operators $\mathcal{O}_1$ and $\mathcal{O}_2$. Since the first operator $\mathcal{O}_1$ is a $1/2$-BPS operator, it is annihilated by 8 supercharges ($Q$'s). In addition, it is a {\it superconformal primary operator} and is annihilated by {\it all} 16 superconformal charges ($S$'s). On the other hand, the second operator $\mathcal{O}_2$ is annihilated by 16 $Q$'s and 8 $S$'s, since the operator is at infinity and the roles of supercharges and superconformal charges are swapped. As a consequence, we have 8 supercharges and 8 superconformal charges which annihilate both $\mathcal{O}_1$ and $\mathcal{O}_2$. Together with the bosonic symmetries we just described, they form a super group $\mathfrak{psu}(2|2)^2$.

There is also an extra symmetry generator $D-J$, where $D$ is the dilatation and $J$ is the U$(1)$ rotation $Z \to e^{i\alpha} Z$. This generator commutes with the $\mathfrak{psu}(2|2)^2$ generators and therefore is a central charge of this group.

Having identified the little group, the last step is to classify the ``internal motions" using the symmetry group. In this case, the internal motions correspond to adding magnons
\begin{align}
\cdots ZZ\cdots \to \cdots Z \textcolor[rgb]{1,0,0}{\Psi} Z \textcolor[rgb]{1,0,0}{D}Z \textcolor[rgb]{1,0,0}{X}Z\cdots \quad \text{etc.}
\end{align}
As it turns out, magnons belong to a bi-fundamental representation of the $\mathfrak{psu}(2|2)^2$ symmetry. To make clear the structure of the group and the representation, it is customary to represent the magnons $\mathcal{X}$ as follows:
\begin{align}\label{magnondef1}
\mathcal{X}=\chi^A\dot{\chi}^{\dot{A}}\,,
\end{align}
Here $\chi$ and $\dot{\chi}$ are the fundamental representations of the left and the right $\mathfrak{psu}(2|2)$ respectively and are given by
\begin{align}\label{magnondef2}
\chi^A = (\varphi^1,\varphi^2, \psi^1,\psi^2)\,,\qquad\dot{\chi}_{\dot{A}} = (\dot{\varphi}^1,\dot{\varphi}^2, \dot{\psi}^1,\dot{\psi}^2)\,.
\end{align}
They are related to the fields in $\mathcal{N}=4$ SYM as
\begin{align}
\begin{aligned}
&\varphi^1\dot{\varphi}^1=X \,,\quad \varphi^1\dot{\varphi}^2=Y\,,\quad\varphi^2\dot{\varphi}^1=\bar{Y}\,,\quad\varphi^2\dot{\varphi}^2=-\bar{X}\,,\\
&\psi^\alpha\dot{\psi}^{\dot{\alpha}}=D^{\alpha\dot{\alpha}}Z\,, \qquad \qquad \psi^{\alpha}\dot{\varphi}^{\dot{a}},\varphi^a\dot{\psi}^{\dot{\alpha}}=\text{fermion} \,.
\end{aligned}
\end{align}
Here $D$ is a covariant derivative and $X$, $\bar{X}$, $Y$ and $\bar{Y}$ are the complex scalars defined in Lecture I.

From the analysis above, one can derive certain constraints on the dynamics of magnons. They however are not strong enough. The main problem is that they know nothing about the dependence on the coupling constant $g$. 

\begin{figure}[h]
\centering
\includegraphics[clip,height=4cm]{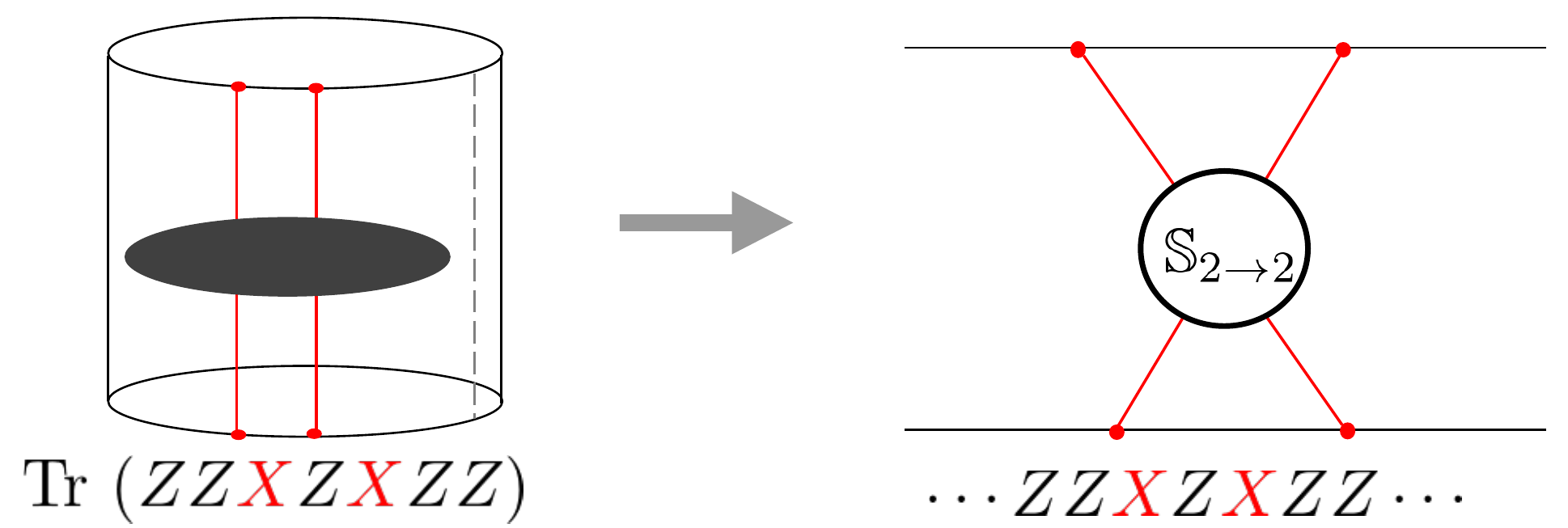}
\caption{The decompactification of the spin chain. At higher loops, the spin-chain Hamiltonian can take quite a complicated form. By considering an infinitely long spin chain and focus on the S-matrix, we can avoid dealing with the Hamiltonian directly and constrain the dynamics by symmetry.\label{fig:cutopen}}
\end{figure}

The key idea to overcome this difficulty is to consider an infinitely long operator and cut open the trace (figure \ref{fig:cutopen}):
\begin{align}
\text{tr} \left(\cdots\right) \to \cdots Z \cdots Z \cdots
\end{align}
There are two advantages in doing this: First, since the spin chain now has an infinite length, we can define the asymptotic states {\it on the spin chain} and study their S-matrix. Focusing on the S-matrix allows us to study the consequences of the symmetry easily, without directly dealing with complicated intermediate processes generated by the Hamiltonian. Second, since the trace is cut open, there exists large gauge transformations, namely the transformation which do not die off at infinity. As is often the case, these transformations should be regarded as a part of the global symmetries. As we see below, the use of the large gauge transformation is crucial for determining the S-matrix at finite coupling.
  
\subsubsection*{``Large gauge" symmetry}
The gauge transformation always comes with the coupling constant as $\delta \Phi \sim g [\Phi, \Lambda]$. This seemingly simple fact leads to a powerful non-perturbative method when combined with other symmetries.

To see this, let us recall the transformation laws of the supersymmetry in $\mathcal{N}=4$ SYM\footnote{Here we only write down the terms necessary for our purposes. The full transformation laws can be found for instance in \cite{BeisertThesis}.}:
\begin{align}
\delta \Phi= i \bar{\epsilon}\Gamma_i \Psi\,, \qquad \delta \Psi = \cdots +\frac{i}{2}g[\Phi_i,\Phi_j]\Gamma^{ij}\epsilon+\cdots \,.
\end{align}
It is straightforward to see that, acting the supersymmetry transformations twice to a fermion, we get a term of the form
\begin{align}
\delta^2 \Psi \sim g[\Phi, \Psi]\,.
\end{align}
This is a (field-dependent) gauge transformation and therefore is proportional to the coupling constant.
In particular, for the supercharges belonging to the $\mathfrak{psu}(2|2)^2$, we have
\begin{align}
Q^2 \mathcal{X} \sim  g [Z,\mathcal{X}] \quad (:= P\cdot \mathcal{X})\,.
\end{align} 
In what follows, we denote this field-dependent transformation as $P$. 

Since $P$ is a gauge transformation, it acts trivially on the gauge-invariant quantities:
\begin{align}
P\cdot \text{tr}\left(\cdots\right)\sim \text{tr}\left([Z,\cdots]\right)=0\,.
\end{align}
It however can act nontrivialy once we cut open the trace and consider an infinitely long operator since it becomes the ``large gauge transformation". For instance, the action on the plane-wave state (of infinite length) reads\footnote{In the second line, we neglected the extra $Z$ field since the operator is infinitely long.}
\begin{align}
\begin{aligned}
P\cdot \left(\sum_n e^{i pn} |\cdots \underbrace{\mathcal{X}}_{n\text{-th site}}\cdots \rangle\right)&\sim g\sum_n e^{i pn} \left(|\cdots Z\underbrace{\mathcal{X}}_{n+1\text{-th}}\cdots \rangle-|\cdots \underbrace{\mathcal{X}}_{n\text{-th}}Z\cdots \rangle\right)\\
&=g(e^{-ip}-1)\sum_n e^{i pn} |\cdots \underbrace{\mathcal{X}}_{n\text{-th}}\cdots \rangle\,.
\end{aligned}
\end{align} 
This shows that the transformation $P$ is the shift (difference) operator of the spin chain.

The symmetry algebra $\mathfrak{psu}(2|2)^2$ contains the superconformal generators $S$'s in addition to supercharges. The study of two-loop dilatation operators \cite{BeisertThesis} suggests that, at finite coupling, $(S)^2$ also produces a field-dependent gauge transformation,
\begin{align}\label{S2=K}
(S)^2 \mathcal{X} \sim  g [Z^{-1},\mathcal{X}] \quad (:= K\cdot \mathcal{X})\,.
\end{align}
Unlike the transformation $P$, the transformation $K$ defined above is purely quantum. Namely, it is not visible at the level of the classical transformation properties of the fields. Their existence is however inevitable consequence of the anti-commutation relation $\{Q,S\}\sim D$: Since $D$ receives the quantum correction (which yields anomalous dimensions), the generators $Q$ and $S$ must also contain the coupling-dependent part. Once one determines the coupling-dependent part of $S$, one can then confirm that $(S)^2$ yields a field-dependent gauge transformation \eqref{S2=K}. An important property of these extra generators $P$ and $K$ is that they commute with all the other generators in $\mathfrak{psu}(2|2)^2$. Namely, they are central charges of the algebra.

\subsubsection*{Centrally-extended $\mathfrak{psu}(2|2)$ }
Based on these observations, Beisert proposed the centrally-extended $\mathfrak{psu}(2|2)^2$ as the algebra governing the scattering on the spin chain for $\mathcal{N}=4$ SYM \cite{dynamic, analytic}. It consists of generators\footnote{Here undotted generators belong to the left $\mathfrak{psu}(2|2)$ and dotted generators belong to the right $\mathfrak{psu}(2|2)$.} of $\mathfrak{psu}(2|2)^2$,
\begin{align}
\begin{aligned}
&\text{Supersymmetry: }Q^{\alpha}{}_{a}\,,\dot{Q}^{\dot{\alpha}}{}_{\dot{a}}\,,\quad&&\text{Superconformal: } S^{a}{}_{\alpha}\,, \dot{S}^{\dot{a}}{}_{\dot{\alpha}}\,,\quad \\
&\text{R-symmetry: }R^{a}{}_{b}\,,\dot{R}^{\dot{a}}{}_{\dot{b}}\,,\quad &&\text{Lorentz: }L^{\alpha}{}_{\beta}\,,  \dot{L}^{\dot{\alpha}}{}_{\dot{\beta}}\,,
\end{aligned}
\end{align}
and three central charges:
\begin{align}
P, \quad K, \quad C\,\,(:=(D-J)/2)\,.
\end{align}
As mentioned above, these central charges appear in the anti-commutators of the fermionic generators as
\begin{shaded}
\begin{align}\label{psucom}
\begin{aligned}
\{Q^{\alpha}{}_{a},Q^{\beta}{}_{b}\}&=\epsilon^{\alpha\beta}\epsilon_{ab}P\,,\\
\{S^{a}{}_{\alpha},S^{b}{}_{\beta}\}&=\epsilon_{\alpha\beta}\epsilon^{ab}K\,,\\
\{Q^{\alpha}{}_{a},S^{b}{}_{\beta}\}&= \delta^{\alpha}_{\beta}R^{b}{}_{a}+\delta^{b}_{a}L^{\alpha}{}_{\beta}+\delta^{b}_{a}\delta^{\alpha}_{\beta}C
\end{aligned}
\end{align}
\end{shaded}
\noindent Here and below we only write down the formulas for the left $\mathfrak{psu}(2|2)$ since the ones for the right $\mathfrak{psu}(2|2)$ are similar\footnote{Namely, we just need to put dots to the $\mathfrak{psu}(2|2)$ generators in \eqref{psucom}.}. 

These generators act on magnons \eqref{magnondef1} as\footnote{The actions of the bosonic generators of $\mathfrak{psu}(2|2)$ are standard: It merely acts on indices.}
\begin{shaded}
\begin{align}
\begin{aligned}
&Q^{\alpha}{}_{a}|\varphi^{b}(u)\rangle =\textbf{a}\, \delta^{b}_{a}|\psi^{\alpha}(u)\rangle\,,\quad &&Q^{\alpha}{}_{a}|\psi^{\beta}(u)\rangle ={\bf b}\epsilon^{\alpha\beta}\epsilon_{ab}|Z\varphi^{b}(u)\rangle\,,\\
&S^{a}{}_{\alpha}|\varphi^{b}(u)\rangle =\textbf{c}\,\epsilon^{ab}\epsilon^{\alpha\beta}|Z^{-1}\psi^{\beta}(u)\rangle\,,\quad &&S^{a}{}_{\alpha}|\psi^{\beta}(u)\rangle =\textbf{d}\,\delta^{\beta}_{\alpha}|\varphi^{a}(u)\rangle\,,
\\
&P|\mathcal{X}^{A\dot{A}}(u)\rangle=g \left(1-\frac{x^{+}(u)}{x^{-}(u)}\right)|Z\mathcal{X}^{A\dot{A}}\rangle\,,\quad &&K|\mathcal{X}^{A\dot{A}}(u)\rangle=g \left(1-\frac{x^{-}(u)}{x^{+}(u)}\right)|Z^{-1}\mathcal{X}^{A\dot{A}}\rangle\,,
\end{aligned}
\end{align}
with
\begin{align}\label{abcddef}
\begin{aligned}
&{\bf a}=\sqrt{g}\gamma\,,\quad {\bf b}=\frac{\sqrt{g}}{\gamma}\left(1-\frac{x^{+}}{x^{-}}\right)\,,\quad {\bf c}=\frac{i \sqrt{g}\gamma}{x^{+}}\,,\quad {\bf d}=\frac{\sqrt{g}x^{+}}{i\gamma}\left(1-\frac{x^{+}}{x^{-}}\right)\,,\\
&\hspace{5cm}\gamma :=\sqrt{i (x^{-}-x^{+})}\,.
\end{aligned}
\end{align}
\end{shaded}
\noindent Here $u$ is the rapidity of the magnon and $Z^{\pm}$ accounts for the change of the length caused by a field-dependent gauge transformation. Note that, for $P$ and $K$, we used the full expression of a magnon, $\mathcal{X}^{A\dot{A}}$, instead of its left/right part since the central charges are shared between two $\mathfrak{psu}(2|2)$ factors and they act both on the left and the right parts. The positions of $Z^{\pm}$ inside the state can be changed according to the rule,
\begin{shaded}
\begin{align}\label{moveZ}
|\mathcal{X}^{A\dot{A}}Z^{\pm}\rangle=\left(\frac{x^{+}}{x^{-}}\right)^{\pm}|Z^{\pm}\mathcal{X}^{A\dot{A}}\rangle\,.
\end{align}
\end{shaded}
\noindent Changing the position of $Z^{\pm}$ corresponds to moving the position of the excitation by one site (forward or backward). This results in a phase shift $e^{\pm ip }=(x^{+}/x^{-})^{\pm}$, which is present on the right hand side of \eqref{moveZ}.

As shown in \eqref{abcddef}, ${\bf b}$ and ${\bf d}$ contain a factor $(1-e^{ip(u)})$ since they correspond to field-dependent gauge tranformations described above\footnote{This can also be understood from the fact that these transformations involve $Z^{\pm}$.}.
\begin{itemize}
\item[] {\bf Exercise 1}: Derive the actions of $P$ and $K$ from actions of fermionic generators and the anti-commutation relations \eqref{psucom}. 
\item[] {\bf Exercise 2}: Check that ${\bf a},{\bf d}\sim O(1)$ and ${\bf b}, {\bf c}\sim O(g)$ at weak coupling. (See also \eqref{Zhukowdef}.) This reflects the fact that the transformations associated with ${\bf b} $ and ${\bf c}$ are gauge transformations whereas others are not. 
\end{itemize}

As shown above, the centrally-extended $\mathfrak{psu}(2|2)$ ``knows" the coupling constant $g$. Therefore, any quantities that are determined by this symmetry group entail the full coupling-constant dependence. For instance, the $2 \to 2$ $S$-matrix of magnons are fixed up to overall scalar factor by imposing the symmetry condition:
\begin{align}
[h, \mathbb{S}_{2\to 2}(u,v)]=0 \quad (h\in \mathfrak{psu}(2|2))\qquad \Rightarrow\qquad \mathbb{S}_{2\to 2}= S_0 \left(\hat{S}(u,v)\otimes \hat{S}(u,v)\right) \,.
\end{align}
Here $\hat{S}$ is the matrix structure determined by the symmetry\footnote{For explicit expressions for $\hat{S}$, see Appendix \ref{ap-a} and also  \cite{dynamic,analytic}.} and $S_0$ is the overall scalar factor. As shown in \cite{Janik,BES}, $S_0$ can be determined by imposing the crossing symmetry of the $S$-matrix.

A notable feature of the two-body $S$-matrix $\mathbb{S}_{2\to 2}$ is that it satisfies the Yang-Baxter equation. Note that this is just a necessary condition for the existence of the integrability and the factorization of the multi-particle $S$-matrices. One can nevertheless proceed by {\it assuming} the existence of integrability and express the multi-particle $S$-matrices in terms of $\mathbb{S}_{2\to 2}$'s. After doing so, one can then follow a standard recipe of solving integrable systems to determine the spectrum: Namely, write down the Bethe equation to determine the asymptotic spectrum and use the Thermodynamic Bethe ansatz to compute the finite-size spectrum. The results turn out to match the direct perturbative computation up to surprisingly high orders, providing strong evidence for the integrability of this model.
\subsection{Symmetries of three-point functions}
Let us now move on to the discussion on the three-point function. 
\subsubsection*{Twisted translations and diagonal $\mathfrak{psu}(2|2)$}
Again the starting point is to consider the ``rest frame" and identify the ``little group". As in the two-point function, we first consider the correlator of three BPS operators
\begin{align}
\langle \underbrace{{\rm tr}(P_1\cdot \Phi)^{L_1} (x_1)}_{\mathcal{O}_1}\,\,\,\,\underbrace{{\rm tr}(P_2\cdot \Phi)^{L_2} (x_2)}_{\mathcal{O}_2}\,\,\,\,\underbrace{{\rm tr}(P_3\cdot \Phi)^{L_3} (x_3)}_{\mathcal{O}_3}\rangle\,,
\end{align}
To discuss the symmetry, it is convenient to perform the conformal $+$ R-symmetry transformations and bring them to the so-called ``twisted translation frame",
\begin{align}
\begin{aligned}
\mathcal{O}_1&= {\rm tr}(\mathfrak{Z}^{L_1}(a_1))\,, \quad \mathcal{O}_2&= {\rm tr}(\mathfrak{Z}^{L_2}(a_2))\,,\quad \mathcal{O}_3&= {\rm tr}(\mathfrak{Z}^{L_3}(a_3))\,,
\end{aligned}
\end{align} 
where $\mathfrak{Z}(a)$ is a twisted-translated scalar defined by
\begin{align}
\mathfrak{Z}(a)= e^{\mathcal{T} a} Z (0)=\left(Z+a^2 \bar{Z}+ a (Y-\bar{Y})\right)(0,a,0,0)
\end{align}
with
\begin{align}
\mathcal{T}:=-i \epsilon_{\alpha\dot{\alpha}}P^{\dot{\alpha}\alpha}+\epsilon_{\dot{a}a}R^{a\dot{a}}\,.
\end{align}

This configuration is invariant under the subgroup $h$ of $\mathfrak{psu}(2|2)^2$ that commute with the twisted-translation generator $\mathcal{T}$, namely $[h,\mathcal{T}]=0$. It turns that that this ``little group" is a diagonal $\mathfrak{psu}(2|2)$ whose generators 
are given by
\begin{shaded}
\begin{align}
\begin{aligned}
&\mathcal{L}^{\alpha}{}_{\beta}:= L^{\alpha}{}_{\beta}+\dot{L}^{\alpha}{}_{\beta}\,,\quad &&\mathcal{R}^{a}{}_{b}:=R^{a}{}_{b}+\dot{R}^{a}{}_{b}\,,\\
&\mathcal{Q}^{\alpha}{}_{a}:=Q^{\alpha}{}_{a}+i\epsilon^{\alpha\dot{\beta}}\epsilon_{a\dot{b}}\dot{S}^{\dot{b}}{}_{\dot{\beta}}\,,\quad 
&&\mathcal{S}^{a}{}_{\alpha}:=S^{a}{}_{\alpha}+i\epsilon^{a\dot{b}}\epsilon_{\alpha\dot{\beta}}\dot{Q}^{\dot{\beta}}{}_{\dot{b}}\,.
\end{aligned}
\end{align}
\end{shaded}
\noindent In addition to these generators, the commutator between $\mathcal{Q}$ and $\mathcal{S}$ yields one central charge $\mathcal{P}:=P-K$:
\begin{shaded}
\begin{align}
\begin{aligned}
&\{\mathcal{Q}^{\alpha}{}_{a},\mathcal{Q}^{\beta}{}_{b}\}=\epsilon^{\alpha\beta}\epsilon_{ab}\mathcal{P}\,,\qquad\{\mathcal{S}^{a}{}_{\alpha},\mathcal{S}^{b}{}_{\beta}\}=-\epsilon_{\alpha\beta}\epsilon^{ab}\mathcal{P}\,,\\
&\hspace{50pt}\{\mathcal{Q}^{\alpha}{}_{a},\mathcal{S}^{b}{}_{\beta}\}=\delta^{\alpha}_{\beta}\mathcal{R}^{b}{}_{a}+\delta^{b}_{a}\mathcal{L}^{\alpha}{}_{\beta}\,.
\end{aligned}
\end{align}
\end{shaded}
\subsubsection*{Adding ``internal motions"}
Having understood the symmetry for BPS three-point functions, the next step is to add the ``internal motion" to describe more general correlators, which involve non-BPS operators. The best way to do this while keeping the underlying symmetries manifest is to first add magnons to the ``vacuum state" and then perform the twisted translation:
\begin{align}
{\rm tr} \left( Z^{L}\right)(0) \,\, \overset{\text{add magnons}}{\longrightarrow} \,\,\underbrace{{\rm tr}\left(\cdots Z \textcolor[rgb]{1,0,0}{X}Z\textcolor[rgb]{1,0,0}{\Psi}Z\cdots \right)(0)}_{\mathcal{O}(0)} \,\, \overset{\text{twisted translate}}{\longrightarrow} \,\, \mathcal{O}(a):= e^{\mathcal{T} a}\cdot \mathcal{O}(0)\,.
\end{align}
When the operator $\mathcal{O}(0)$ is a conformal primary and is highest weight in R-symmetry, the space-time dependence of the three-point function is fixed to be as follows:
\begin{align}
\langle\mathcal{O}_1(a_1)\mathcal{O}_2(a_2)\mathcal{O}_3 (a_3)\rangle&= \frac{\mathbb{C}_{123}}{(a_1-a_2)^{\delta_{12|3}}(a_2-a_3)^{\delta_{23|1}}(a_3-a_1)^{\delta_{31|2}}}\,.
\end{align}
Here $\delta_{ij|k}$ is given by
\begin{align}
\delta_{ij|k}=2 (C_i+C_j-C_k)\,,
\end{align}
where $C_i$ is the central charge
\begin{align}
C_i = (\Delta_i -J_i)/2\,.
\end{align}
What the hexagon approach computes is the proportionality factor\footnote{In general, the factor $\mathbb{C}_{123}$ contains a tensor structure when the operators have spin. What is important however is that it does not depend on the positions $a_i$.} $\mathbb{C}_{123}$.
\subsection{Constraining the hexagon form factors from symmetries}
We are now in a position to constrain the hexagon form factor 
\begin{align}\label{defhpsi}
\mathcal{H}(\Psi)=\langle \mathcal{H}|\Psi\rangle\,
\end{align}
using the diagonal $\mathfrak{su}(2|2)$ (which we denote as $\mathfrak{psu}_D(2|2)$ below). In \eqref{defhpsi}, $|\Psi\rangle $ is the spin-chain state for which we compute the form factor. In this representation, the symmetry constraints read
\begin{align}
\langle \mathcal{H}| h =0\,, \qquad h\in \mathfrak{su}_D(2|2)
\end{align} 
\subsubsection*{One-particle form factor}
Let us first consider the case with a single magnon. 

As stated above, each magnon belongs to a bi-fundamental representation of $\mathfrak{psu}(2|2)^2$ and thus is denoted by $\chi^{A}\dot{\chi}^{\dot{A}}$. By imposing the bosonic part of $\mathfrak{su}_D(2|2)$, we can show
\begin{align}
\begin{aligned}
&\langle \mathcal{H}|\psi^{\alpha}\psi^{\dot{\beta}}\rangle=\sqrt{\mu(u)}\,\epsilon^{\alpha\dot{\beta}}\,,\\
&\langle \mathcal{H}|\varphi^{a}\psi^{\dot{\alpha}}\rangle=\langle \mathcal{H}|\psi^{\alpha}\varphi^{\dot{a}}\rangle=0\,,\\
&\langle \mathcal{H}|\varphi^{a}\varphi^{\dot{b}}\rangle=\sqrt{\nu(u)}\,\epsilon^{a\dot{b}}\,,
\end{aligned}
\end{align}
where $\mu$ and $\nu$ are functions of the rapidity $u$, which are not fixed by the symmetry. (We are defining them with square-roots for later convenience.) The derivation is pretty much straightforward. For instance by imposing $\langle \mathcal{H}|\mathcal{R}^{1}{}_{2}=0$, we get
\begin{align}
\begin{aligned}
0&=\langle \mathcal{H}| \mathcal{R}^{1}{}_{2}|\varphi^{1}\varphi^{2}\rangle=\langle \mathcal{H}| \left(R^{1}{}_2+\dot{R}^{\dot{1}}{}_{\dot{2}} \right)|\varphi^{1}\varphi^{2}\rangle=\langle \mathcal{H}|\varphi^{1}\varphi^{\dot{2}}\rangle+\langle \mathcal{H}|\varphi^{2}\varphi^{\dot{1}}\rangle\,,\\
&\Leftrightarrow \qquad \langle \mathcal{H}|\varphi^{1}\varphi^{\dot{2}}\rangle=-\langle \mathcal{H}|\varphi^{2}\varphi^{\dot{1}}\rangle\,.
\end{aligned}
\end{align}

We next consider the constraint from the central charge $\mathcal{P}$. The Ward identity for $\mathcal{P}$ reads
\begin{align}
0=\langle \mathcal{H}|\mathcal{P}|\chi^{A}\chi^{\dot{A}}\rangle=g \left(1-e^{ip}\right)\langle \mathcal{H}|Z\chi^{A}\chi^{\dot{A}}\rangle -g \left(1-e^{-ip}\right)\langle \mathcal{H}|Z^{-}\chi^{A}\chi^{\dot{A}}\rangle\,.
\end{align}
The ``square-root" of this relation leads to
\begin{align}
\langle\mathcal{H}|Z^{\pm}\chi^{A}\chi^{\dot{A}}\rangle=\left(i U^{-1}\right)^{\pm}\langle\mathcal{H}|\chi^{A}\chi^{\dot{A}}\rangle
\end{align}
with $U=e^{ip/2}$.

Finally we have contraints coming from fermionic charges. The action of $\mathcal{Q}^{\beta}{}_{b}$ on $|\varphi^{a}\psi^{\dot{\alpha}}\rangle$ is given by\footnote{Recall that $\mathcal{Q}^{\alpha}{}_{a}$ is given by a linear combination $Q^{\alpha}{}_{a}+i\epsilon^{\alpha\dot{\beta}}\epsilon_{a\dot{b}}\dot{S}^{\dot{b}}{}_{\dot{\beta}}$.}
\begin{align}
\mathcal{Q}^{\beta}{}_{b} |\varphi^{a}\psi^{\dot{\alpha}}\rangle = {\bf a}\delta^{a}_{c} |\psi^{\beta}\psi^{\dot{\alpha}}\rangle+i{\bf d}\epsilon^{\beta \dot{\alpha}}\epsilon_{b\dot{c}}|\varphi^{a}\varphi^{\dot{c}}\rangle\,.
\end{align}
The Ward identity associated with this action gives
\begin{shaded}
\begin{align}\label{munu}
\sqrt{\nu(u)}=\frac{{\bf a}}{i{\bf d}}\sqrt{\mu(u)} =-i \sqrt{\mu(u)}\,.
\end{align}
\end{shaded}
\begin{itemize}
\item[] {\bf Exercise}: Derive \eqref{munu} from $\langle \mathcal{H}|\mathcal{Q}^{\beta}{}_{b}|\varphi^{a}\psi^{\dot{\alpha}}\rangle =0 $.
\end{itemize}
\noindent We can also consider the Ward identity associated with $\mathcal{S}^{b}{}_{\beta}$, but it turns out that they also yield the same relation.
\begin{itemize}
\item[] {\bf Exercise}: Check that the Ward identity for $\mathcal{S}^{b}{}_{\beta}$ also gives \eqref{munu}.
\end{itemize}
\begin{figure}[t]
\centering
\includegraphics[clip,height=3.5cm]{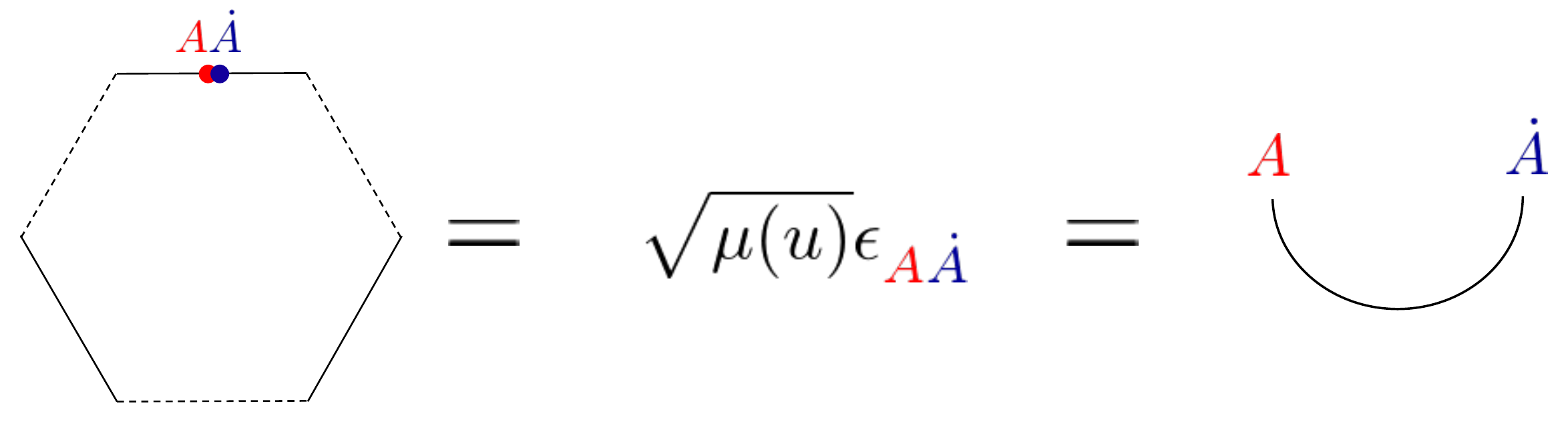}
\caption{A pictorial representation of the one-particle hexagon form factor. The line connecting two indices denotes a product of the measure factor and the invariant tensor (denoted by $\epsilon_{A\dot{A}}$ in the figure). \label{fig:onephex}}
\end{figure}

In summary, the one-particle form factor is constrained by the symmetries as
\begin{shaded}
\begin{align}
\begin{aligned}
&\langle \mathcal{H}|\psi^{\alpha}\psi^{\dot{\beta}}\rangle=\sqrt{\mu(u)}\,\epsilon^{\alpha\dot{\beta}}\,,\\
&\langle \mathcal{H}|\varphi^{a}\psi^{\dot{\alpha}}\rangle=\langle \mathcal{H}|\psi^{\alpha}\varphi^{\dot{a}}\rangle=0\,,\\
&\langle \mathcal{H}|\varphi^{a}\varphi^{\dot{b}}\rangle=-i\sqrt{\mu(u)}\,\epsilon^{a\dot{b}}\,.
\end{aligned}
\end{align}
\end{shaded}
\noindent The overall factor $\mu (u)$ is often called the {\it measure factor}.

It is convenient to represent this structure of the one-particle form factor as in figure \ref{fig:onephex}. In this pictorial representation, a line connecting two indices denotes a product of the {\it measure factor} $\sqrt{\mu}$ and the invariant tensor of $\mathfrak{psu}(2|2)$ ($\epsilon^{\alpha\dot{\beta}}$ and $\epsilon^{a\dot{b}}$).
\subsubsection*{Two-particle form factor}
We can also determine the matrix structure of two-particle form factors using the symmetry. Since the analysis is straightforward but more complicated than one-particle cases, here we simply state the final result:
\begin{shaded}
\noindent {\it Two-particle form factor}:\\
The two-particle form factor consists of the {\it matrix part} $\mathcal{H}_{\text{mat}}$, which is constrained by the symmetry,  and the {\it dynamical factor} $h$, which is an overall scalar factor and not fixed by the symmetry:
\begin{align}
\langle \mathcal{H}|\Big(\chi^{A_1}\chi^{\dot{A}_1}\Big)(u_1)\Big(\chi^{A_2}\chi^{\dot{A}_2}\Big)(u_2)\rangle=h(u_1,u_2)\mathcal{H}_{\text{mat}} (u_1,u_2)\,.
\end{align}
The matrix part $\mathcal{H}_{\rm mat}$ can be computed as follows:
\begin{enumerate}
\item First rearrenge the indices $(A_1\dot{A}_1)(A_2\dot{A}_2)$ to $(A_1A_2)(\dot{A}_1\dot{A}_2)$. This rearrangement produces a sign factor $(-1)^{|\dot{A}_1||A_2|}$, where $|A|=0$ for $\varphi$'s and $|A|=1$ for $\psi$'s.
\item After the rearrangement, we obtain the state
\begin{align}
|\chi^{A_1}(u_1)\chi^{A_2}(u_2)\rangle \otimes |\chi^{\dot{A}_1}(u_1)\chi^{\dot{A}_2}(u_2)\rangle\,.
\end{align}
Then apply the S-matrix $\hat{S}$ to the left $\mathfrak{psu}(2|2)$ sector, and get the state of the form
\begin{align}
\hat{S}_{12}|\chi^{A_1}(u_1)\chi^{A_2}(u_2)\rangle \otimes |\chi^{\dot{A}_1}(u_1)\chi^{\dot{A}_2}(u_2)\rangle\,.
\end{align}
\item After doing so, contract the resulting state with the {\it gluing vertex} $\langle \mathcal{G}|$, which produces a product of one-point functions. (The precise definition will be given shortly).
\end{enumerate}
\end{shaded}
\noindent In summary, it can be expressed as follows:
\begin{align}\label{twoparticleexplicit}
\begin{split}
&\langle \mathcal{H}| \Big(\chi^{A_1}\chi^{\dot{A}_1} \Big)\Big(\chi^{A_1}\chi^{\dot{A}_1} \Big)\rangle\\
&=(-1)^{|\dot{A}_1||A_2|}h(u_1,u_2)\langle \mathcal{G}|\left(\hat{S}_{12}|\chi^{A_1}(u_1)\chi^{A_2}(u_2)\rangle \otimes |\chi^{\dot{A}_1}(u_1)\chi^{\dot{A}_2}(u_2)\rangle\right)
\end{split}
\end{align}

The contraction between the gluing vertex and a multi-magnon state is defined by
\begin{shaded}
\noindent {\it Definition of the gluing vertex}:
\begin{align}\label{gluingdef1}
&\langle \mathcal{G}| \left( |\chi^{A_M}\cdots \chi^{A_2}\chi^{A_1}\rangle\otimes|\chi^{\dot{A}_1}\chi^{\dot{A}_2}\cdots \chi^{\dot{A}_M}\rangle\right):=\prod_{j=1}^{M}\langle \mathcal{H}|\chi^{A_1}\chi^{\dot{A}_1}\rangle\,,\\
&\begin{aligned}\label{gluingdef2}
&\langle \mathcal{G}| \left( |Z\chi^{A_M}\cdots \chi^{A_2}\chi^{A_1}\rangle\otimes|\chi^{\dot{A}_1}\chi^{\dot{A}_2}\cdots \chi^{\dot{A}_M}\rangle\right)\\
&:=\left(\prod_{j=1}^{M}\frac{i}{U_i}\right)\langle \mathcal{G}| \left( |\chi^{A_M}\cdots \chi^{A_2}\chi^{A_1}\rangle\otimes|\chi^{\dot{A}_1}\chi^{\dot{A}_2}\cdots \chi^{\dot{A}_M}\rangle\right)\,,\\
&\langle \mathcal{G}| \left( |\chi^{A_M}\cdots \chi^{A_2}\chi^{A_1}\rangle\otimes|Z\chi^{\dot{A}_1}\chi^{\dot{A}_2}\cdots \chi^{\dot{A}_M}\rangle\right)\\
&:=\left(\prod_{j=1}^{M}\frac{i}{U_i}\right)\langle \mathcal{G}| \left( |\chi^{A_M}\cdots \chi^{A_2}\chi^{A_1}\rangle\otimes|\chi^{\dot{A}_1}\chi^{\dot{A}_2}\cdots \chi^{\dot{A}_M}\rangle\right)\,,
\end{aligned}
\end{align}
\end{shaded}
\noindent where $U_i$ is defined by $U_i= e^{ip_i/2}$.

To discuss the symmetry property of the two-particle hexagons, it is convenient to first study the symmetry of the gluing vertex. Firstly the definition \eqref{gluingdef2} guarantees that $\langle \mathcal{G}|$ is annihilated by the central charge $\mathcal{P}$. Secondly, it is annihilated by all the generators of $\mathfrak{psu}(2|2)_D$:
 It is relatively easy to see that $\langle \mathcal{G}|$ is annihilated by $\mathcal{R}$'s and $\mathcal{L}$'s. Since these transformations act locally in the spin chain (namely they do not produce and $Z$'s), the factorized form of the gluing vertex \eqref{gluingdef1} guarantees the invariance under these symmetry transformations.

Therefore, we just need to worry about $\mathcal{Q}$'s and $\mathcal{S}$'s. As it turns out that, owing to the factorized form of our definition \eqref{gluingdef1}, the only nontrivial cases we need to check are
\begin{align}\label{nontrivialcheck}
\langle \mathcal{G}| \mathcal{Q}\left(|\chi^{A_1}\cdots \psi^{A_N}\cdots \rangle\otimes|\cdots \varphi^{\dot{A}_N}\cdots \chi^{\dot{A}_1}\rangle\right)=0\,,\\
\langle \mathcal{G}| \mathcal{S}\left(|\chi^{A_1}\cdots \varphi^{A_N}\cdots \rangle\otimes|\cdots \psi^{\dot{A}_N}\cdots \chi^{\dot{A}_1}\rangle\right)=0\,.\label{nontrivialcheck2}
\end{align}
In the first case \eqref{nontrivialcheck}, the action of $\mathcal{Q}$ produces a state of the form
\begin{align}
{\bf c}(u_N)|\cdots Z\varphi^{B_N}\cdots \rangle\otimes|\cdots \varphi^{\dot{A}_N}\cdots \rangle+{\bf d}(u_N)|\cdots \psi^{B_N}\cdots \rangle\otimes|\cdots Z^{-1}\psi^{\dot{A}_N}\cdots \rangle
\end{align}
with $u_N$ being the rapidity of $N$-th magnon. We can then use the rules to move and remove $Z$'s (\eqref{moveZ} and \eqref{gluingdef1}) to show that they vanish when contracted with $\langle \mathcal{G}|$:
\begin{align}
\begin{aligned}
&{\bf c}(u_N)\langle \mathcal{G}|\left(|\cdots Z\varphi^{B_N}\cdots \rangle\otimes|\cdots \varphi^{\dot{A}_N}\cdots \rangle\right)\\
&={\bf c}(u_N)\frac{i \prod^{N-1}_{k=1}U_i}{\prod^{M}_{i=N}U_i}\langle \mathcal{G}|\left(|\cdots \varphi^{B_N}\cdots \rangle\otimes|\cdots \varphi^{\dot{A}_N}\cdots \rangle\right)\,.\\
&{\bf d}(u_N)\langle \mathcal{G}|\left(|\cdots \psi^{B_N}\cdots \rangle\otimes|\cdots Z^{-1}\psi^{\dot{A}_N}\cdots \rangle\right)\\
&={\bf d}(u_N)\frac{ \prod^{N}_{k=1}U_i}{i\prod^{M}_{i=N+1}U_i}\langle \mathcal{G}|\left(|\cdots \psi^{B_N}\cdots \rangle\otimes|\cdots \psi^{\dot{A}_N}\cdots \rangle\right)\,.
\end{aligned}
\end{align}
Then, the sum can be expressed as
\begin{align}
\begin{aligned}
&\frac{\prod_{i=1}^{N-1}U_i}{\prod_{i=N+1}^{M}U_i}\left[\frac{i {\bf c}(u_N)}{U_N}\langle \mathcal{G}|\left(|\cdots \varphi^{B_N}\cdots \rangle\otimes|\cdots \varphi^{\dot{A}_N}\cdots \rangle\right) \right.\\
&\hspace{70pt}\left.+\frac{{\bf d}(u_N)U_N}{i}\langle \mathcal{G}|\left(|\cdots \psi^{B_N}\cdots \rangle\otimes|\cdots \psi^{\dot{A}_N}\cdots \rangle\right)\right]\,.
\end{aligned}
\end{align}
Importantly, all the nonlocal effects (dependence on the momenta of other excitations) can be factorized out and the prefactors inside the square-brackets only depend on quantum numbers of the $N$-th excitation. We can therefore recycle the proof for the one-particle case and easily confirm that the gluing vertex is annihilated by $\mathcal{Q}$.
\begin{itemize}
\item[] {\bf Exercise}: Show the second identity \eqref{nontrivialcheck2} in a similar manner.
\end{itemize} 

Once the symmetry of the gluing vertex is understood, it is straightforward to check that the two-particle hexagon (in particular the matrix part $\mathcal{H}_{\rm mat}$) satisfies the symmetry constraints: Since the two particle hexagon is defined by the two-particle $S$-matrix $\hat{S}$, which commutes with $\mathfrak{psu}(2|2)$, and the gluing vertex, which is annihilated by $\mathfrak{psu}(2|2)_D$, it is obviously annihilated by $\mathfrak{psu}(2|2)_D$.

This argument also provides a quick proof for the uniqueness of the matrix part $\mathcal{H}_{\rm mat}$: Without loss of generality, a general two-particle form factor can be expressed as
\begin{align}
\begin{split}
&\langle \mathcal{H}| \Big(\chi^{A_1}\chi^{\dot{A}_1} \Big)\Big(\chi^{A_1}\chi^{\dot{A}_1} \Big)\rangle\\
&=(-1)^{|\dot{A}_1||A_2|}\langle \mathcal{G}|\left(\hat{F}_{12}|\chi^{A_1}(u_1)\chi^{A_2}(u_2)\rangle \otimes |\chi^{\dot{A}_1}(u_1)\chi^{\dot{A}_2}(u_2)\rangle\right)
\end{split}
\end{align}
where $\hat{F}$ is some operator acting on the left $\mathfrak{psu}(2|2)$ state. Owing to the symmetry property of the gluing vertex, the operator $\hat{F}$ has to satisfy
\begin{align}\label{symrelF}
[\hat{F},h]=0\,,\qquad h\in \mathfrak{psu}(2|2) \text{ $+$ central charges}\,.
\end{align}
(It follows from the condition $\langle \mathcal{H}| h_D =0$ with $h_D\in\mathfrak{psu}(2|2)_D$.) As discussed in \cite{dynamic,analytic}, the only two-particle tensor which satisfies the property \eqref{symrelF} (and swaps the order of magnons) is the $S$-matrix $\hat{S}_{12}$. This completes the proof for the uniqueness.

\begin{figure}[t]
\centering
\includegraphics[clip,height=3.5cm]{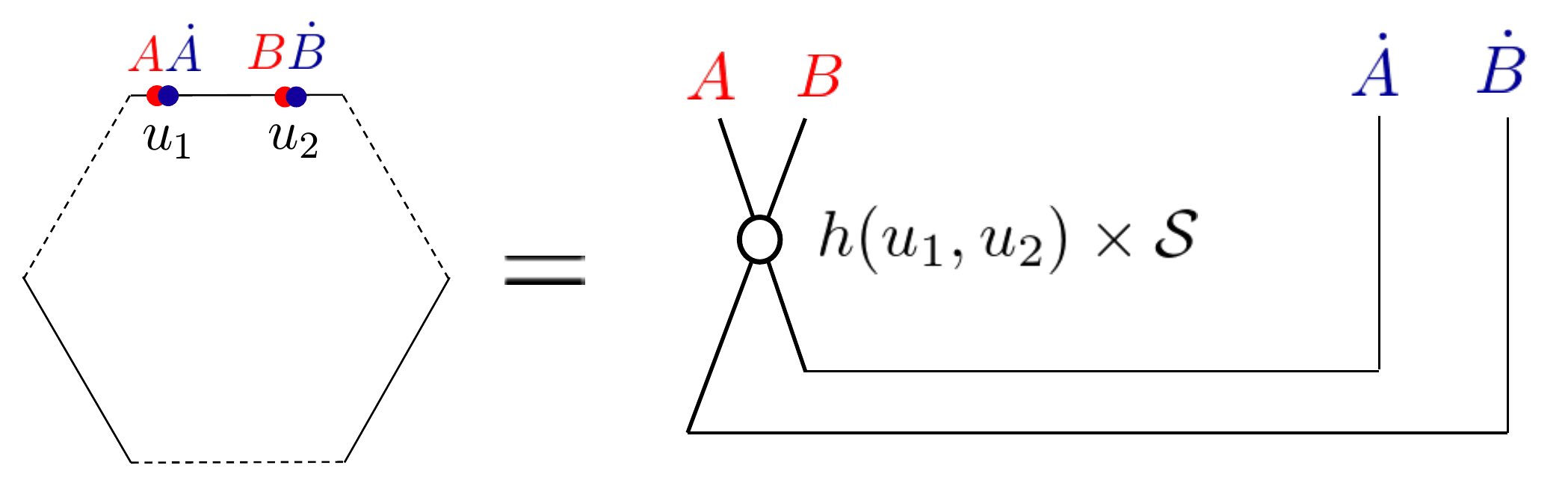}
\caption{A pictorial representation of the two-particle hexagon form factor. As in figure \ref{fig:onephex}, the line connecting indices denotes a product of the measure factor and the invariant tensor. The circle dot in the middle represents the action of the $\mathfrak{psu}(2|2)$ S-matrix and the multiplication of the dynamical factor $h$. \label{fig:twophex}}
\end{figure}
The two-particle hexagon also admits a pictorial representation as shown in figure \ref{fig:twophex}. In the figure, the intersection of two lines denotes a product of the $S$-matrix $\hat{S}_{12}$ and a dynamical factor $h$, and the contraction of two indices corresponds to a product of the measure factor $\sqrt{\mu}$ and the invariant tensor. As alluded to above, the dynamical factor is not constrained by the symmetry analysis we performed here. To determine its structure, one has to use the integrability as we see shortly. 
\subsubsection*{Multi-particle form factor}
Unlike the two-particle form factor, the structures of the multi-particle form factors are not determined by the symmetries. In principle, they can have very complicated tensor structures and their dependence on the rapidities can be highly nontrivial. This is in fact the case for general form factors in integrable models and is what hinders the application of integrability beyond the spectral problem.

\begin{figure}[t]
\centering
\includegraphics[clip,height=3.5cm]{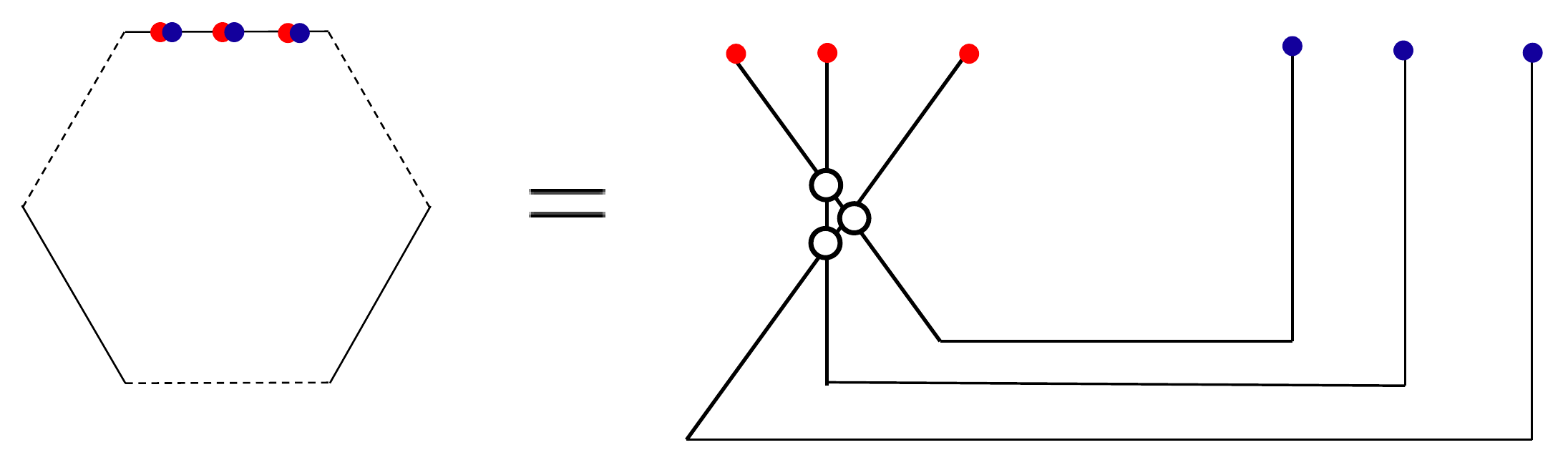}
\caption{A pictorial representation of the multi-particle hexagon form factor. Notations are the same as in previous figures \label{fig:multiphex}}
\end{figure}

However, in the case at hand, we know that the multi-particle form factor is no more complicated than the two-particle form factors at least in the weak-coupling regime. Namely, they are given by a product of two-particle form factors as we saw in \eqref{multihextree}:
\begin{align}
\mathcal{H}(\{u_j\})\,\,\overset{\text{weak coupling}}{=}\,\,\prod_{i<j}h(u_i,u_j)
\end{align}
Motivated by this simple structure, we conjecture that the multi-particle form factor at finite coupling is also given by a ``product" of two-particle form factors: 
\begin{shaded}
\noindent {\it Multi-particle form factors}:\\
The multi-particle form factor is given by a product of the {\it matrix part} $\mathcal{H}_{\text{mat}}$ and the {\it dynamical part} $\mathcal{H}_{\text{dyn}}$ as
\begin{align}\label{multiconjecture}
\langle \mathcal{H}|\Big(\chi^{A_1}\chi^{\dot{A}_1}\Big)(u_1)\cdots \Big(\chi^{A_M}\chi^{\dot{A}_M}\Big)(u_M)\rangle=\mathcal{H}_{\text{dyn}}(\{u_j\})\mathcal{H}_{\text{mat}} (\{u_j\})\,.
\end{align}
The dynamical part is a simple product of the two-particle dynamical factors,
\begin{align}
\mathcal{H}_{\text{dyn}}(\{u_j\})=\prod_{i<j} h(u_i,u_j)\,,
\end{align}
whereas the matrix part is a natural generalization of the matrix part for the two-particle form factor:
\begin{enumerate}
\item First rearrange the indices $(A_1\dot{A}_1)\cdots(A_M\dot{A}_M)$ to $(A_1A_2\cdots A_M)(\dot{A}_1\dot{A}_2\cdots \dot{A}_M)$. This rearrangement produces a sign factor $(-1)^{\sum_{i<j}|\dot{A}_i||A_j|}$.
\item After the rearrangement, we obtain the state
\begin{align}
|\chi^{A_1}(u_1)\cdots \chi^{A_M}(u_M)\rangle \otimes |\chi^{\dot{A}_1}(u_1)\cdots\chi^{\dot{A}_M}(u_M)\rangle\,.
\end{align}
Then apply the S-matrix $\hat{S}$ to the left $\mathfrak{psu}(2|2)$ sector to reverse the order of the particles,
\begin{align}
\begin{aligned}
&\prod_{j<k}\hat{S}_{jk}|\chi^{A_1}(u_1)\cdots \chi^{A_M}(u_M)\rangle \otimes |\chi^{\dot{A}_1}(u_1)\cdots \chi^{\dot{A}_M}(u_M)\rangle\\
&\sim |\chi^{B_M}(u_M)\cdots \chi^{B_1}(u_1)\rangle \otimes |\chi^{\dot{A}_1}(u_1)\cdots \chi^{\dot{A}_M}(u_M)\rangle\,.
\end{aligned}
\end{align}
\item After doing so, contract the resulting state with the {\it gluing vertex} $\langle \mathcal{G}|$, which produces a product of one-point functions.
\end{enumerate}
\end{shaded}
The structure of the multi-particle form factor is summarized neatly in figure \ref{fig:multiphex}. We should emphasize again that this structure is just a conjecture and is not dictated by the symmetry. However, as we will see later, it nicely realizes the constraints from integrability and also reproduces various perturbative data.
\subsection{Constraints from integrability}
In the previous subsection, we determined the matrix part of the hexagon form factor by imposing the symmetry and presented some conjecture for the multi-particle hexagons. However, as alluded to above, the overall scalar factors $\sqrt{\mu}$ and $h(u,v)$ cannot be fixed just from the symmetry analysis. To determine them, one has to take into account constraints from integrability, or more precisely, one has to impose the so-called {\it form factor axioms} of the integrable quantum field theories.
\subsubsection*{Hexagon as the branch-point twist operator}
Let us first clarify how we should think of the hexagon from the point of view of integrable quantum field theories.

\begin{figure}[h]
\centering
\includegraphics[clip,height=4cm]{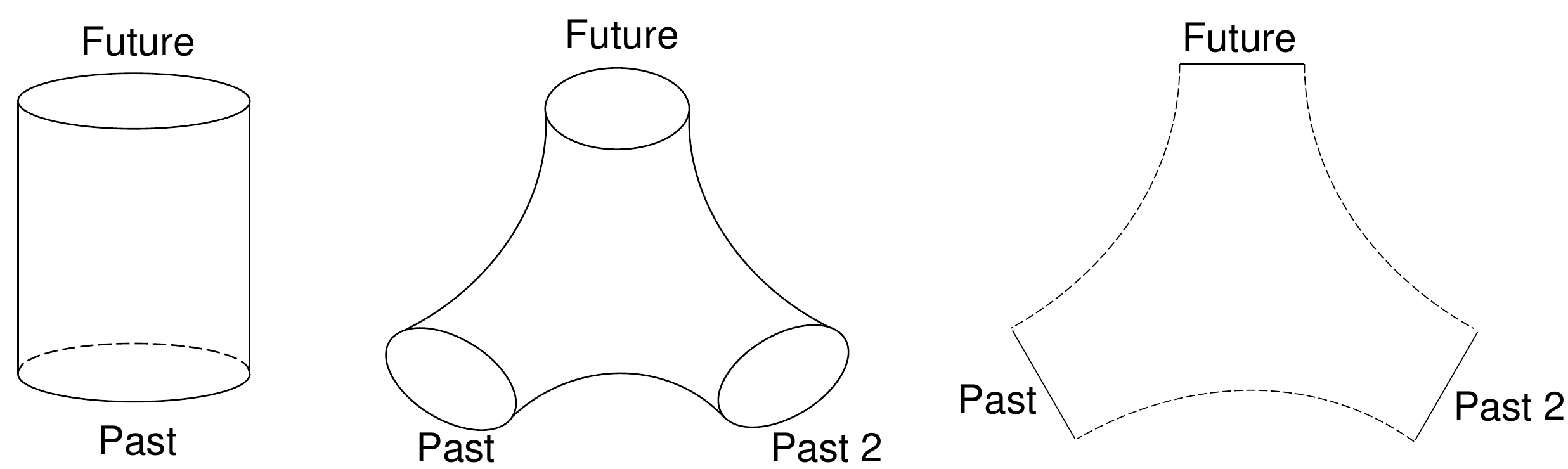}
\caption{Asymptotic regions of a cylinder, a pair of pants and a hexagon.\label{fig:asregion}}
\end{figure}

As should be clear from the discussions so far, the three-point function is described by an integrable quantum field theory on a pair of pants. A distinctive feature of the pair of pants is that it has three asymptotic regions; one future and two pasts. The structure is also inherited by the hexagon as shown in figure \ref{fig:asregion}. This should be contrasted with the cylindrical geometry used in the study of the finite-size spectrum, which only contains two asymptotic regions; one future and one past.

\begin{figure}[t]
\centering
\includegraphics[clip,height=5cm]{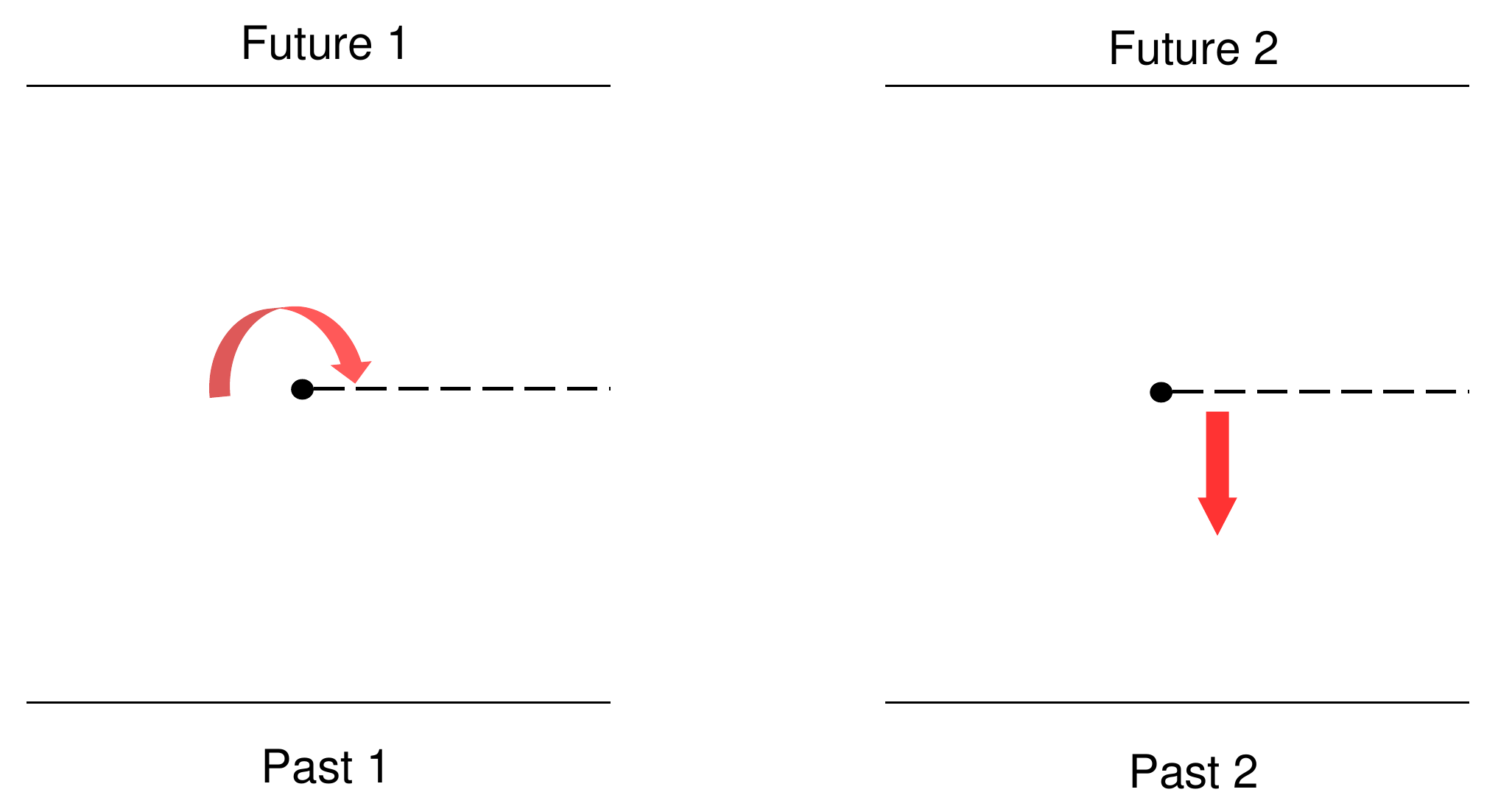}
\caption{The branch point twist operator for the replica number $n=2$. As shown above it has two futures and two pasts.\label{fig:branchpoint}}
\end{figure}

Although a quantum field theory on a geometry with more than two asymptotic regions may sound peculiar, a similar object actually shows up in the literature (but in a slightly different context); namely in the study of the Renyi entropy. In $1+1$ dimensions, the Renyi entropy  can be computed as the correlation functions of the so-called branch-point twist operators \cite{Cardy}. The branch-point twist operator is an operator which introduces an conical excess $2\pi n$ (with $n$ being the replica number) and creates an ``extra" spacetime. As depicted in figure \ref{fig:branchpoint}, the resulting geometry contains $2 n$ asymptotic regions; $n$ futures and $n$ pasts.
Thus, as far as the counting goes, the hexagon corresponds to a twist operator with the replica number $n=3/2$.

An advantage of regarding the hexagon as a sort of twist operators is that it allows us to study them using integrability. In massive quantum field theories, the correlators of the twist operators can be decomposed into form factors;
\begin{align}
\langle 0 | \mathcal{O}_{\rm twist} \mathcal{O}_{\rm twist}|0\rangle=\sum_{\psi} \underbrace{\langle 0 | \mathcal{O}_{\rm twist}|\psi\rangle}_{\text{form factor}} \underbrace{\langle \psi | \mathcal{O}_{\rm twist}|0\rangle}_{\text{form factor}}\,. 
\end{align}
 If the theory is integrable, one can then constrain the form factors of such twist operators by the so-called form factor axioms, which were written down in \cite{CCD}. In what follows, we impose similar axioms and determine the hexagon form factors explicitly.
\subsubsection*{Watson equation}
\begin{figure}[t]
\centering
\includegraphics[clip,height=4.5cm]{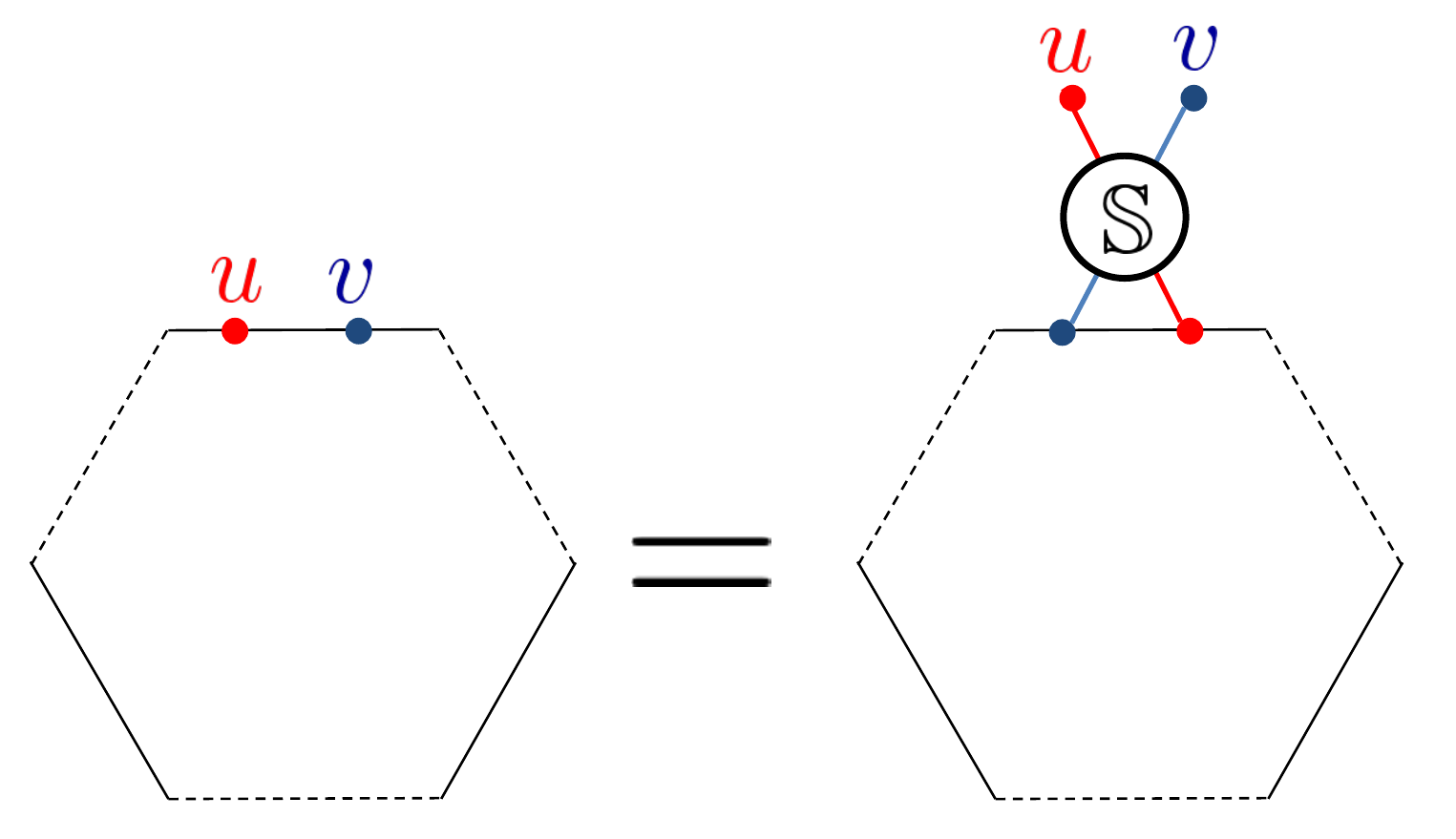}
\caption{Watson equation. The Watson equation states that the form factor does not change if one first changes the order of the particle by acting the S-matrix and then compute the form factor.\label{fig:watson}}
\end{figure}

The first constraint is what is called the Watson equation. It essentially tells us that the form factor does not change if one first acts the S-matrix and then computes the form factor. More precisely it can be expressed as the following equality:
\begin{align}\label{watsoneq}
\langle \mathcal{H}|\mathcal{X}^{A_1\dot{A}_1}(u_1)\cdots\mathcal{X}^{A_M\dot{A}_M}(u_M)\rangle=\langle \mathcal{H}|\mathbb{S}_{i,i+1}|\mathcal{X}^{A_1\dot{A}_1}(u_1)\cdots\mathcal{X}^{A_M\dot{A}_M}(u_M)\rangle
\end{align}
Here $\mathbb{S}_{i,i+1}$ is the S-matrix which swaps the order of $\mathcal{X}^{A_i\dot{A}_i}$ and $\mathcal{X}^{A_{i+1}\dot{A}_{i+1}}$. It can also be expressed pictorially as shown in figure \ref{fig:watson}. The equation \eqref{watsoneq} is in principle a highly nontrivial matrix equality since the $S$-matrix $\mathbb{S}$ has quite a complicated structure. However, thanks to our conjecture on the matrix structure of the multi-particle form factor \eqref{multiconjecture}, it boils down to a single scalar equality for the dynamical factor $h(u,v)$.

To see this in the simplest example, let us consider the two-particle form factor. A crucial observation is that the gluing vertex defined in \eqref{gluingdef1} and \eqref{gluingdef2} has a nice property that it ``commutes" with the $S$-matrix:
\begin{align}\label{niceGS}
\langle \mathcal{G} | \left( \hat{S}_{12}|\chi^{A}_1\chi^{B}_2\rangle \otimes |\chi^{\dot{A}}_1\chi^{\dot{B}}_2\rangle\right)=\langle \mathcal{G} | \left( |\chi^{A}_1\chi^{B}_2\rangle \otimes \hat{S}_{12}|\chi^{\dot{A}}_1\chi^{\dot{B}}_2\rangle\right)\,.
\end{align}
This relation can be verified by computing both sides explicitly. 

\begin{figure}[t]
\centering
\includegraphics[clip,height=3.5cm]{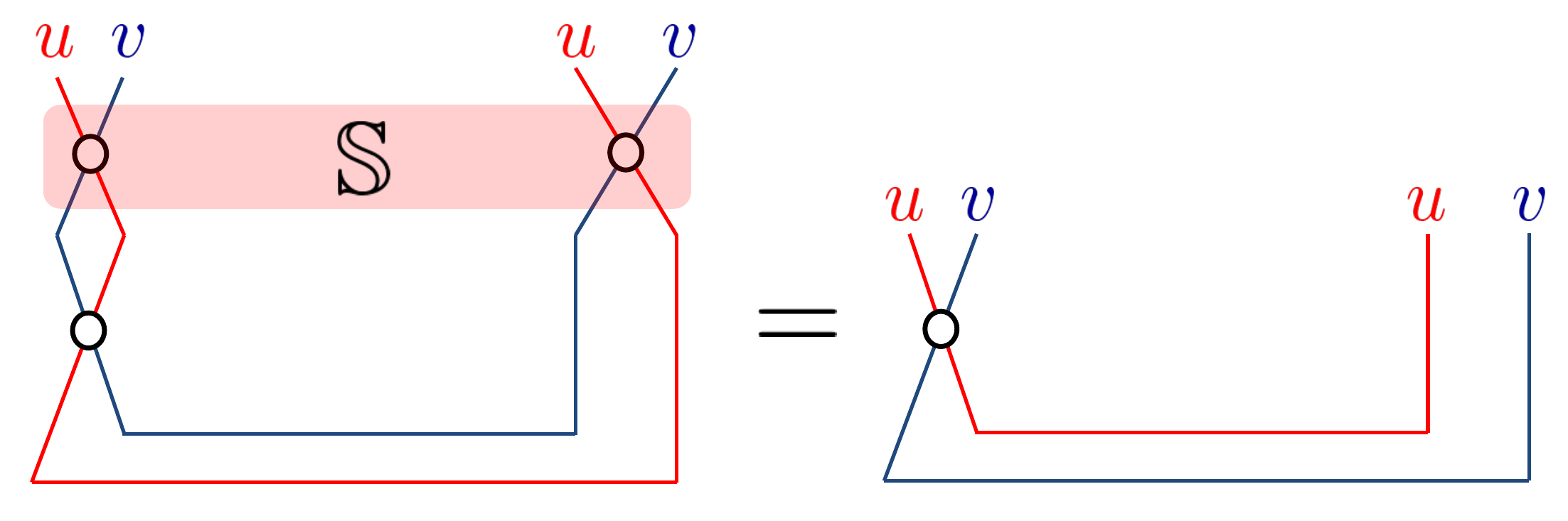}
\caption{A pictorial explanation for how the Watson equation is realized in the hexagon form factor. The left hand side denotes the form factor with the action of the S-matrix. By using the unitarity of the S-matrix, we can eliminate two circle dots and obtain the right figure, which is the form factor without the action of the S-matrix. \label{fig:watson2}}
\end{figure}
Now, using the definition of the hexagon form factor \eqref{twoparticleexplicit}, we can express the right hand side of \eqref{watsoneq} (for two particles) as
\begin{align}
\begin{aligned}
&\langle \mathcal{H}|\textcolor[rgb]{1,0,0}{\mathbb{S}_{12}}|\mathcal{X}^{A_1\dot{A}_1}(u)\mathcal{X}^{A_2\dot{A}_2}(v)\rangle\\
& = (-1)^{|\dot{A}_1||A_2|}h(v,u) \textcolor[rgb]{1,0,0}{S_0 (u,v)}\langle \mathcal{G}|\left( \hat{S}_{21}\textcolor[rgb]{1,0,0}{\hat{S}_{12}} |\chi^{A_1}\chi^{A_2}\rangle \otimes \textcolor[rgb]{1,0,0}{\hat{S}_{12}}|\chi^{\dot{A}_1}\chi^{\dot{A}_2}\rangle\right)\,.
\end{aligned}
\end{align}
Here we denoted the factors associated with the action of $\mathbb{S}_{12}$ in red. We can then use the unitarity relation $\hat{S}_{21}\hat{S}_{12}=\textbf{1} $ and the relation \eqref{niceGS} to rewrite it further as
\begin{align}
\langle \mathcal{H}|\mathbb{S}_{12}|\mathcal{X}^{A_1\dot{A}_1}(u)\mathcal{X}^{A_2\dot{A}_2}(v)\rangle &= (-1)^{|\dot{A}_1||A_2|}h(v,u) S_0 (u,v)\langle \mathcal{G}|\left( \hat{S}_{12} |\chi^{A_1}\chi^{A_2}\rangle \otimes |\chi^{\dot{A}_1}\chi^{\dot{A}_2}\rangle\right)\nonumber\\
&=\frac{h(v,u)}{h(u,v)}S_0(u,v)\langle \mathcal{H}|\mathcal{X}^{A_1\dot{A}_1}(u)\mathcal{X}^{A_2\dot{A}_2}(v)\rangle\,.
\end{align}
In the second line, we used the defintion of the two-particle form factor \eqref{twoparticleexplicit}.
Therefore, the Watson equation boils down to a single equality
\begin{shaded}
\begin{align}\label{Watson}
\frac{h(u,v)}{h(v,u)}=S_0 (u,v)\,.
\end{align}
\end{shaded}
\noindent There is also a pictorial way to understand how the Watson equation works. See figure \ref{fig:watson2}.
\subsubsection*{Crossing/Mirror transformations}
\begin{figure}[t]
\centering
\includegraphics[clip,height=6cm]{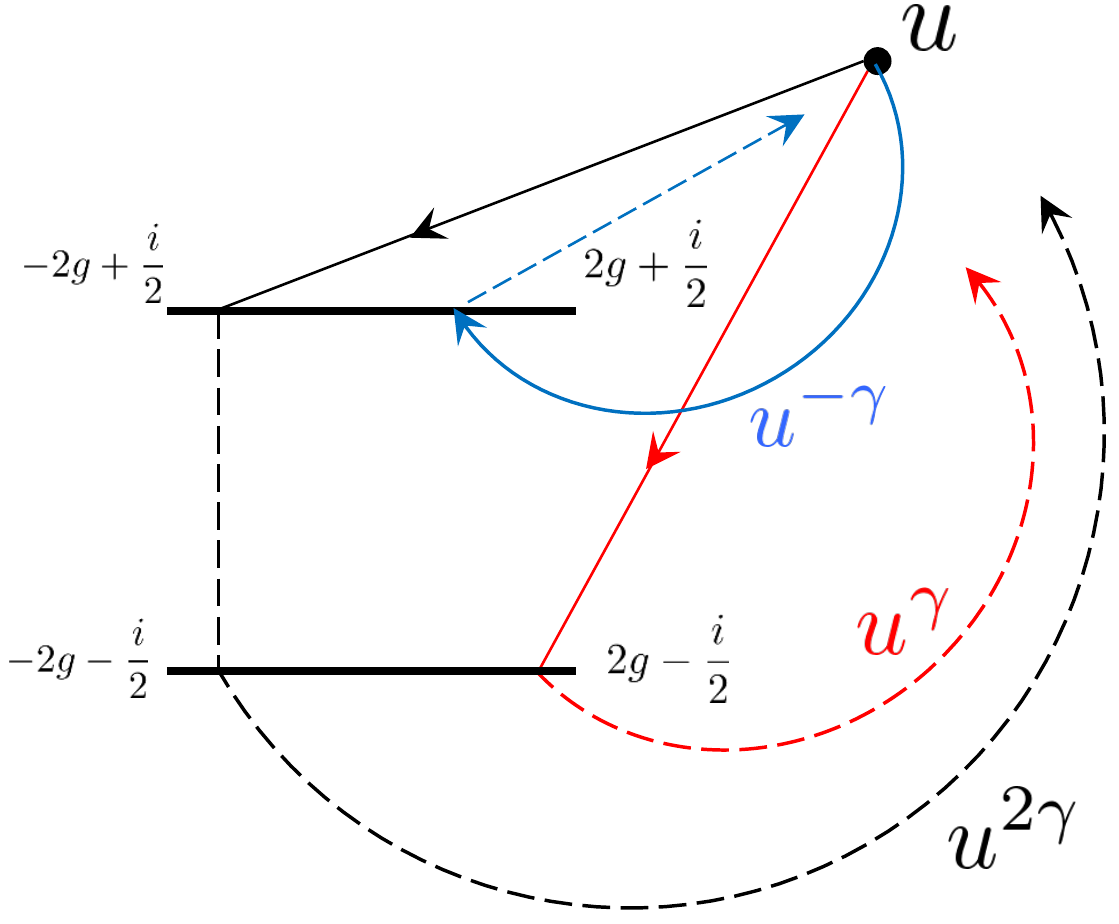}
\caption{Analytic structure of the magnon energy/momentum and the crossing and the mirror transformations. The energy and the momentum of the magnon has two cuts $[-2g \pm i/2, 2g \pm i/2]$. The upper and the lower cuts correspond to the cuts of $x^{-}$ and $x^{+}$ respectively. The crossing and the mirror transformations are defined as analytic continuations along the paths depicted in the figure. \label{fig:crossing}}
\end{figure}

\begin{figure}[t]
\centering
\includegraphics[clip,height=4cm]{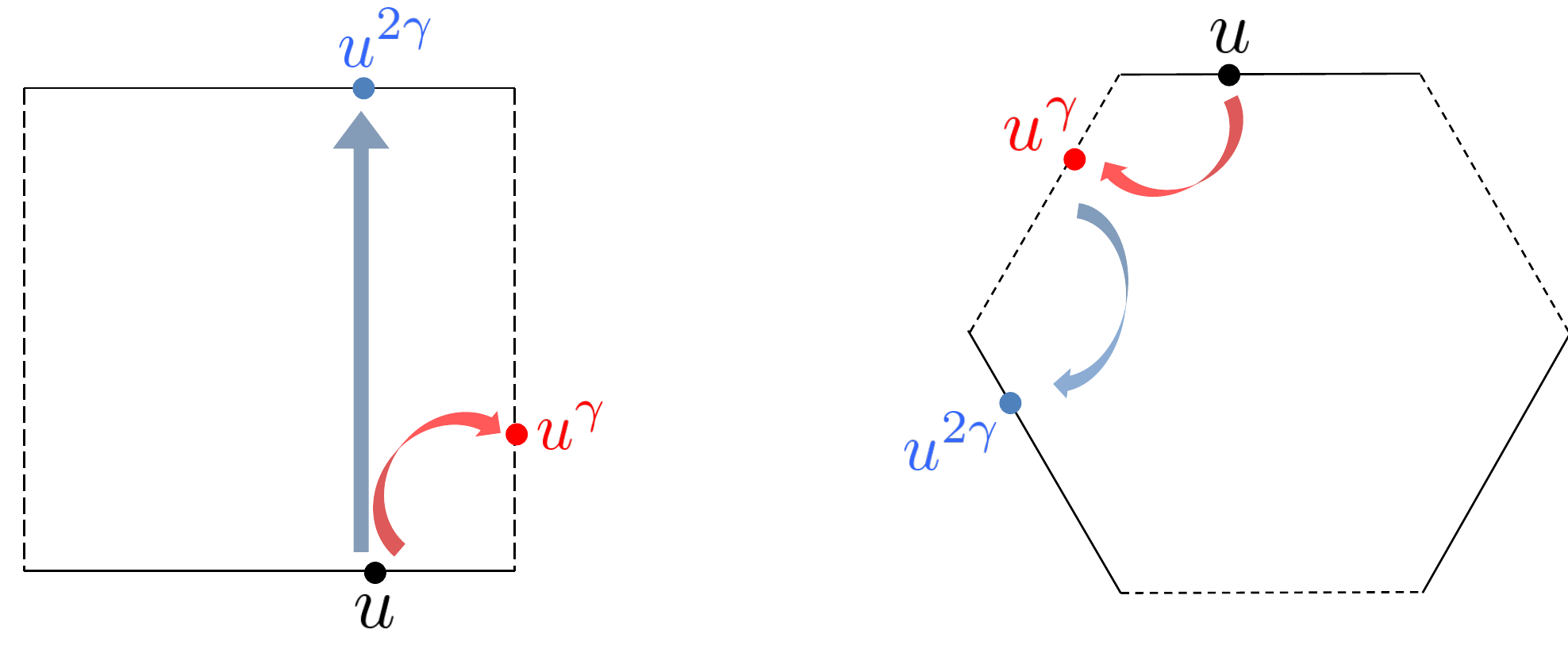}
\caption{Physical interpretation of the crossing and the mirror transformations. For two-point functions, the crossing transformation sends an anti-particle in the past to a particle in the future while the mirror transformation rotates the position of the particle by $90$ degrees and brings it to a theory with space and time swapped. The same transformation can be performed in the three-point functions. In that case, we can move a particle inside a hexagon by successive applications of mirror transformations \label{fig:crossing2}}
\end{figure}

Before considering the other constraint, it is useful to introduce the notion of the crossing and mirror transformations. In terms of Zhukowski variables introduced in \eqref{Zhukowdef}, the energy and momentum of a magnon at finite coupling read as
\begin{align}\label{energyandmomenta}
e^{i p}=\frac{x^{+}(u)}{x^{-}(u)} \,, \qquad E=\frac{1}{2}\frac{1+\frac{1}{x^{+}x^{-}}}{1-\frac{1}{x^{+}x^{-}}}\,.
\end{align}
Since the Zhukowski variables have cuts as a function of the rapidity $u$,
\begin{align}
x (u)=\frac{u+\sqrt{u^2-4g^2}}{2g}\,,
\end{align}
the functions \eqref{energyandmomenta} live on a complex plane with two branch cuts (see figure \ref{fig:crossing}).
Since we are equipped with a complex plane with cuts, a natural question to ask is what happens if we pass through a cut and go to a different sheet. It turns out that the effect of this analytic continuation on the Zhukowski variable is quite simple:
\begin{align}
x(u) \overset{\text{analytic continuation}}{\to}\frac{1}{x (u)}\,.
\end{align}
From this transformation property, we can immediately conclude that going through two cuts in \eqref{energyandmomenta} (which we denote by $2\gamma$) lead to
\begin{align}\label{simplestcrossing}
e^{ip} \overset{u\to u^{2\gamma}}{\to} e^{-ip}\,,\quad E \overset{u\to u^{2\gamma}}{\to} -E\,.
\end{align}
\begin{itemize}
\item[] {\bf Exercise}: Check \eqref{simplestcrossing}.
\end{itemize} 
As can be seen from \eqref{simplestcrossing}, the transformation $2\gamma$ inverts the signs of the energy and the momentum. Physically this should be interpreted as a crossing transformation which takes a particle in the past and converts it into an anti-particle in the future. See figure \ref{fig:crossing2} for a pictorial explanation. We therefore call it the {\it crossing transformation}.

We can also consider a ``half-crossing" transformation, by going through only one of two cuts:
\begin{align}
\begin{aligned}
&e^{i p} \overset{u\to u^{\gamma}}{\to}\frac{1}{x^{+}x^{-}}\,,\qquad E\overset{u\to u^{\gamma}}{\to}\frac{1}{2}\frac{1+\frac{x^{+}}{x^{-}}}{1-\frac{x^{+}}{x^{-}}}\,,\\
&e^{i p} \overset{u\to u^{-\gamma}}{\to}x^{+}x^{-}\,,\qquad E\overset{u\to u^{-\gamma}}{\to}\frac{1}{2}\frac{1+\frac{x^{-}}{x^{+}}}{1-\frac{x^{-}}{x^{+}}}\,,
\end{aligned}
\end{align}
Here $\gamma$ denotes the transformation which inverts $x^{+}$ while the transformation $-\gamma$ inverts $x^{-}$. 
To understand its physical meaning, let us recall how the crossing transformation is realized in ordinary relativistic 2d QFT"s. In relativistic 2d QFT's, the momenta of particles can be labeled by the rapidity $\theta$ given by
\begin{align}
E= m\cosh \theta \,, \quad p= m\sinh \theta\,.
\end{align}
Then, the crossing transformation corresponds to adding $i\pi$ to $\theta$:
\begin{align}
E\overset{\theta\to \theta + i\pi}{\to}-E\,,\quad p\overset{\theta\to \theta+i\pi}{\to}-p\,.
\end{align}
On the other hand, the half-crossing transformation, namely adding $i\pi/2$ to $\theta$, leads to
\begin{align}
E\overset{\theta\to \theta+i\pi/2}{\to} ip\,,\quad p\overset{\theta\to \theta+i\pi/2}{\to}i E\,.
\end{align}
This means that addition by $i\pi/2$ exchanges the space and the time directions. Such a space-time exchanged theory is often called a {\it mirror} theory. 

Let us now go back to our problem. Also in our case, it is known\footnote{A more detailed explanation from the point of view of string theory in AdS$_5$ is given in \cite{Arutyunov}.} that the half-crossing transformation (to be called the mirror transformation below) is interpreted as a map from the original theory to a theory with space and time swapped:
\begin{align}
\begin{aligned}
&e^{i p} \overset{u\to u^{\gamma}}{\to}e^{-\tilde{E}}=\frac{1}{x^{+}x^{-}}\,,\qquad E\overset{u\to u^{\gamma}}{\to}i\tilde{p}=\frac{1}{2}\frac{1+\frac{x^{+}}{x^{-}}}{1-\frac{x^{+}}{x^{-}}}\,,\\
&e^{i p} \overset{u\to u^{-\gamma}}{\to}e^{\tilde{E}}=x^{+}x^{-}\,,\qquad E\overset{u\to u^{-\gamma}}{\to}-i\tilde{p}=\frac{1}{2}\frac{1+\frac{x^{-}}{x^{+}}}{1-\frac{x^{-}}{x^{+}}}\,,
\end{aligned}
\end{align}
Here $\tilde{E}$ and $\tilde{p}$ are the energy and the momentum in the mirror theory. Unlike the relativistic QFT's, the dispersion relation changes under the mirror transformation.
\begin{itemize}
\item[] {\bf Exercise}: Compute the mirror energy and momentum at weak coupling and show 
\begin{align}
\tilde{E}=\frac{g^2}{u^2+1/4}+O(g^4) \,,\qquad \tilde{p}=u+O(g^2)\,.
\end{align}
\end{itemize} 
For the case of two-point functions, the mirror transformation moves the excitations from one edge to a neighboring edge, see figure \ref{fig:crossing2}. This is also true for three-point functions: By applying the mirror transformation, one can move magnons from one edge to another\footnote{There is one subtle point that we did not discuss in this lecture. When we perform the crossing transformation inside a hexagon, the excitations sometime change their flavors: For instance, a derivative along one direction gets transformed into a derivative along another direction. The detail  of this transformation rule can be found in Appendix D of \cite{BKV}. Its symmetry origin is recently discussed in Appendix A of \cite{hexagonalization}.} inside a hexagon, see figure \ref{fig:crossing2}.

\subsubsection*{Decoupling}
\begin{figure}[t]
\centering
\includegraphics[clip,height=4.5cm]{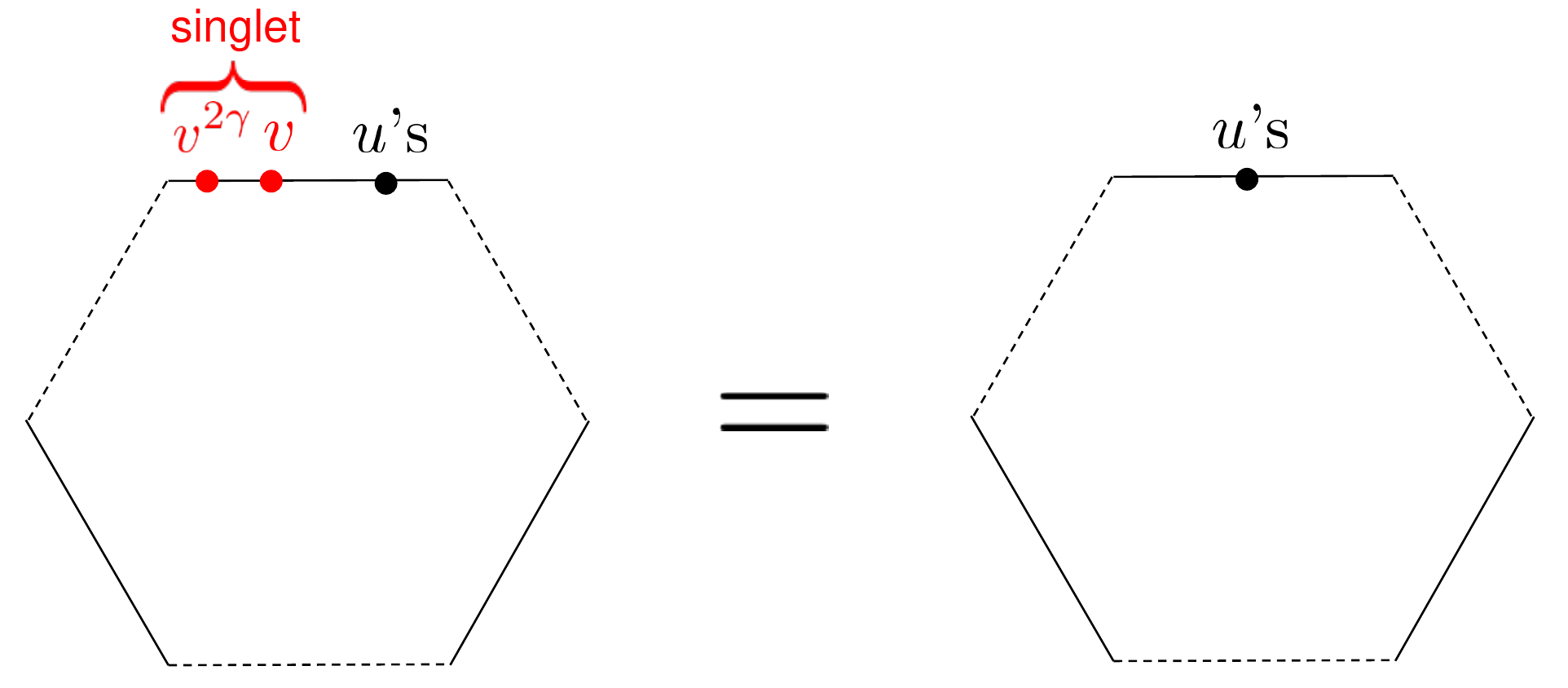}
\caption{Decoupling condition. The addition of a particle-antiparticle pair should not change the form factor. \label{fig:decoupling}}
\end{figure}

The other constraint from integrability is called the decoupling relation\footnote{In the context of form factor axioms in integrable quantum field theories, this relation is sometimes called the {\it kinematical pole axiom}.}. Physically it means that the addition of a particle-antiparticle pair, which can be generated by the vacuum fluctuation, does not affect the hexagon form factor, see figure \ref{fig:decoupling}. 
It can be expressed in formulae as
\begin{align}\label{decouplingeq}
-i\underset{u= v}{\text{Res}}\left[ \langle \mathcal{H}|\underbrace{\mathcal{X}(u^{2\gamma})\mathcal{X}(v)}_{\text{singlet}}\mathcal{X}(u_1)\cdots \mathcal{X}(u_M)\rangle\right]= \langle \mathcal{H}|\mathcal{X}(u_1)\cdots \mathcal{X}(u_M)\rangle\,.
\end{align}
Here we suppressed writing indices for simplicity.

\begin{figure}[t]
\centering
\includegraphics[clip,height=3cm]{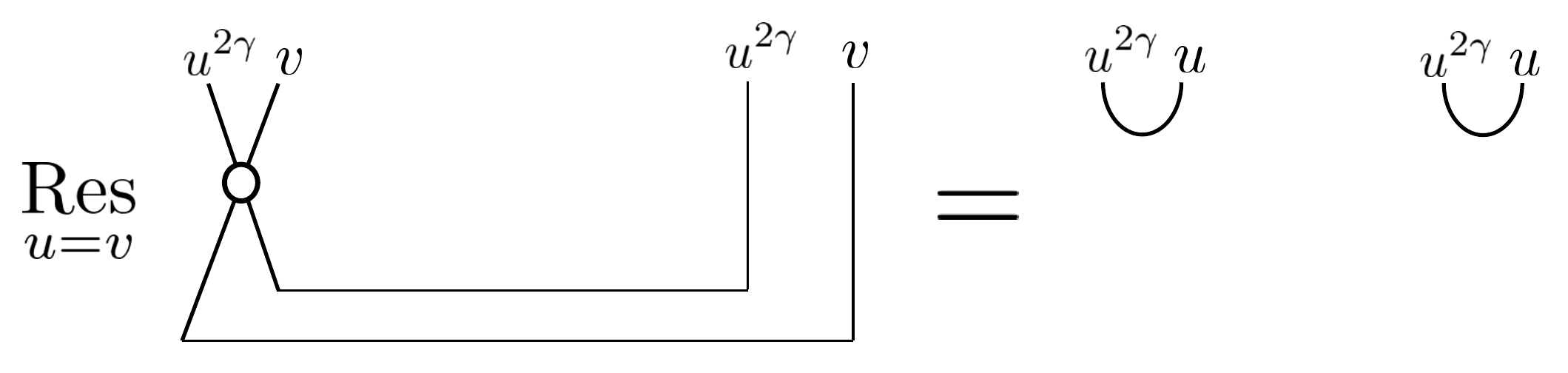}
\caption{A pictorial explanation of the decoupling condition with no extra particles. When the rapidities $u$ and $v$ are set equal, the matrix part simplifies and we obtain the two disconnected lines. We are thus left with the condition on the measure factor given in \eqref{crossing1}. \label{fig:decoupling2}}
\end{figure}

Let us first consider the simplest case, where there are no extra particles. In this case, \eqref{decouplingeq} boils down to
\begin{align}\label{decouplingeq2particle}
-i\underset{u= v}{\text{Res}}\left[ \langle \mathcal{H}|\underbrace{\mathcal{X}(u^{2\gamma})\mathcal{X}(v)}_{\text{singlet}}\rangle\right]= 1\,.
\end{align}
The simplest choice of the particle-antiparticle pair is given by a pair of derivatives,
\begin{align}\label{samplestate}
|D_{1\dot{2}}(u^{2\gamma})D_{2\dot{1}}(v)\rangle\,.
\end{align}
Inserting this state in the equation \eqref{decouplingeq2particle}, one can compute the left hand side as
\begin{align}\label{simplifylefthand}
\begin{aligned}
-i\underset{u= v}{\text{Res}}\left[ \langle \mathcal{H}|D_{1\dot{2}}(u^{2\gamma})D_{2\dot{1}}(v)\right]&=-i\sqrt{\mu(v)\mu(u^{2\gamma})}\underset{u= v}{\text{Res}}\left[h(u^{2\gamma},v)\mathcal{H}_{\rm mat}(u^{2\gamma},v)\right]\\
&=-i \mu(u) \underset{u=v}{\text{Res}}\left[h(u^{2\gamma},v)\mathcal{H}_{\rm mat}(u^{2\gamma},v)\right]\,.
\end{aligned}
\end{align}
Here $\mathcal{H}_{\rm mat}$ is the matrix part for the state given by \eqref{samplestate}. In the second line of \eqref{simplifylefthand}, we used the fact that the measure is invariant under the crossing transformation. (This is because two hexagons, one with $u$ and one with $u^{2\gamma}$, are related by the rotation by $120$ degrees.). The matrix part $\mathcal{H}_{\rm mat}$ can be computed by the definition of the two-particle form factor and the $S$-matrix given in Appendix \ref{ap-a} (See also \cite{dynamic, analytic}).
\begin{itemize}
\item[] {\bf Exercise}: Using the $S$-matrix given in Appendix \ref{ap-a}, derive the following expression for the matrix part for \eqref{samplestate}:
\begin{align}\label{matrixpartforD12}
\mathcal{H}_{\rm mat}(u^{2\gamma},v)=\frac{v-u+i}{u-v}\frac{(1-1/x_u^{-}x_{v}^{-})(1-1/x_u^{+}x_{v}^{+})}{(1-1/x_u^{-}x_{v}^{+})(1-1/x_u^{+}x_{v}^{-})}\,.
\end{align}
Here $x_{u}=x(u)$ and $x_v=x(v)$.
\end{itemize} 
The equation \eqref{matrixpartforD12} shows that there is a simple pole at $u=v$ in the matrix part. Since this is the only singularity we expect, we can conclude that the dynamical factor $h(u^{2\gamma},v)$ is not singular in the limit $u\to v$. Therefore, one can rewrite the decoupling condition \eqref{decouplingeq2particle} as
\begin{shaded}
\begin{align}\label{crossing1}
\mu(u)=\frac{1}{h(u^{2\gamma},u)}\frac{(1-1/x^{+}x^{-})^2}{(1-1/(x^{-})^2)(1-1/(x^{+})^2)}\,.
\end{align}
\end{shaded}
\noindent In terms of the pictorial representation of the matrix part, the decoupling condition we just discussed can be interpreted as resolution of intersecting lines. See figure \ref{fig:decoupling2}. 

\begin{figure}[t]
\centering
\includegraphics[clip,height=7cm]{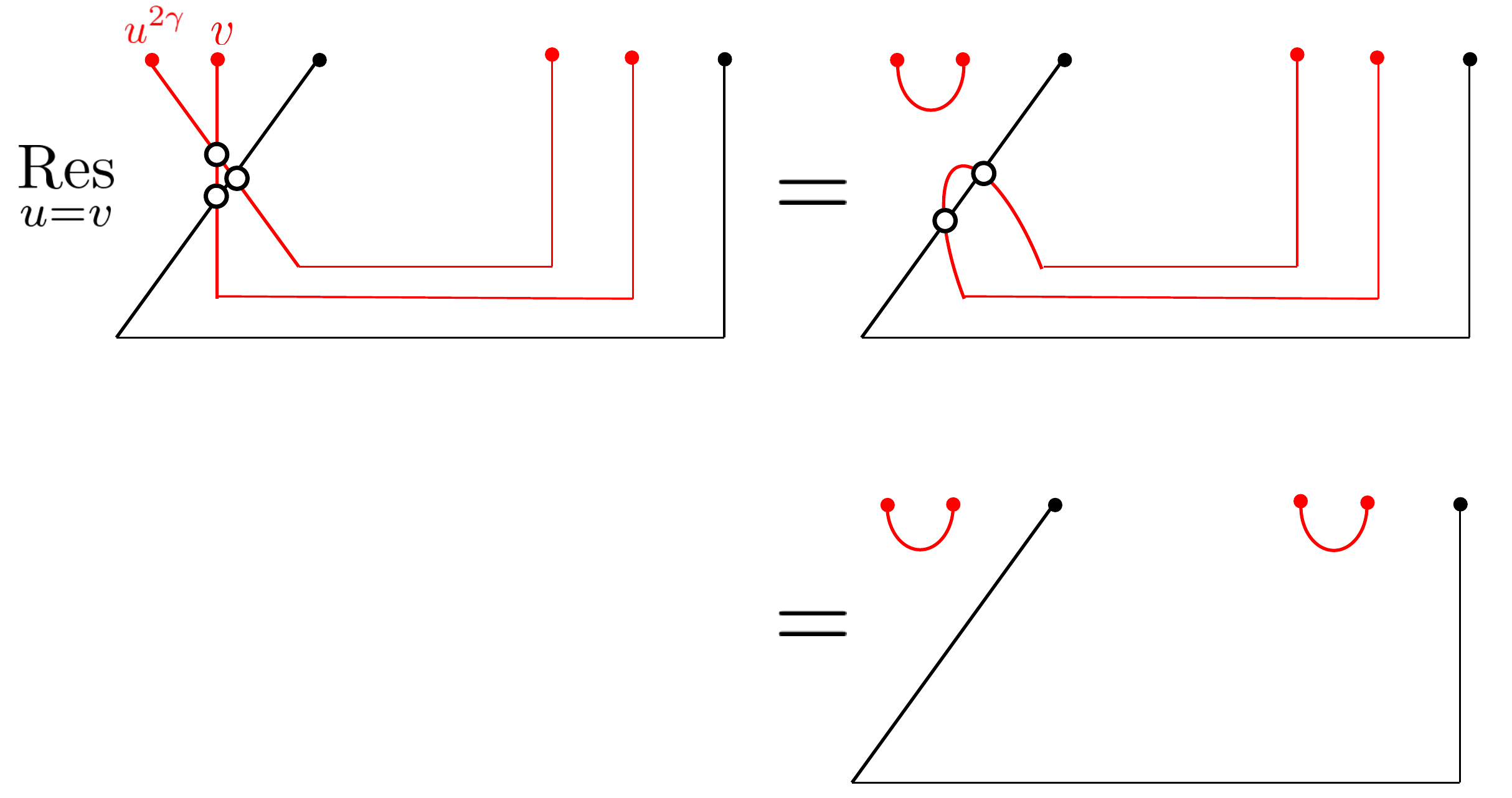}
\caption{A pictorial explanation of the decoupling condition with one extra particle. By setting the rapidities $u$ and $v$ to be equal, we can resolve the crossing of the lines and transform it into the second figure above. Then, the decoupling condition reduces to the second equality in the figure. \label{fig:decoupling3}}
\end{figure}

Let us now discuss the next complicated case, the case with a single spectator excitation. 
\begin{align}\label{decouplingeq3particle}
-i\underset{u= v}{\text{Res}}\left[ \langle \mathcal{H}|\underbrace{\mathcal{X}(u^{2\gamma})\mathcal{X}(v)}_{\text{singlet}}\mathcal{X}(w)\rangle\right]= \langle \mathcal{H}|\mathcal{X}(w)\rangle\,.
\end{align}
Since working it out in full detail will take up too much space, we will resort to the pictorial representation at the cost of accuracy. The three-particle hexagon form factor can be depicted as the first figure in figure \ref{fig:decoupling3}. Since the three-particle form factor is essentially a product of two-particle form factors, taking the residue at $u=v$ corresponds to locally resolving the scattering of the particles $u$ and $v$ as shown in the second figure in figure \ref{fig:decoupling3}. Then, the decoupling condition boils down to the equality relating the second and the third figures in figure \ref{fig:decoupling3}. This relation is reminiscent of a version of the crossing equation for the S-matrix which was reformulated by Beisert \cite{analytic}. It physically means that the scattering of the singlet against a test particle is trivial. The only difference from the usual $S$-matrix story is that we are now dealing with only one copy of $\mathfrak{psu}(2|2)$ $S$-matrix. This case, however, was also discussed already in the paper \cite{analytic}, and, as was shown there, it produces the following constraint on the dynamical factors:
\begin{shaded}
\begin{align}\label{crossing2}
h(u^{2\gamma},w)h(u,w)=\frac{x^{-}_{u}-x^{-}_{w}}{x^{-}_{u}-x^{+}_{w}}\frac{1-1/x^{+}_{u}x^{-}_{w}}{1-1/x^{+}_{u}x^{+}_{w}}\,.
\end{align}
\end{shaded}

In principle, we can go on to more complicated cases where there are several spectator excitations. However, it turns out that they do not yield any new constraints. Thus, the constraints coming from decoupling are exhausted by \eqref{crossing1} and \eqref{crossing2}.
\begin{itemize}
\item[] {\bf Exercise}: Using the pictorial representation of the hexagon form factor and studying the cases with more than one spectator excitations, convince yourself that there are no other constraints. 
\end{itemize} 

\subsubsection*{A solution to the constraints}
Let us now present a solution to the constraints coming from integrability. We will not discuss its uniqueness, but instead we show {\it a posteriori} that the solution satisfies all the constraints and also matches with the weak-coupling expressions.
\begin{shaded}
\noindent {\it Definitions of the dynamical part and the measure}:\\
The dynamical part $h(u,v)$ and the measure $\mu(u)$ are given by
\begin{align}
\begin{aligned}
h(u,v)&=\frac{x_{u}^{-}-x_{v}^{-}}{x^{-}_{u}-x^{+}_{v}}\frac{1-1/x^{-}_{u}x^{+}_{v}}{1-1/x^{+}_{u}x^{+}_{v}}\frac{1}{\sigma (u,v)}\,,\\
\mu(u)&=\frac{(1-1/x_u^{+}x_u^{-})^2}{(1-1/(x^{+})^2)(1-1/(x^{-})^2)}\,.
\end{aligned}
\end{align}
\end{shaded}
\begin{itemize}
\item[] {\bf Exercise 1}: Check that they satisfy all the constraints \eqref{Watson}, \eqref{crossing1} and \eqref{crossing2} using the properties of the dressing phase:
\begin{align}
\sigma(u,u)=1\,,\qquad \sigma(u^{2\gamma},v)\sigma(u,v)=\frac{(1-1/x_u^{+}x_{v}^{+})(1-x_u^{-}/x_{v}^{+})}{(1-1/x_u^{+}x_{v}^{-})(1-x_u^{-}/x_{v}^{-})}\,.
\end{align}
\item[] {\bf Exercise 2}: Show that the hexagon form factor in the SU(2) sector is given in \eqref{conjecturesu2}. (One has to use the Beisert S-matrix given in Appendix \ref{ap-a}).
\end{itemize} 
\subsection{Finite-size correction}
The conjecture given in section \ref{sec:asymptotic3pt}  is valid only when all the bridge lengths are large. In general, it also receives extra contributions coming from the fact that the bridge lengths are finite. In what follows, we discuss very briefly how to compute such finite-size corrections\footnote{This material was not covered at all during the summer school, and the explanation given below is not self-contained by any means. For details, see the original article \cite{BKV}.}.

Let us first explain why there can be finite-size corrections from the point of view of the gauge theory. As mentioned below figure \ref{fig:oneloop3pt}, one has to insert a operator at the splitting points in order to compute loop corrections to the three-point function. Owing to general properties of the loop expansions in the planar limit, we expect that the operator at $n$ loops acts on $n$ consecutive spin sites. Therefore, if the bridge length is smaller than $2n$, two operators (the operator inserted at the front and at the back) start to talk to each other. This indicates that one has to take into account some new effect when the bridge length is small. In addition, Feynman diagrams corresponding to such processes generally yield complicated numbers including zeta functions, and it is rather easy to see that such transcendental numbers cannot arise from our conjecture on the asymptotic three-point function. 
\begin{figure}[t]
\centering
\includegraphics[clip,height=5.5cm]{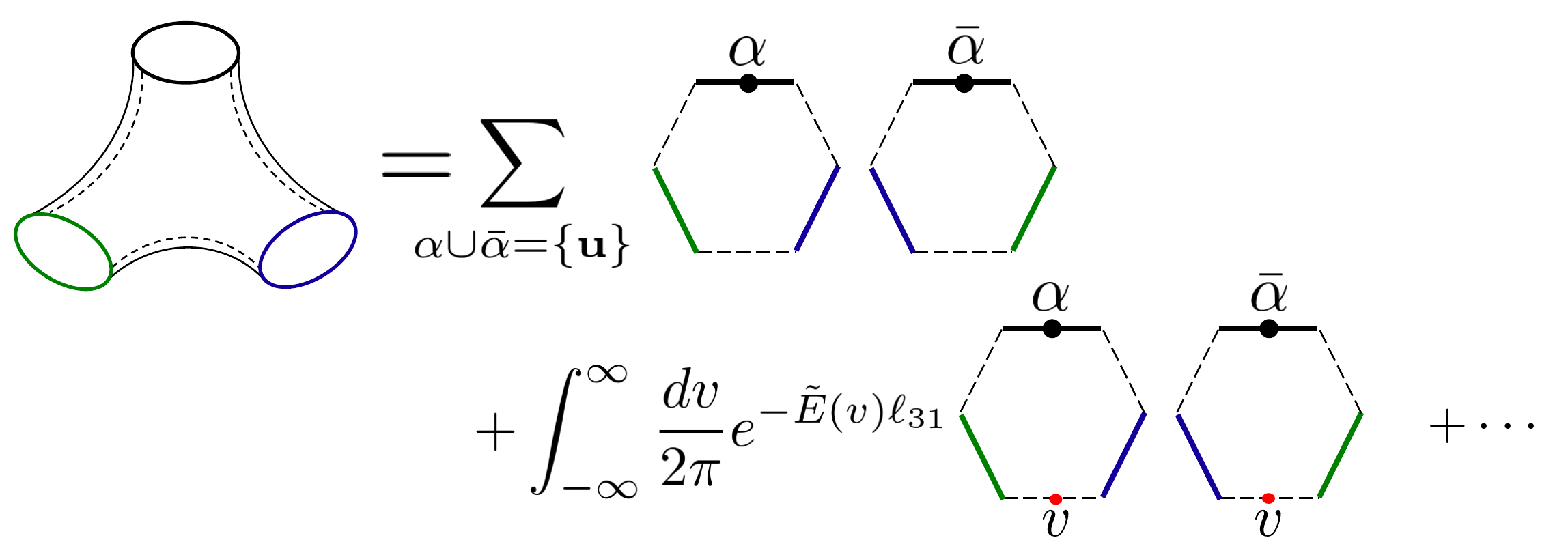}
\caption{The mirror particle corrections to the hexagon approach. When we cut the pair of pants into two, we need to insert a complete basis of states on the ``seams" (dashed lines in thei figure). This leads to a series expansion in terms of the number of mirror particles in each edge. Because of the propagation factor $e^{-\tilde{E}\ell}$, these corrections are suppressed when the operators are long. \label{fig:mirrorpic}}
\end{figure}

In the spectral problem, similar effects, called the wrapping effects, were discussed already. From the integrability point of view, those wrapping effects were understood as the corrections arising from mirror particles (magnons living in the mirror theory). Also in the present case, we propose that the mirror particles are responsible for these finite-size corrections. More precisely, we claim that when we cut the pair of pants into hexagons according to our proposal, we need to insert a complete set of states along the ``seams" (dashed line in figure \ref{fig:mirrorpic}), and hexagons with such nontrivial states on the seams yield the finite-size corrections. 

The states appearing on the ``seams" are the states in the mirror theory. Therefore, they are also labelled by the number and the rapidities of magnons as well as their flavor indices. Thus, the series we obtain from this procedure looks like (see also figure \ref{fig:mirrorpic})
\begin{align}
C_{123}\sim \sum_{\alpha\cup \bar{\alpha}=\{\textbf{u}\}} w_{\alpha,\bar{\alpha}}\left( \mathcal{H}(\alpha)\mathcal{H}(\bar{\alpha})+\int \frac{dv}{2\pi}e^{-\tilde{E}(v)\ell_{31}} \sum_{\text{flavor}}\mathcal{H}(\alpha;v)\mathcal{H}(\bar{\alpha};v)+\cdots\right)\,.
\end{align}
Here $w_{\alpha,\bar{\alpha}}$ is the partition-dependent weight factor ($e^{i p \ell}$ and a product of S-matrices). We only wrote down the leading correction from one mirror particle living on the bottom edge (see figure \ref{fig:mirrorpic}). An important point in this series expression is that each mirror particle come with a factor $e^{-\tilde{E}\ell}$. Since $e^{-\tilde{E}}$ is of order $g^2$, the mirror particle correction is suppressed when the bridge lengths are large, as expected. $\mathcal{H}(\alpha;v)$ and $\mathcal{H}(\bar{\alpha};v)$ are the hexagon form factors with one mirror particle at the bottom edge, and they depend on the flavors of the mirror particle (although the dependence was not written explicitly). In principle, they can be computed by utilizing the crossing/mirror transformations and collecting all the particles at the top edge, but depending on the flavor of the mirror magnon, the expression can be quite complicated in practice.  

\begin{figure}[t]
\centering
\includegraphics[clip,height=4.5cm]{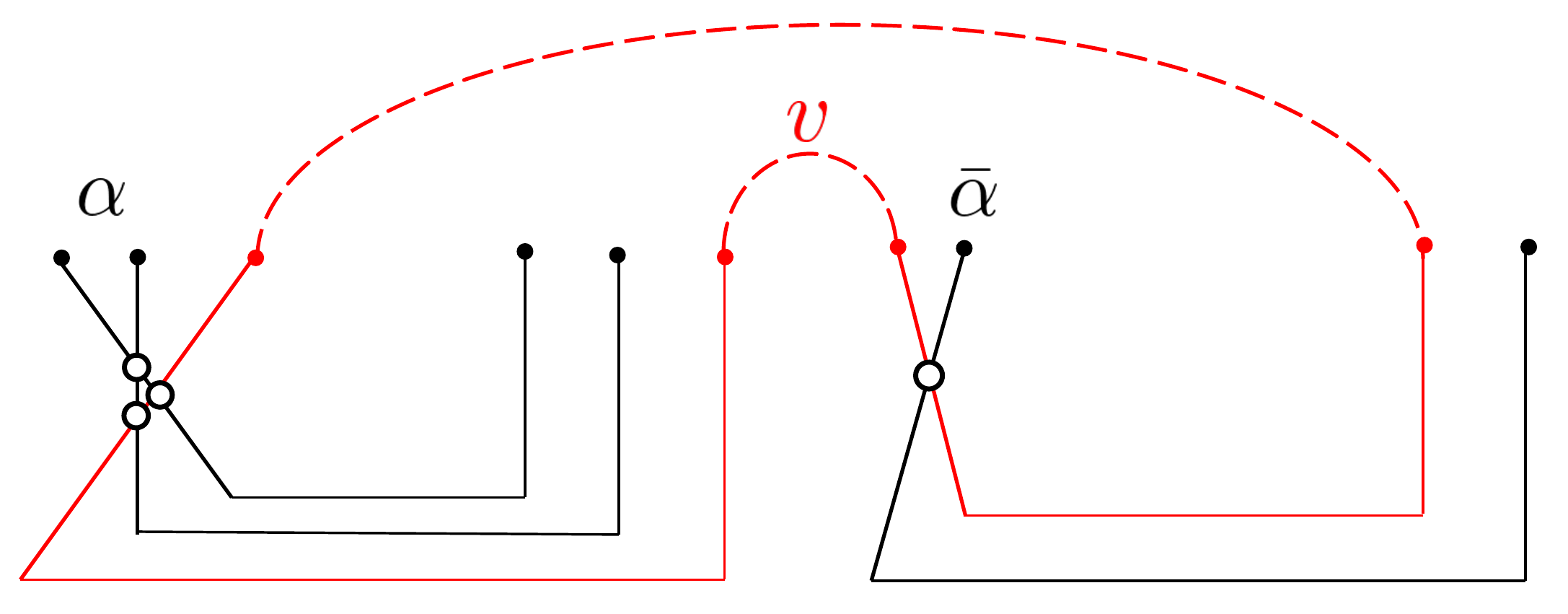}
\caption{The mirror particle corrections and the transfer matrix. The red line in the figure denotes the contribution from the mirror particle. If we sum over the flavor indices of the mirror particle, it effectively adds dashed curves shown above and transform the mirror-particle constribution into a transfer matrix. \label{fig:mirrorpic2}}
\end{figure}

However, something nice happens when we sum over the flavor indices. As shown in figure \ref{fig:mirrorpic2}, if we sum over the flavor indices of the mirror magnon, the line corresponding to the mirror particle forms a loop in the pictorial representation of the hexagon form factor. This means that the matrix part can be simply expressed as a transfer matrix. Since the transfer matrix is a well-studied object, this feature greatly simplifies the actual computation. In fact, utilizing such observations, the hexagon formalism was tested at two \cite{BKV}, three \cite{Gluing,ES} and four loops \cite{wrappingpaper}.

An extremely interesting, but challenging question to ask is whether we can resum such finite-size corrections. In the case of the spectrum, such resummation led to the thermodynamic Bethe ansatz, which was later transformed into an elegant formalism called the quantum spectral curve. The ultimate goal in the study of three-point functions in $\mathcal{N}=4$ SYM would be to find an analogue of the quantum spectral curve, although it is not clear at this moment whether such a notion exists or not.
\section{Conclusion and Prospects}
In these lectures, we discussed how to compute the three-point function using the integrability. Let us summarize the basic strategy we took to solve the problem:
\begin{enumerate}
\begin{shaded}
\item Perform the weak-coupling computation.
\item Identify the basic objects.
\item Constrain the 1- and 2-particle contributions using symmetries.
\item Guess the multi-particle structure (using the weak-coupling results).
\item Formulate and solve the crossing equation by imposing the decoupling of a pair of a particle and an anti-particle.
\item Include the finite-size correction using the mirror transformation.
\end{shaded}
\end{enumerate}
As it turns out, the same strategy works also for the two-point functions. In this sense, the procedure outlined above may be viewed as a universal recipe for solving $\mathcal{N}=4$ SYM using the integrability.

One point we wished to emphasize in these lectures is the importance of the gauge symmetry. In a sense, the gauge symmetry is a mixed blessing: It is, on the one hand, a source of complication since it inevitably introduces the length-changing processes to the spin-chain description. On the other hand, a seemingly simple fact that the gauge transformation is always accompanied by the coupling constant proves to be very powerful when combined with educated guesses, and allows us to solve the problem completely.

To conclude the lectures, let us indicate several future directions. Firstly it would be interesting to study other observables following the strategy summarized above. For instance, a similar idea may be applicable to the one-point function in the presence of the domain wall, which is discussed recently in \cite{one-point}. Another interesting direction is to study the non-planar dilatation operator. In the so-called BMN limit, it was shown that there exists a relation between the matrix element of the non-planar dilatation operator and the string vertex \cite{GMR,GMP}. Given that the string vertex is studied recently using the integrability \cite{BJ}, it would be worthwhile to revisit this problem now using the full machinery of integrability. 

Secondly, as mentioned in the second lecture the building blocks of our construction can be interpreted as the form factor of the twist operator. A similar idea of using the twist operator for studying the string interaction showed up previously in the IR limit of the matrix string theory\cite{Motl,DVV}. Since the matrix string theory describes the type IIA string theory whereas the theory discussed here corresponds to the type IIB string theory, we do not expect direct physical connection between the two. Nevertheless the intriguing similarity certainly calls for further investigation. 

Thirdly some of the ideas discussed in these lectures, such as the idea of constraining the quantities using the combination of the global symmetry and the ``large gauge" symmetry, do not rely strongly on the existence of the integrability. It would thus be interesting to see how much of such ideas can be applied to a broader class of theories including non-integrable ones. 
\subsection*{Acknowledgement}
First of all,  I would like to thank the organizers of Les Houches Summer School, {\it Integrability: From statistical systems to gauge theory}, for the invitation to give lectures about the three-point functions. I also thank the participants of the school for attending the lectures and giving me feedbacks. I am indebted to Benjamin Basso and Pedro Vieira for the collaboration which led to the main result in the Lecture II. My gratitude goes also to Naoki Kiryu, Ho Tat Lam and Takuya Nishimura; the discussion with them was quite helpful for preparing the Lecture I. I would also like to thank Luc\'{i}a C\'{o}rdova, Frank Coronado and Minkyoo Kim for the careful reading and the valuable comments on the draft. Lastly, I thank the mountains in Les Houches and Chamonix for enjoyable hikes during the summer school.  This research was supported in part by Perimeter Institute for Theoretical Physics. Research at Perimeter Institute is supported by the Government of Canada through the Department of Innovation, Science and Economic Development Canada and by the Province of Ontario through the Ministry of Research, Innovation and Science.
\appendix
\section{$\mathfrak{psu}(2|2)$ S-matrix\label{ap-a}}
In this Appendix, we summarize the action of the $\mathfrak{psu}(2|2)$ S-matrix determined by Beisert \cite{dynamic, analytic}. In the expressions below, $x_{1,2}$ mean $x (u_1)$ and $x(u_2)$ while $\gamma_{i}$ denotes
\begin{align}
\gamma_i \equiv \sqrt{i (x_i^{-}-x_i^{+})}\period
\end{align}
\begin{shaded}
\begin{align}
\begin{aligned}
\text{\bf Spin chain}\qquad &\\
\hat{S}_{12}\ket{\phi^{a}_1\phi^{b}_2}&=A_{12}\ket{\phi_2^{\{a}\phi_1^{b\}}}
+B_{12}\ket{\phi_2^{[a}\phi_1^{b]}}+\frac{1}{2}C_{12}\epsilon^{ab}\epsilon_{\alpha\beta}\ket{Z^{-}\psi^{\alpha}_2\psi^{\beta}_1}\comma\\\hat{S}_{12}\ket{\psi^{\alpha}_1\psi^{\beta}_2}&=D_{12}\ket{\psi_2^{\{\alpha}\psi_1^{\beta\}}}
+E_{12}\ket{\psi_2^{[\alpha}\psi_1^{\beta]}}+\frac{1}{2}F_{12}\epsilon^{\alpha\beta}\epsilon_{ab}\ket{Z\phi_2^{a}\phi_1^{b}}\comma\\
\hat{S}_{12}\ket{\phi^{a}_1\psi^{\beta}_2}&=G_{12}\ket{\psi^{\beta}_2\phi_1^{a}}
+H_{12}\ket{\phi_2^{a}\psi^{\beta}_1}\comma\\
\hat{S}_{12}\ket{\psi^{\alpha}_1\phi^{b}_2}&=K_{12}\ket{\psi_2^{\alpha}\phi_1^{b}}
+L_{12}\ket{\phi_2^{b}\psi_1^{\alpha}}
\end{aligned}\label{spinframeS1}
\end{align}
\begin{align}
\begin{aligned}
A_{12}&=\frac{x_2^{+}-x_1^{-}}{x_{2}^{-}-x_1^{+}}\comma\\
B_{12}&=\frac{x_2^{+}-x_1^{-}}{x_{2}^{-}-x_1^{+}}\left(1-2\frac{1-1/(x_2^{-}x_1^{+})}{1-1/(x_2^{+}x_1^{+})}\frac{x_2^{-}-x_1^{-}}{x_2^{+}-x_1^{-}} \right)\comma\\
C_{12}&=\frac{2\gamma_1 \gamma_2}{ x_1^{+}x_2^{+}}\frac{1}{1-1/(x_1^{+}x_2^{+})}\frac{x_2^{-}-x_1^{-}}{x_2^{-}-x_1^{+}}\comma\\
D_{12}&=-1\comma\\
E_{12}&=-\left( 1-2\frac{1-1/(x_2^{+}x_1^{-})}{1-1/(x_2^{-}x_1^{-})}\frac{x_2^{+}-x_1^{+}}{x_2^{-}-x_1^{+}}\right)\comma\\
F_{12}&=-\frac{2(x_1^{+}-x_1^{-})(x_2^{+}-x_2^{-})}{\gamma_1\gamma_2 x_1^{-}x_2^{-}}\frac{1}{1-1/(x_1^{-}x_2^{-})}\frac{x_2^{+}-x_1^{+}}{x_2^{-}-x_1^{+}}\comma\\
G_{12}&=\frac{x_2^{+}-x_1^{+}}{x_2^{-}-x_1^{+}}\comma\\
H_{12}&=\frac{\gamma_1}{\gamma_2}\frac{x_2^{+}-x_2^{-}}{x_2^{-}-x_1^{+}}\comma\\
K_{12}&=\frac{\gamma_2}{\gamma_1}\frac{x_1^{+}-x_1^{-}}{x_2^{-}-x_1^{+}}\comma\\
L_{12}&=\frac{x_2^{-}-x_1^{-}}{x_2^{-}-x_1^{+}}\period
\end{aligned}
\end{align}
\end{shaded}
\noindent Note that the S-matrix presented here is the S-matrix for a single $\mathfrak{psu}(2|2)$. The full worldsheet S-matrix is given by a product of two such S-matrices together with the overall factor $S_0$.

\end{document}